\documentclass[12pt]{article}
\usepackage{amsmath,amssymb,array,calc,rotating,epsfig,psfrag,amscd, cite}
\usepackage{epsfig}
\usepackage{amsmath}
\makeatletter
\renewcommand\section{\@startsection {section}{1}{\z@}%
                                   {-3.5ex \@plus -1ex \@minus -.2ex}
                                   {2.3ex \@plus.2ex}%
                                   {\normalfont\large\bfseries}}
\renewcommand\subsection{\@startsection{subsection}{2}{\z@}%
                                     {-3.25ex\@plus -1ex \@minus -.2ex}%
                                     {1.5ex \@plus .2ex}%
                                     {\normalfont\bfseries}}
\makeatother

\newenvironment{acknowledgments}{%
\vskip 3.25ex
\noindent {\bf Acknowledgments}
}

\newcommand{\be}{\begin{equation}}
\newcommand{\ee}{\end{equation}}
\newcommand{\bea}{\begin{eqnarray}}
\newcommand{\eea}{\end{eqnarray}}

\newcommand{\al}{\alpha}
\renewcommand{\d}{\delta}

\newcommand{\G}{\Gamma}

\newcommand{\m}{\mu}
\newcommand{\n}{\nu}

\newcommand{\s}{\sigma}

\newcommand{\vp}{\varphi}
\newcommand{\whg}{{\widehat{g}}}
\newcommand{\whB}{{\widehat{B}}}
\newcommand{\wtf}{{\widetilde{f}}}
\newcommand{\wtH}{{\widetilde{H}}}
\newcommand{\whn}{\widehat{\nabla}}

\newcommand{\hlf}{\frac{1}{2}}

\newcommand{\p}{\partial}
\renewcommand{\P}{\mathbb{P}}

\newcommand{\R}{\mathbb{R}}
\newcommand{\rr}{\rightarrow}

\newcommand{\w}{\wedge}
\newcommand{\Z}{\mathbb{Z}}

\newcommand{\tr}{\operatorname{tr}}

\newcommand{\U}{\operatorname{U}}

\newcommand{\lp}{\left(}
\newcommand{\rp}{\right)}
\newcommand{\ls}{\left[}
\newcommand{\rs}{\right]}

\newcommand{\hph}[1]{{\hphantom{#1}}}
\newcommand{\non}{\nonumber}

\begin{document}
\begin{titlepage}

\begin{center}

\today
\hfill         MIFPA-14-03\phantom{xxx}

\vskip 2 cm
{\Large \bf Higher Derivative Corrections to O-plane Actions: NS-NS Sector}\\
\vskip 1.25 cm { Daniel Robbins\footnote{email address:
D.G.Robbins@uva.nl}$^{a}$ and Zhao Wang\footnote{email
address:
 dsuym@tamu.edu}}$^{b}$\\
{\vskip 0.5cm $^{a}$ \it Institute for Theoretical Physics, University of Amsterdam, \\ Science Park 904, Postbus 94485, 1090 GL, Amsterdam, The Netherlands \\}

{\vskip 0.5cm $^{b}$ \it Depatment of Physics, Texas A{\&}M University, \\ College Station, TX 77843, USA \\}

\end{center}
\vskip 2 cm

\begin{abstract}
\baselineskip=18pt
We classify all possible two- and four-derivative couplings of bulk NS-NS sector fields to a single O$p$-plane which are compatible with diffeomorphism invariance and $B$-field gauge invariance.  This is applicable to type IIA or IIB superstrings or to the bosonic string.  We then consider this general action in various classes of backgrounds that admit a $U(1)$ isometry and determine the constraints on the couplings from consistency with T-duality.  We show that this consistency requires the two-derivative action to vanish, and the entire non-linear four-derivative action is fixed up to one overall constant which can be determined by comparison with a two-point scattering amplitude.  The resulting action is consistent with all previously computed couplings.

\end{abstract}

\end{titlepage}

\pagestyle{plain}
\baselineskip=19pt
\section{Introduction}

One of the most important tools used in exploring string theory, its vacuum structure, and its dynamics, is the low-energy effective action.  For many purposes, it is enough to use only the lowest order pieces in this action, but sometimes it turns out that we need to go to higher orders, either in a derivative expansion ($\al'$ expansion), or in the string coupling ($g_s$ expansion).  In fact, there are situations where the higher order terms are crucial to correctly determine the vacuum structure.

For example, consider M-theory on $\R^{1,2}\times X$, where $X$ is a Calabi-Yau four-fold with a Ricci-flat metric.  This is certainly a valid solution of eleven-dimensional supergravity, which is the leading part of the low-energy effective action of M-theory.  However, once one also incorporates the leading (eight-derivative) corrections to the effective action~\cite{Becker:1996gj}, then it is no longer a solution, and in fact there is a topological obstruction (unless $X$ has vanishing Euler number, $\chi(X)=0$).  To find solutions, we must include internal fluxes or space-filling M$2$-branes.

If $X$ is elliptically fibered with a section, then there is a dual IIB compactification to four dimensions on the base $B$ of the fibration with D$7$-branes and O$7$-planes located at points where the fiber degenerates~\cite{Dasgupta:1999ss}.  In this situation, the topological obstruction arises from higher-derivative corrections that are localized on the D$7$-branes and O$7$-planes and that have the form (neglecting an order one dimensionless coefficient)
\be
\label{eq:D7AnomalyCoupling}
T_7(\al')^2\int_{D7/O7}C_4\w\ls\tr\lp R_T\w R_T\rp-\tr\lp R_N\w R_N\rp\rs,
\ee
where $R_T$ and $R_N$ are the tangent and normal curvature two-forms on the brane, and $T_7$ is the tension.  This gives a contribution to the tadpole for $C_4$ in the $\R^{1,3}$ directions arising from integrating the couplings above over the four-cycle in $B$ wrapped by the D$7$s and O$7$s.  What this teaches us is the importance of understanding the leading order higher-derivative corrections to effective actions, including those that are localized on D-branes or O-planes.

Note that this coupling is only one piece of the full action at this order in derivatives.  In more general backgrounds, one expects that additional couplings involving $H$-flux and other fields will be important, and may in fact lead to induced charges like in the situation above~\cite{McOrist:2012yc,Maxfield:2013wka,MMQRSinprogress}.  In those cases, a proper understanding of the higher derivative corrections will again be crucial to correctly understand the vacuum structure.

There are many approaches which can be used to determine these corrections.  The specific couplings above were predicted using anomaly cancellation~\cite{Green:1996dd,Cheung:1997az,Scrucca:1999uz}, K-theoretic considerations~\cite{Minasian:1997mm}, and verified by direct scattering amplitude calculations~\cite{Stefanski:1998yx,Craps:1998fn}.  In the current work we will follow a different route, using constraints from T-duality to determine the full non-linear (in the bulk fields) couplings of a type II O$p$-plane to the NS-NS sector bulk fields.

There are many different perspectives available on T-duality.  On the world-sheet, it is a duality which, if one of the world-sheet scalars is compact, exchanges Neumann boundary conditions with Dirichlet boundary conditions, and exchanges momentum modes with winding modes.  In the target space, where we will be focusing, T-duality arises for backgrounds that admit a $\U(1)$ isometry, i.e.\ a circle fibration.  Consider the sector of the low-energy theory in which no fields have dependence on the coordinate of this isometry.  If we Ka{\l}u\.za-Klein reduce on this circle, then T-duality acts as a $\Z_2$ symmetry of the reduced theory\footnote{Note that the low-energy theory does not include winding modes on the circle (these would masses that scaled like $R/\al'$, where $R$ is the radius of the circle).  By restricting to the sector with no dependence on the circle coordinate we are also dropping the momentum modes, which is why T-duality can act as a symmetry.}.  Of course, since this is only a $\Z_2$ symmetry, there are many potential couplings of the reduced theory fields which would be invariant, obtained by simply adding a candidate coupling together with its image under T-duality.  However, we have the additional information that the theory has been reduced from a covariant, gauge-invariant theory in one dimension higher.  It is the combination of this knowledge with T-duality invariance which is surprisingly powerful.

Thus to use T-duality to constrain the leading order\footnote{A modified procedure could also be used to constrain the action beyond leading order, but it gets more convoluted.  The reason is that the action of T-duality itself (i.e.\ the Buscher rules) can receive corrections.  At leading order, this implies we should combine the uncorrected Buscher rules acting on the leading correction to the action with the corrections to the Buscher rules acting on the two-derivative action.  But the latter contributions will clearly be proportional to the variations of the two-derivative action with respect to the fields (since the Buscher rules act on the fields), i.e.\ the lowest-order equations of motion.  As such, their effect can be removed by a field redefinition.} higher-derivative corrections, an unsophisticated brute-force approach would be to write down all possible generally covariant, gauge-invariant couplings in the bulk theory, with arbitrary coefficients, and at the first non-vanishing order in the derivative expansion.  Next, make an ansatz that there is a $\U(1)$ isometry and reduce the theory on the circle.  This reduced theory now has a set of couplings parameterized by the coefficients of the parent theory (and in particular they are not the most general possible couplings).  Finally, demanding that T-duality is a symmetry of the reduced theory will put constraints on those couplings.

This procedure was followed for the bosonic string, or equivalently for the NS-NS sector of the superstring, for the two-derivative action in~\cite{Becker:2010ij} and in a related approach for the bosonic string to order $\al'$ in~\cite{Godazgar:2013bja} (see also~\cite{Meissner:1996sa} and~\cite{Hohm:2013jaa}).  At linearized order in the Buscher rules, some terms were obtained in the order $(\al')^3$ superstring action in~\cite{Garousi:2012yr}, and similar techniques have been recently exploited by~\cite{Liu:2013dna} to obtain some more of the type II couplings at order $(\al')^3$.  One would like to pursue the full unsophisticated brute-force approach to continue the work of these latter papers, but unfortunately this becomes quite difficult, owing to the huge number of covariant and gauge-invariant couplings which one would have to consider at eight-derivative order.  Instead, we would prefer to work in a situation where the leading corrections come in at a lower order in derivatives, like in the bosonic string example of~\cite{Godazgar:2013bja}.

Fortunately, this is the case for the actions which localize at D-branes and O-planes, for which, even in the superstring, corrections start at order $(\al')^2$, which is four derivatives in the bulk fields.  There is a complication however, since T-duality exchanges a direction along one of these localized objects with a direction transverse (for D-branes this is simply the statement above that T-duality exchanges Neumann and Dirichlet boundary conditions), in other words exchanging a $p$-brane wrapping the circle with a $(p-1)$-brane localized on the circle.  A priori it's not clear that the localized action on the former should be related in a simple way to the latter - the couplings could have explicit dependence on the brane dimension $p$.  However, it is a remarkable fact that, when written in string frame fields, all known brane couplings are universal in this sense.  We will take this as an assumption.  We consider the fact that we will find a unique four-derivative action on the O-plane, and that this action is consistent with all previously known couplings, to be a fairly strong check on this assumption.

Our procedure will be similar to that outlined above for bulk couplings.  We will write down all possible consistent (covariant, gauge-invariant) brane couplings at leading order in the derivative expansion which might mix under T-duality and assume that they have the same arbitrary coefficients (in string frame) for all $p$.  Then we will make the ansatz of a $\U(1)$ isometry in the bulk and demand that the reduced action for the $p$-brane wrapping the circle gets mapped into the action for the $(p-1)$-brane transverse to the circle.  In this way we will put constraints on our couplings.

For D-branes, even though the corrections begin at four-derivatives, the full procedure remains prohibitively difficult, because the combinations of world-volume and bulk fields, and tangent and normal indices, lead to a very large number of potential couplings.  Nonetheless, by working to linearized order in the Buscher rules and the fields, many restrictions can be put on some of the higher derivative couplings~\cite{Garousi:2009dj,Becker:2010ij,Garousi:2010rn,Becker:2011ar,McOrist:2012yc,Garousi:2013gea}.  

The situation is most tractable for O-planes.  In this case, there are no world-volume fields\footnote{We are considering a single O-plane with no coincident D-branes, not even fractional D-branes.  Most of the subtleties of O-plane taxonomy (plus or minus, tilde or no tilde) will not be relevant here.  The only information we need to use is that there are no world-volume degrees of freedom and that the orientifold projection acts on the bulk fields in the usual way as detailed in section~\ref{sec:Classifying}.}, and many couplings get removed by the orientifold projection.  We can reduce the number of couplings even further by restricting to terms with no R-R fields (the Buscher rules act linearly on R-R fields, so they will not mix couplings with different numbers of R-R fields).  This is the arena where we would like to implement our unsophisticated brute-force approach to constraining the higher-derivative corrections.

In section~\ref{sec:Classifying} and appendix~\ref{app:Classification}, we classify all possible couplings that we need to consider up to four derivatives in the bulk fields, and assign coefficients to the terms that can appear.  The next step is to reduce these couplings in the presence of a $\U(1)$ isometry.  Unfortunately, even our simplified situation can get cumbersome if we work with the most general $\U(1)$-isometry ansatz, largely because of the need to commute covariant derivatives on a general curved base of our circle fibration.  For this reason, we will consider not the most general circle bundle ansatz, but a pair of simplified classes of backgrounds.  The first class has a flat base metric and no off-diagonal components between base and fiber for either the metric or $B$-field, but allows the dilaton and circle radius to have arbitrary profile over the base.  We call this the warped product.  The second class has again a flat base metric, a constant dilaton and radius, but arbitrary off-diagonal components of the metric and $B$-field, which become a pair of vectors on the base (and are interchanged under T-duality).  We call this the twisted product.  In each case we get a set of constraints on our list of coefficients.  Neither of our two classes is broad enough to determine all the coefficients, but by combining the results from the two classes, we get our final result,
\begin{multline}
\label{eq:IntroResult}
\mathcal{L}=T_p'\frac{\pi^2\lp\al'\rp^2}{96}\sqrt{-g}e^{-\Phi}\left\{ -\nabla^a\Phi\nabla_a\Phi\nabla^b\Phi\nabla_b\Phi+2\nabla^a\Phi\nabla^b\Phi H_a^{\hph{a}ci}H_{bci}+\frac{1}{4}H^{abi}H_{ab}^{\hph{ab}j}H^{cd}_{\hph{cd}i}H_{cdj}\right.\\
\left.-\frac{1}{4}H^{abi}H_{ab}^{\hph{ab}j}H_i^{\hph{i}k\ell}H_{jk\ell}-\frac{1}{6}H^{abi}H_a^{\hph{a}cj}H_{bc}^{\hph{bc}k}H_{ijk}+\frac{1}{8}H^{abi}H_a^{\hph{a}cj}H_{b\hph{d}j}^{\hph{b}d}H_{cdi}+\frac{1}{24}H^{ijk}H_i^{\hph{i}\ell m}H_{j\ell}^{\hph{j\ell}n}H_{kmn}\right.\\
\left.+3\nabla^a\Phi\nabla_a\Phi\nabla^b_{\hph{b}b}\Phi-\nabla^{ab}\Phi H_a^{\hph{a}ci}H_{bci}-\frac{3}{2}\nabla^{ij}\Phi H^{ab}_{\hph{ab}i}H_{abj}+\hlf\nabla^{ij}\Phi H_i^{\hph{i}k\ell}H_{jk\ell}-2\nabla^a_{\hph{a}a}\Phi\nabla^b_{\hph{b}b}\Phi\right.\\
\left.+2\nabla^{ij}\Phi\nabla_{ij}\Phi+2\nabla^a\Phi\nabla_a\Phi R^{bc}_{\hph{bc}bc}-2\nabla^a\Phi\nabla^b\Phi R_{a\hph{c}bc}^{\hph{a}c}-R^{ab\hph{a}c}_{\hph{ab}a}H_b^{\hph{b}di}H_{cdi}+R^{abij}H_{ab}^{\hph{ab}k}H_{ijk}\right.\\
\left.+2R^{abij}H_{a\hph{c}i}^{\hph{a}c}H_{bcj}-\frac{3}{2}R^{ai\hph{a}j}_{\hph{ai}a}H^{bc}_{\hph{bc}i}H_{bcj}+\hlf R^{ai\hph{a}j}_{\hph{ai}a}H_i^{\hph{i}k\ell}H_{jk\ell}-2\nabla^a_{\hph{a}a}\Phi R^{bc}_{\hph{bc}bc}+4\nabla^{ij}\Phi R^a_{\hph{a}iaj}\right.\\
\left.-2R^{ab\hph{a}c}_{\hph{ab}a}R_{b\hph{d}cd}^{\hph{b}d}+R^{abcd}R_{abcd}-R^{abij}R_{abij}+2R^{ai\hph{a}j}_{\hph{ai}a}R^b_{\hph{b}ibj}+4\nabla^a\Phi H_a^{\hph{a}bi}\nabla^cH_{bci}-2\nabla^aH_a^{\hph{a}bi}\nabla^cH_{bci}\right.\\
\left.-\hlf\nabla^aH^{bci}\nabla_aH_{bci}-\frac{1}{6}\nabla^aH^{ijk}\nabla_aH_{ijk}\right\}.
\end{multline}

This is the main result of the current paper.  We also point out an alternative formulation in (\ref{eq:DilatonFreeResult}), related to (\ref{eq:IntroResult}) by a field redefinition, in which the dilaton only appears through the factor of $e^{-\Phi}$.  This result agrees with all previously known couplings.

The plan of the paper is as follows.  In section~\ref{sec:Classifying}, we classify all couplings which can appear in our orientifold plane action.  We proceed very carefully, making clear precisely where we use field redefinitions, integrations by parts, or Bianchi identities.  It is not really necessary to spell out all of these details (and most of them are in fact relegated to appendix~\ref{app:Classification}), but we do so with an eye towards computerizing the task for related calculations in future work.  Section~\ref{sec:Strategy} explains the procedure outlined above in more technical detail.  Section~\ref{sec:Warped} performs the computation for the warped product, with the horrible details appearing in appendix~\ref{app:Warped}, and section~\ref{sec:Twisted} and appendix~\ref{app:Twisted} do the same for the twisted product.  The final results and the reformulation alluded to above appear in section~\ref{sec:Combined}.

\section{Classifying allowed couplings}
\label{sec:Classifying}


In its basic construction, an orientifold plane (O$p$-plane, in the case that the world-volume is $(p+1)$-dimensional, or O-plane in general) in type II or bosonic string theory arises from a $\Z_2$ quotient of the theory combining a worldsheet orientation reversal with an involution on the spacetime manifold.  The fixed point locus of the involution is called an orientifold plane.  Away from this locus, the quotient relates fields at two different points in spacetime, and at the O-plane itself, the quotient acts as a projection on the fields which we will discuss below.  In its most elementary form, there are no perturbative degrees of freedom localized at the O-plane\footnote{There are other flavors of O$p$-planes which do host open string degrees of freedom.  These can often be thought of (at least as a guide to our intuition) as combinations of O$p$-planes and (possibly fractional) D$p$-branes.  But in this paper we shall focus on the simplest case with no localized fields living on the O$p$-plane.}.  However, there will still be interactions in the spacetime effective theory which are localized at the O-plane, and which can be captured by an action which is an integral over the orientifold world-volume of a local Lagrangian, constructed from bulk fields that have been pulled back to the world-volume.

In this section we would like to enumerate all the possible couplings that can appear in this action up to four derivatives.  We will demand consistency with general covariance, gauge invariance (for the $B$-field), and the orientifold projection.  We will take careful account of all the relations between couplings arising from integrations by parts, Bianchi identities, and field redefinitions, so that we arrive at a consistent linearly independent basis of physical couplings.

\subsection{Ingredients}
\label{subsec:Ingredients}

In this paper we will focus only on the part of the action that has no R-R fields (since T-duality acts linearly on R-R fields, it will not mix pieces of the action with different numbers of R-R fields)\footnote{In work in preparation~\cite{RWinprogress} we will repeat this exercise for terms with R-R fields.}.  We will also focus on the bosonic sector (again T-duality will not mix purely bosonic couplings with couplings that involve fermions).  As such we restrict to the NS-NS sector of type II, and the bulk fields consist only of the dilaton $\Phi$, the metric $G_{\m\n}$, and the NS-NS antisymmetric tensor $B_{\m\n}$.  We could also consider our set-up to be in a bosonic string context; the classification of couplings is the same.  However, in that case the bulk action gets corrected already at order $\al'$, and we have not been careful to keep track of the consequences of this in later sections, so we will focus primarily on type II superstrings.

To simplify our lives, we will work in local coordinates in which the involution is simply reflection in the final $D-p-1$ coordinates which we denote $x^i$, $i=p+1,\cdots,D-1$ ($D=10$ for type II, $D=26$ if we want to consider O-planes in the bosonic string theory).  This means that the orientifold is located at the point $x^i=0$, and its world-volume can be parameterized by the first $p+1$ coordinates $x^a$, $a=0,\cdots,p$.  In these local coordinates, the pull-backs of our bulk fields are simply given by restriction to $x^i=0$.  We will use $x^\m$, $\m=0,\cdots,D-1$, to denote the full set of $D$ coordinates.

Under orientation reversal, $B_{\m\n}$ changes sign, while $\Phi$ and $G_{\m\n}$ are invariant.  Combining with the involution, it means that $\Phi$, $G_{ab}$, $G_{ij}$, and $B_{ai}$ can be non-vanishing at the O-plane, while $G_{ai}$, $B_{ab}$, and $B_{ij}$ are projected out.  Furthermore, we can of course have derivatives acting on these fields, and each normal derivative brings an extra minus sign from the involution.  Thus the rule is that $\Phi$, $G_{ab}$, $G_{ij}$, and $B_{ai}$ can appear with any number of derivatives along world-volume directions and an even number of normal derivatives, while $G_{ai}$, $B_{ab}$, and $B_{ij}$ can have any number of world-volume derivatives and must carry an odd number of normal derivatives (and in particular not zero).

Now in order to ensure invariance under $B$-field gauge transformations, $\d B_{\m\n}=2\p_{[\m}\Lambda_{\n]}$, the $B$-field should only appear in the action via its field strength\footnote{One might imagine the possibility of Chern-Simons type terms, but such parity-odd terms are intrinsically dimension-dependent.  Thus, by our assumption (discussed further in section~\ref{sec:Strategy}) that the string frame couplings are the same for all $p$, these terms are disallowed.} $H=dB$, or $H_{\m\n\rho}=3\p_{[\m}B_{\n\rho]}$.  The rule for projection of $H$ is then that $H_{abi}$ and $H_{ijk}$ can appear with an even number of normal derivatives, while $H_{abc}$ and $H_{aij}$ require an odd number of normal derivatives.

Similarly, consistency with general covariance requires that all derivatives be covariant derivatives $\nabla_a$ or $\nabla_i$, and that explicit derivatives of the metric only be packaged inside of the bulk Riemann tensor.  The projection means that $R_{abcd}$, $R_{abij}$, $R_{aibj}$ can appear with even numbers of normal derivatives, while $R_{abci}$ and $R_{aijk}$ require an odd number of normal derivatives.  Additionally, each covariantly constructed coupling should be integrated with the proper world-volume measure $\sqrt{-g}$, where $g=\det(G_{ab})$ is the determinant of the pull-back of the bulk metric.

We also need to confront the fact that covariant derivatives do not commute, and any commutator of covariant derivatives can be replaced by terms involving the Riemann tensor.  To eliminate this freedom, we will use the convention that whenever more than one covariant derivative hits a field, we will only take the completely symmetrized combination of derivatives.  We will write this using a single nabla with multiple indices, so for example,
\be
\nabla_{abi}H_{cjk}:=\nabla_{(a}\nabla_b\nabla_{i)}H_{cjk}=\frac{1}{3}\nabla_{(a}\nabla_{b)}\nabla_iH_{cjk}+\frac{1}{3}\nabla_{(a}\nabla_{|i|}\nabla_{b)}H_{cjk}+\frac{1}{3}\nabla_i\nabla_{(a}\nabla_{b)}H_{cjk},
\ee
or
\be
\nabla^{a\hph{a}b}_{\hph{a}a\hph{b}b}\Phi:=\frac{1}{3}\nabla^a\nabla_a\nabla^b\nabla_b\Phi+\frac{1}{3}\nabla^a\nabla^b\nabla_a\nabla_b\Phi+\frac{1}{3}\nabla^a\nabla^b\nabla_b\nabla_a\Phi.
\ee

Finally, using basic symmetries (antisymmetry of $H_{\m\n\rho}$, symmetrization of the covariant derivatives discussed above, $R_{abcd}=-R_{abdc}=R_{cdab}$, and exchange of identical fields) we will always order the indices lexicographically when possible.  The first step in our classification is then, at a fixed derivative order (where $\Phi$ counts zero, $H$ counts one, $R$ counts two, and each extra $\nabla$ counts one more), to list all possible scalars which can be constructed using these ingredients.  Clearly we can always include an arbitrary function $f(\Phi)$ in front of our coupling, and apart from this we need only consider appearances of $\Phi$ which have been hit by at least one derivative.  Thus for each scalar we can build out of $\nabla\Phi$, $H$, $R$, and extra covariant derivatives, subject to the orientifold projections above, we have a potential coupling whose coefficient is a function of $\Phi$.  At a given derivative order there are a finite number of such couplings and we can think of them as forming a vector space $V$.  A candidate Lagrangian is specified by a vector of $\Phi$-dependent coefficients in this vector space (we will see below in section \ref{subsec:TrivialProduct} that T-duality fixes every one of these functions to be proportional to $e^{-\Phi}$, so we will only be dealing with constant vectors in coupling space times this overall function of $\Phi$).

\subsection{Redundancies}
\label{subsec:Redundancies}

Next we need to discuss the possible redundancies which reduce the number of physically independent couplings.  In other words, rather than the vector space $V$ of couplings constructed in section \ref{subsec:Ingredients}, we are interested in the vector space $U$ of physically independent couplings, which will be given by a quotient $U=V/K$, where $K$ is a subspace of $V$ spanned by combinations of couplings that are not physically relevant; i.e.\ which do not contribute to physical amplitudes.  These redundancies come from three sources: Bianchi identities, total derivatives, and bulk equations of motion.

\subsubsection{Bulk equations of motion}
\label{subsubsec:BulkEOM}

Our general perspective on the full spacetime effective theory is to consider the O-plane action as being a small perturbation to the bulk action (probe limit).  In that case, the bulk equations of motion should be taken essentially as identities for the purpose of the O-plane action, and any scalars that we can form by contracting those equations of motion with combinations of other fields and derivatives will not be physically relevant couplings, and hence will represent vectors in $K$.  Another perspective on this is that we can really imagine this action as a source of extra vertices for Feynman diagrams describing scattering of bulk fields.  Any vertices which are proportional to the lowest order equations of motion will give vanishing contributions to the amplitude in exactly the same way as they would for bulk vertices, even if the usual arguments regarding field redefinitions are no longer as clean (since they would seem to require redefinitions which were localized on the O-plane).

Let us recall what the (string frame) equations of motion for the NS-NS fields in type II,
\bea
0 &=& R+4\nabla^\m_{\hph{\m}\m}\Phi-4\nabla^\m\Phi\nabla_\m\Phi-\frac{1}{12}H^{\m\n\rho}H_{\m\n\rho}+\cdots,\\
0 &=& R_{\m\n}+2\nabla_{\m\n}\Phi-\frac{1}{4}H_\m^{\hph{\m}\rho\s}H_{\n\rho\s}+\cdots,\\
0 &=& \nabla^\rho H_{\m\n\rho}-2\nabla^\rho\Phi H_{\m\n\rho}+\cdots.
\eea
Here $\cdots$ represent terms involving the R-R fields, as well as higher derivative corrections starting at order $(\al')^3$.

There are many ways we could choose to eliminate this redundancy.  For reasons that we will discuss in section \ref{sec:Strategy} below, our choice will be to eliminate any coupling in which two normal indices are contracted within a single field (including the derivatives acting on that field).  In other words, we will use
\bea
\nabla^i_{\hph{i}i}\Phi &=& 2\nabla^a\Phi\nabla_a\Phi-\frac{1}{4}H^{abi}H_{abi}-\frac{1}{12}H^{ijk}H_{ijk}-\nabla^a_{\hph{a}a}\Phi,\\
R_{a\hph{i}bi}^{\hph{a}i} &=& \hlf H_a^{\hph{a}ci}H_{bci}-2\nabla_{ab}\Phi-R_{a\hph{c}bc}^{\hph{a}c},\\
R_{i\hph{k}jk}^{\hph{i}k} &=& \frac{1}{4}H^{ab}_{\hph{ab}i}H_{abj}+\frac{1}{4}H_i^{\hph{i}k\ell}H_{jk\ell}-2\nabla_{ij}\Phi-R^a_{\hph{a}iaj},\\
\nabla^jH_{aij} &=& -2\nabla^b\Phi H_{abi}+\nabla^bH_{abi}.
\eea
Note that we have made use of the projections to eliminate certain terms, and that we have dropped the extra $\cdots$ terms from the equations of motion.  Note also that, through the use of Bianchi identities we can do something similar for any expression that involves contraction of normal indices within a field.  For example,
\be
\nabla^iR_{abci}=-\nabla_aR_{b\hph{i}ci}^{\hph{b}i}+\nabla_bR_{a\hph{i}ci}^{\hph{a}i},
\ee
and we can then rewrite the right hand side using the previous expressions.

\subsubsection{Bianchi identities}

Some combinations that don't contribute come simply from Bianchi identities which might have caused us to overcount the number of terms.  For instance, from the definition of $H_{\m\n\rho}$ in terms of $B_{\m\n}$, it follows that $dH=0$, i.e.\ that
\be
\nabla_{[\m}H_{\n\rho\s]}=0.
\ee
This means that although we might have, in a preliminary enumeration of terms, included separately couplings
\be
\nabla^aH^{bci}\nabla_aH_{bci},\qquad\nabla^aH^{bci}\nabla_bH_{aci},\qquad\mathrm{and}\qquad\nabla^aH^{bci}\nabla_iH_{abc},
\ee
the Bianchi identity means that the combination
\be
4\nabla^aH^{bci}\nabla_{[a}H_{bci]}=\nabla^aH^{bci}\nabla_aH_{bci}-2\nabla^aH^{bci}\nabla_bH_{aci}-\nabla^aH^{bci}\nabla_iH_{abc},
\ee
vanishes and hence sits in $K$.

Similar considerations apply to the two types of Bianchi identity obeyed by the Riemann tensor,
\be
R_{[\m\n\rho]\s}=0,\qquad\mathrm{and}\qquad\nabla_{[\m}R_{\n\rho]\s\tau}=0.
\ee
Any of these three Bianchi identities ($\nabla H$, $R$, and $\nabla R$) can be contracted with other fields or derivatives, including potentially derivatives acting on the Bianchi identity itself (for example $H^{abi}\nabla^c\nabla_{[a}H_{bci]}=0$) to get a scalar, and the resulting combinations of couplings will all be vectors in $K$.

\subsubsection{Total derivatives}

Similarly, any combinations of couplings which is a total divergence on the world-volume will correspond to a vector in $K$.  In other words, any combination of couplings that can be written in the form
\be
\p_a\lp\sqrt{-g}\chi^a\rp=\sqrt{-g}\nabla_a\chi^a,
\ee
for any vector $\chi^a$ constructed from the fields and derivatives will be in the subspace $K$.

We will follow the strategy of eliminating the couplings described in \ref{subsubsec:BulkEOM} by hand, and we will use $V$ to refer only to the space of remaining couplings.  Then the subspace $K$ will be given by the span of all vectors arising from Bianchi identities and total derivatives.

As an example, if we are considering only two derivative couplings, then we would need to find all possible combinations of fields with one free world-volume index, and which is first order in derivatives.  Since the Riemann tensor starts at second order in derivatives, and since there is no way to contract the indices of an $H$-field appropriately, the only possibility is
\be
\chi^a=f(\Phi)\nabla^a\Phi,
\ee
where $f(\Phi)$ is an arbitrary function of $\Phi$.

\subsection{Lexicography}
\label{subsec:Lexicography}

To facilitate comparisons, it will be necessary to have an explicit ordering, to ensure that we always write terms and expressions in the same way.  To this end, we will make use of the following rules that give an unambiguous (though certainly not canonical) ordering of the couplings which we can construct.

Couplings\footnote{In this section and almost all the rest of the paper, except where noted, we have already used the bulk equations of motion to remove any couplings in which two normal indices are contracted within a single field and its derivatives.} (i.e.\ vectors in $V$) are built from linear combinations of monomials, which in turn are made up of a product of fields and derivatives, which we call letters, subject to the orientifold projections, and whose indices are completely contracted to make a scalar.

To order these monomials, we first put an order on the letters.  We order them first by derivative order, and at a given derivative order we list $\Phi$ first, then $R$, then $H$.  In other words, the ordered list of possible letters is
\be
\nabla\Phi,H,\nabla^2\Phi,R,\nabla H,\nabla^3\Phi,\nabla R,\nabla^2H,\cdots,\nabla^n\Phi,\nabla^{n-2}R,\nabla^{n-1}H,\nabla^{n+1}\Phi,\cdots.
\ee
This ordering corresponds roughly to the complexity of the resulting expressions that come when we reduce in a circle bundle background.  For aesthetic reasons, within a monomial we will write all the $\Phi$ letters first, in increasing derivative order, then all the $R$ letters, then all the $H$ letters.

Now to compare two different monomials, we will first compare their largest letters.  If one has a letter that is larger than the other, then it will appear later in our list.  In case of a tie, we proceed to compare the next largest letters, and so on.  Thus, schematically (i.e.\ before worrying about possible distributions of indices and contractions), the full ordered list of two derivative monomials is
\be
\lp\nabla\Phi\rp^2,H^2,\nabla^2\Phi,R.
\ee
At four derivatives, the analogous ordered list is
\begin{multline}
\label{eq:FourDSchematicList}
\lp\nabla\Phi\rp^4,\lp\nabla\Phi\rp^2H^2,H^4,\lp\nabla\Phi\rp^2\nabla^2\Phi,\nabla^2\Phi H^2,\lp\nabla^2\Phi\rp^2,\lp\nabla\Phi\rp^2R,RH^2,\nabla^2\Phi R,R^2,\\
\nabla\Phi H\nabla H,\lp\nabla H\rp^2,\nabla\Phi\nabla^3\Phi,\nabla\Phi\nabla R,H\nabla^2H,\nabla^4\Phi,\nabla^2R.
\end{multline}

Next we must turn to the distribution of indices.  We first write down all the possible assignments of world-volume and normal indices which is consistent with the orientifold projection.  

For example, consider terms which are schematically $\nabla^2\Phi H^2$.  Using $A$ to represent a world-volume index and $I$ to represent a normal index, the possibilities consistent with the projection are
\begin{multline}
\label{eq:ExampleList1}
\nabla^{AA}\Phi H^{AAI}H^{AAI},\nabla^{AA}\Phi H^{AAI}H^{III},\nabla^{AA}\Phi H^{III}H^{III},\\
\nabla^{II}\Phi H^{AAI}H^{AAI},\nabla^{II}\Phi H^{AAI}H^{III},\nabla^{II}\Phi H^{III}H^{III}.
\end{multline}
Take the first case, $\nabla^{AA}\Phi H^{AAI}H^{AAI}$.  We have three pairs of world-volume indices and one pair of normal indices.  Without taking account of symmetries, there are fifteen ways of doing the world-volume contractions and one way of doing the normal index contraction:
\begin{multline}
\nabla^a_{\hph{a}a}\Phi H^{b\hph{b}i}_{\hph{b}b} H^c_{\hph{c}ci},\quad\nabla^a_{\hph{a}a}\Phi H^{bci}H_{bci},\quad\nabla^a_{\hph{a}a}\Phi H^{bci}H_{cbi},\quad\nabla^{ab}\Phi H_{ab}^{\hph{ab}i}H^c_{\hph{c}ci},\quad\nabla^{ab}\Phi H_a^{\hph{a}ci}H_{bci},\\
\nabla^{ab}\Phi H_a^{\hph{a}ci}H_{cbi},\quad\nabla^{ab}\Phi H_{ba}^{\hph{ba}i}H^c_{\hph{c}ci},\quad\nabla^{ab}\Phi H_b^{\hph{b}ci}H_{aci},\quad\nabla^{ab}\Phi H_b^{\hph{b}ci}H_{cai},\quad\nabla^{ab}\Phi H^{c\hph{a}i}_{\hph{c}a}H_{bci},\\
\nabla^{ab}\Phi H^{c\hph{a}i}_{\hph{c}a}H_{cbi},\quad\nabla^{ab}\Phi H^{c\hph{b}i}_{\hph{c}b}H_{aci},\quad\nabla^{ab}\Phi H^{c\hph{b}i}_{\hph{c}b}H_{cai},\quad\nabla^{ab}\Phi H^{c\hph{c}i}_{\hph{c}c}H_{abi},\quad\nabla^{ab}\Phi H^{c\hph{c}i}_{\hph{c}c}H_{bai}.
\end{multline}
Now we take symmetries into account, namely that the indices of $H$ are all antisymmetric and the covariant derivatives acting on $\Phi$ are symmetric.  We can also use the fact that interchanging the two $H$'s is a symmetric operation as well.  For each term, we can look at all of its images under these symmetries, relabeling the dummy indices into lexicographic order.  In some cases, the starting term will appear again among the images, but with a minus sign from antisymmetry, thus indicating that the term is in fact zero.  For instance, in the list above, this eliminates the first, fourth, seventh, fourteenth, and fifteenth terms.  The remaining terms will fall into orbits of the symmetry group.  In the list above, there are two such orbits - one of order two comprising the second and third terms, and another of order eight comprising the remaining ones (fifth, sixth, eighth, ninth, tenth, eleventh, twelfth, and thirteenth).  From each orbit we will select the representative with the lexicographically earliest distribution of indices, read from left to right.  So in the case at hand, we would select
\be
\nabla^a_{\hph{a}a}\Phi H^{bci}H_{bci},\qquad\mathrm{and}\qquad\nabla^{ab}\Phi H_a^{\hph{a}ci}H_{bci}.
\ee

Repeating that exercise for the other possibilities in (\ref{eq:ExampleList1}), we extract nothing from the second and fifth entries on the list, while from the others we find one orbit each, selecting terms
\be
\nabla^a_{\hph{a}a}\Phi H^{ijk}H_{ijk},\quad\nabla^{ij}\Phi H^{ab}_{\hph{ab}i}H_{abj},\quad\nabla^{ij}\Phi H_i^{\hph{i}k\ell}H_{jk\ell}.
\ee
Note that we remove by hand possibilities such as
\be
\nabla^i_{\hph{i}i}\Phi H^{abj}H_{abj},
\ee
that include contractions of normal indices within a field.  The five surviving terms, again in the order corresponding to lexicographic distribution of indices, are the ones which appear in the list in appendix \ref{app:Classification}.  By repeating this with each of the structures in (\ref{eq:FourDSchematicList}), we generate the full list of terms in the appendix.

\subsection{List}
\label{subsec:List}

For two derivative terms, the possible terms we can write down are
\be
\nabla^a\Phi\nabla_a\Phi,\qquad H^{abi}H_{abi},\qquad H^{ijk}H_{ijk},\qquad\nabla^a_{\hph{a}a}\Phi,\qquad R^{ab}_{\hph{ab}ab}.
\ee
There are no Bianchi identities to worry about in this case (they all start at least at two derivatives and are not scalars), but there is one term which can be removed by integration by parts, since
\be
\nabla^a\lp\nabla_a\Phi\rp=\nabla^a_{\hph{a}a}\Phi.
\ee
So the space of physical couplings at two derivatives consists of four terms, each of which can have an arbitrary function of $\Phi$,
\be
S_2=\int d^{p+1}x\sqrt{-g}\ls f_1(\Phi)\nabla^a\Phi\nabla_a\Phi+f_2(\Phi)H^{abi}H_{abi}+f_3(\Phi)H^{ijk}H_{ijk}+f_4(\Phi)R^{ab}_{\hph{ab}ab}\rs.
\ee

In appendix \ref{app:Classification} we repeat this exercise for the four derivative action.  Instead of four physical couplings, we find after eliminating redundancies due to Bianchi identities and total derivatives, forty-eight couplings,
\begin{itemize}
\item $f_1(\Phi)\nabla^a\Phi\nabla_a\Phi\nabla^b\Phi\nabla_b\Phi$,
\item $f_2(\Phi)\nabla^a\Phi\nabla_a\Phi H^{bci}H_{bci}$, $f_3(\Phi)\nabla^a\Phi\nabla_a\Phi H^{ijk}H_{ijk}$, $f_4(\Phi)\nabla^a\Phi\nabla^b\Phi H_a^{\hph{a}ci}H_{bci}$,
\item $f_5(\Phi)H^{abi}H_{abi}H^{cdj}H_{cdj}$, $f_6(\Phi)H^{abi}H_{abi}H^{jk\ell}H_{jk\ell}$, $f_7(\Phi)H^{abi}H_{ab}^{\hph{ab}j}H^{cd}_{\hph{cd}i}H_{cdj}$,\\ $f_8(\Phi)H^{abi}H_{ab}^{\hph{ab}j}H_i^{\hph{i}k\ell}H_{jk\ell}$, $f_9(\Phi)H^{abi}H_{a\hph{c}i}^{\hph{a}c}H_b^{\hph{b}dj}H_{cdj}$, $f_{10}(\Phi)H^{abi}H_a^{\hph{a}cj}H_{bc}^{\hph{bc}k}H_{ijk}$,\\ $f_{11}(\Phi)H^{abi}H_a^{\hph{a}cj}H_{b\hph{d}j}^{\hph{b}d}H_{cdi}$, $f_{12}(\Phi)H^{ijk}H_{ijk}H^{\ell mn}H_{\ell mn}$, $f_{13}(\Phi)H^{ijk}H_{ij}^{\hph{ij}\ell}H_k^{\hph{k}mn}H_{\ell mn}$,\\ $f_{14}(\Phi)H^{ijk}H_i^{\hph{i}\ell m}H_{j\ell}^{\hph{j\ell}n}H_{kmn}$,
\item $f_{15}(\Phi)\nabla^a\Phi\nabla_a\Phi\nabla^b_{\hph{b}b}\Phi$,
\item $f_{16}(\Phi)\nabla^a_{\hph{a}a}\Phi H^{bci}H_{bci}$, $f_{17}(\Phi)\nabla^a_{\hph{a}a}\Phi H^{ijk}H_{ijk}$, $f_{18}(\Phi)\nabla^{ab}\Phi H_a^{\hph{a}ci}H_{bci}$, $f_{19}(\Phi)\nabla^{ij}\Phi H^{ab}_{\hph{ab}i}H_{abj}$, $f_{20}(\Phi)\nabla^{ij}\Phi H_i^{\hph{i}k\ell}H_{jk\ell}$,
\item $f_{21}(\Phi)\nabla^a_{\hph{a}a}\Phi\nabla^b_{\hph{b}b}\Phi$, $f_{22}(\Phi)\nabla^{ij}\Phi\nabla_{ij}\Phi$,
\item $f_{23}(\Phi)\nabla^a\Phi\nabla_a\Phi R^{bc}_{\hph{bc}bc}$, $f_{24}(\Phi)\nabla^a\Phi\nabla^b\Phi R_{a\hph{c}bc}^{\hph{a}c}$,
\item $f_{25}(\Phi)R^{ab}_{\hph{ab}ab}H^{cdi}H_{cdi}$, $f_{26}(\Phi)R^{ab}_{\hph{ab}ab}H^{ijk}H_{ijk}$, $f_{27}(\Phi)R^{ab\hph{a}c}_{\hph{ab}a}H_b^{\hph{b}di}H_{cdi}$, $f_{28}(\Phi)R^{abcd}H_{ab}^{\hph{ab}i}H_{cdi}$, $f_{29}(\Phi)R^{abij}H_{ab}^{\hph{ab}k}H_{ijk}$, $f_{30}(\Phi)R^{abij}H_{a\hph{c}i}^{\hph{a}c}H_{bcj}$, $f_{31}(\Phi)R^{ai\hph{a}j}_{\hph{ai}a}H^{bc}_{\hph{bc}i}H_{bcj}$, $f_{32}(\Phi)R^{ai\hph{a}j}_{\hph{ai}a}H_i^{\hph{i}k\ell}H_{jk\ell}$, $f_{33}(\Phi)R^{aibj}H_{a\hph{c}i}^{\hph{a}c}H_{bcj}$, $f_{34}(\Phi)R^{ijk\ell}H_{ij}^{\hph{ij}m}H_{k\ell m}$,
\item $f_{35}(\Phi)\nabla^a_{\hph{a}a}\Phi R^{bc}_{\hph{bc}bc}$, $f_{36}(\Phi)\nabla^{ij}\Phi R^a_{\hph{a}iaj}$,
\item $f_{37}(\Phi)R^{ab}_{\hph{ab}ab}R^{cd}_{\hph{cd}cd}$, $f_{38}(\Phi)R^{ab\hph{a}c}_{\hph{ab}a}R_{b\hph{d}cd}^{\hph{b}d}$, $f_{39}(\Phi)R^{abcd}R_{abcd}$, $f_{40}(\Phi)R^{abij}R_{abij}$,\\ $f_{41}(\Phi)R^{ai\hph{a}j}_{\hph{ai}a}R^b_{\hph{b}ibj}$, $f_{42}(\Phi)R^{aibj}R_{aibj}$, $f_{43}(\Phi)R^{ijk\ell}R_{ijk\ell}$,
\item $f_{44}(\Phi)\nabla^a\Phi H_a^{\hph{a}bi}\nabla^cH_{bci}$,
\item $f_{45}(\Phi)\nabla^aH_a^{\hph{a}bi}\nabla^cH_{bci}$, $f_{46}(\Phi)\nabla^aH^{bci}\nabla_aH_{bci}$, $f_{47}(\Phi)\nabla^aH^{ijk}\nabla_aH_{ijk}$,\\ $f_{48}(\Phi)\nabla^iH^{ajk}\nabla_iH_{ajk}$.
\end{itemize}

\section{Strategy}
\label{sec:Strategy}

T-duality can be characterized in many different ways in string theory, either from a world-sheet perspective or a target space perspective.  In this paper, we emphasize the latter point of view.  For our purposes, T-duality is a process which takes as input a solution to the low-energy effective theory of string theory which admits a $\U(1)$ isometry, and generates a new solution which also admits a $\U(1)$ isometry.  The mapping between the two solutions is provided by the Buscher rules~\cite{Buscher:1987sk}.

Equivalently, in the presence of a $\U(1)$ isometry, we can dimensionally reduce the low energy theory to obtain a new theory in one fewer dimension.  Then T-duality, as encoded by the Buscher rules, should act as a symmetry of this reduced theory.

If we were trying to constrain the higher derivative corrections to the bulk action, this describes precisely how we could proceed.  First, we would parameterize all of the possible physically independent couplings which could arise.  Then we would then make an assumption of a $\U(1)$ isometry and we would dimensionally reduce our theory; the couplings parameterizing the corrections to the higher dimensional theory would map into couplings of the reduced theory.  Finally, we would demand that the reduced theory is symmetric under application of T-duality, thus constraining the couplings.  One might be concerned that the Buscher rules themselves get corrections at a given order in the derivative expansion, but at leading order, such corrections won't matter; the extra terms that would result would always be proportional to the leading order equations of motion, and hence will not affect the space of physical couplings.  At higher orders this will no longer be true, and modifications to the Buscher rules may become important.

In the presence of localized sources such as D-branes or O-planes, the story changes somewhat.  We will focus on the case of O-planes, leaving the analysis with D-branes for future work.  We will be working in the probe limit, in which we are given a bulk solution that admits an orientifold involution, and we wish to know the form of the action localized at the resulting orientifold plane, without worrying about any backreaction effects.  Now, if the bulk solution also admits a $\U(1)$ isometry, then we can apply T-duality.  If the orientifold involution acts as a reflection on the isometry direction (so that the O-plane is localized on the T-duality circle), then T-duality will generate a solution in which the isometry direction is invariant under the orientifold involution (so that the O-plane wraps the circle), and vice versa.  Thus an O$p$-plane wrapping the circle gets mapped to an O$(p-1)$-plane transverse to the circle.

A key assumption that we will be making is that the string frame action localized to the orientifold plane is independent of the dimension $p$ of the O$p$-plane.  Though this seems like a strong assumption, it holds for all known couplings, both leading order and higher derivative\footnote{Note, however, that it does not hold for couplings written in Einstein frame.  For example, in Einstein frame the leading order dilaton couplings on D-branes or O-planes are all proportional to $(p-3)$, and hence the dilaton decouples from the action on a D3-brane or O3-plane, but this is not true in string frame.}.  We shall also see that the current work provides a solid test of this assumption, since the couplings we will derive will pass several consistency checks.

Given this assumption, we can imagine performing the following procedure.  We first enumerate and parameterize all the possible physical couplings which could correct the O-plane action at a given order.  Then, making an ansatz of an isometry along the O-plane world-volume, we can dimensionally reduce to get a new action in terms of our parameters.  On the other hand, we can make an ansatz of a bulk isometry transverse to the O-plane and again perform a dimensional reduction.  The Buscher rules should then map one reduced action into the other.  Since both actions are written in terms of the same parameters, this will constrain the possible couplings.

Though straight-forward in principle, this procedure can be difficult to implement in practice.  The first hurdle is in enumerating the possible couplings, but we have actually accomplished that for the NS-NS sector of O-plane actions already in section~\ref{sec:Classifying}.  Our lives were simplified by the fact that the leading corrections appear already at four-derivative order (contrast this with the corrections to the type II bulk actions, which do not arise until eight derivatives), and by the fact that the orientifold projection effectively halves the number of allowed fields.  The second source of difficulty comes from implementing the dimensional reduction for a general background with $\U(1)$ isometry.  In particular, if the base of the circle fibration is curved, then one has to be very careful with commuting covariant derivatives in the reduced theory, which makes comparing terms potentially quite tedious.

To elide the second difficulty, we will follow a slightly lazier procedure.  Rather than reduce the theory in the most general background admitting an isometry, we will reduce the action in various simplified backgrounds.  In each case we will get a set of constraints on our parameters that will not be the most general constraints, but by combining this procedure on different backgrounds, we will find that the constraints are, in fact, sufficient to reduce the allowed corrections to a single parameter (which can be thought of as $\al'$).

A key point regarding this strategy is that it was essential that we chose to use the bulk equations of motion in such a way that cleanly divided the space of possible couplings in two, and that in particular, the subspace of physically irrelevant couplings generated by total derivatives and Bianchi identities did not mix these two sets of couplings.  This means that in the reduced theory we again only have to worry about total derivatives and Bianchi identities, and not about equations of motion.  If we had to include the latter, we would lose a lot of information, since in our simplified backgrounds, solving the equations of motion is very restrictive (for instance there are essentially no non-trivial solutions of the Einstein equation for a warped product of a circle and flat space).

\subsection{Generalities}
\label{subsec:Generalities}

A general background with a $\U(1)$ isometry can always be put into the following form\footnote{In this section, capital letters $M$, $N$ represent ten-dimensional indices, while $\m$ and $\n$ represent the nine-dimensional base of the circle fibration, which will in turn be separated into $a$, $b$, etc.\ for indices parallel to the O-plane, and $i$, $j$, etc.\ for indices perpendicular to the O-plane.  The isometry direction is always denoted by $y$.},
\be
g_{MN}=\lp\begin{matrix} \whg_{\m\n}+e^\vp a_\m a_\n & e^\vp a_\n \\ e^\vp a_\m & e^\vp\end{matrix}\rp,\qquad g^{MN}=\lp\begin{matrix} \whg^{\m\n} & -a^\n \\ -a^\m & e^{-\vp}+a^\rho a_\rho\end{matrix}\rp.
\ee
\be
B_{\m\n}=\whB_{\m\n}-\hlf a_\m b_\n+\hlf a_\n b_\m,\qquad B_{\m y}=b_\m,
\ee
where we have split our space into a circle parameterized by $y$ fibered over a base with coordinates $x^\m$.  In other words, our nine-dimensional fields are encoded by a base metric $\whg_{\m\n}$, a base $B$-field $B_{\m\n}$, two vectors $a_\m$ and $b_\m$, and two scalars $\Phi$ and $\vp$.  Note that $\m$ and $\n$ indices are raised and lowered using $\whg_{\m\n}$.  The isometry means that nothing depends on the coordinate $y$, only on the base coordinates $x^\m$.

By restricting to ten-dimensional diffeomorphisms and $B$-field gauge transformations that preserve our isometry (i.e.\ the gauge parameters are independent of $y$), we generate diffeomorphisms and $B$-field transfomations of $\whg$ and $\whB$ on the base, as well as gauge transformations of the vectors $a_\m$ and $b_\m$ (generated by ten-dimensional diffeomorphisms $\xi^y(x^\m)$ and $B$-field gauge parameters $\Lambda_y(x^\m)$ respectively).  Any covariant scalar couplings of the ten-dimension fields, when written in terms of the base fields, must be invariant under these gauge transformations, so should only depend on the field strengths
\be
f_{\m\n}=2\p_{[\m}a_{\n]},\qquad\wtf_{\m\n}=2\p_{[\m}b_{\n]}.
\ee
Note however that the field $\whB_{\m\n}$, as well as its naive field strength $\widehat{H}=d\widehat{B}$, are not invariant under these gauge transformations.  For the $B$-field potential, this is simply an unavoidable tradeoff; the decomposition of $B_{MN}$ which has nice behavior under T-duality is not invariant under these gauge transformations.  For the field strength, however, there is a fix.  We can define
\be
\wtH_{\m\n\rho}=3\p_{[\m}\widehat{B}_{\n\rho]}-\frac{3}{2}a_{[\m}\wtf_{\n\rho]}-\frac{3}{2}b_{[\m}f_{\n\rho]},
\ee
or
\be
H_{\m\n\rho}=\wtH_{\m\n\rho}+3a_{[\m}\wtf_{\n\rho]},\qquad H_{\m\n y}=\wtf_{\m\n}.
\ee
With this definition, $\wtH$ is invariant under gauge transformations of $b_\m$, as well as gauge transformations of $\widehat{B}_{\m\n}$.  It is also invariant under gauge transformations of $a_\m$, since under a ten-dimensional diffeomorphism generated by $\xi^y(x)$, we have
\be
\wtH_{\m\n\rho}=H_{\m\n\rho}-3a_{[\m}H_{\n\rho]y}\rightarrow\lp H_{\m\n\rho}+3\p_{[\m}\xi^yH_{\n\rho]y}\rp-3\lp a_{[\m}+\p_{[\m}\xi^y\rp H_{\n\rho]y}=\wtH_{\m\n\rho}.
\ee

Under T-duality then, the Buscher rules leave $\whg$, $\whB$, and $\wtH$ invariant, and the other fields transform as
\be
a_\m\leftrightarrow b_\m,\qquad\vp\rightarrow -\vp,\qquad\Phi\rightarrow\Phi-\hlf\vp.
\ee
The price we pay for having a field strength $\wtH$ which is both gauge invariant and behaves nicely under T-duality, is that it now has a non-trivial Bianchi identity,
\be
\whn_{[\m}\wtH_{\n\rho\s]}=-\frac{3}{2}f_{[\m\n}\wtf_{\rho\s]}.
\ee

Now, for reference, we list all the reductions we need from ten-dimensional expressions to expressions on the base, though in practice we will make simplifying assumptions about the base geometry that will lead to simpler expressions than those listed below.

\bea
\nabla_\m\Phi &=& \whn_\m\Phi,\non\\
\nabla_y\Phi &=& 0.
\eea

\bea
\nabla_{\m\n}\Phi &=& \whn_{\m\n}\Phi+\hlf e^\vp\ls\whn^\rho\Phi\whn_\rho\vp\, a_\m a_\n-2\whn^\rho\Phi\, a_{(\m} f_{\n)\rho}\rs,\non\\
\nabla_{\m y}\Phi &=& \hlf e^\vp\ls\whn^\n\Phi\whn_\n\vp\, a_\m-\whn^\n\Phi f_{\m\n}\rs,\\
\nabla_{yy}\Phi &=& \hlf e^\vp\whn^\m\Phi\whn_\m\vp.\non
\eea

\bea
R_{\m\n\rho\s} &=& \widehat{R}_{\m\n\rho\s}+\hlf e^\vp\ls -\whn_{[\m}\vp\whn_{|\rho|}\vp\, a_{\n]}a_\s+\whn_{[\m}\vp\whn_{|\s|}\vp\, a_{\n]}a_\rho-2\whn_{[\m}\vp\, a_{\n]}f_{\rho\s}\right.\non\\
&& \quad\left.-\whn_{[\m}\vp\, a_{|\rho|}f_{\n]\s}+\whn_{[\m}\vp\, a_{|\s|}f_{\n]\rho}+\whn_\rho\vp\, a_{[\m}f_{\n]\s}-\whn_\rho\vp\, a_\s f_{\m\n}-\whn_\s\vp\, a_{[\m}f_{\n]\rho}\right.\non\\
&& \quad\left. +\whn_\s\vp\, a_\rho f_{\m\n}-f_{\m\n}f_{\rho\s}-f_{[\m|\rho|}f_{\n]\s}-2\whn_{[\m|\rho|}\vp\, a_{\n]}a_\s+2\whn_{[\m|\s|}\vp\, a_{\n]}a_\rho\right.\non\\
&& \quad\left.+2a_{[\m}\whn_{\n]}f_{\rho\s}+a_\rho\whn_\s f_{\m\n}-a_\s\whn_\rho f_{\m\n}\rs+\hlf e^{2\vp}\ls a_{[\m}a_{|\rho|}f_{\n]}^{\hph{\n]}\tau}f_{\s\tau}-a_{[\m}a_{|\s|}f_{\n]}^{\hph{\n]}\tau}f_{\rho\tau}\rs,\non\\
R_{\m\n\rho y} &=& \hlf e^\vp\ls -\whn_{[\m}\vp\whn_{|\rho|}\vp\, a_{\n]}+\whn_{[\m}\vp f_{\n]\rho}-\whn_\rho\vp f_{\m\n}-2\whn_{[\m|\rho|}\vp\, a_{\n]}-\whn_\rho f_{\m\n}\rs\non\\
&& \qquad-\hlf e^{2\vp}a_{[\m}f_{\n]}^{\hph{\n]}\s}f_{\rho\s},\\
R_{\m y\n y} &=& \frac{1}{4}e^\vp\ls -\whn_\m\vp\whn_\n\vp-2\whn_{\m\n}\vp\rs+\frac{1}{4}e^{2\vp}f_\m^{\hph{\m}\rho}f_{\n\rho}.\non
\eea

\bea
H_{\m\n\rho} &=& \wtH_{\m\n\rho}+3a_{[\m}\wtf_{\n\rho]},\non\\
H_{\m\n y} &=& \wtf_{\m\n}.
\eea

\bea
\nabla_\m H_{\n\rho\s} &=& -\frac{3}{2}\whn_\m\vp\, a_{[\n}\wtf_{\rho\s]}-\frac{3}{2}\whn_{[\n}\vp\, a_{|\m|}\wtf_{\rho\s]}+\frac{3}{2}f_{\m[\n}\wtf_{\rho\s]}+3a_{[\n}\whn_{|\m|}\wtf_{\rho\s]}+\whn_\m\wtH_{\n\rho\s}\non\\
&& \qquad +\frac{3}{2}e^\vp\ls 2a_\m a_{[\n}f_\rho^{\hph{\rho}\tau}\wtf_{\s]\tau}+\whn^\tau\vp\, a_\m a_{[\n}\wtH_{\rho\s]\tau}-a_\m f_{[\n}^{\hph{[\n}\tau}\wtH_{\rho\s]\tau}-a_{[\n}f_{|\m|}^{\hph{|\m|}\tau}\wtH_{\rho\s]\tau}\rs,\non\\
\nabla_\m H_{\n\rho y} &=& -\hlf\whn_\m\vp\wtf_{\n\rho}+\whn_\m\wtf_{\n\rho}+\hlf e^\vp\ls 2a_\m f_{[\n}^{\hph{[\n}\s}\wtf_{\rho]\s}+\whn^\s\vp\, a_\m\wtH_{\n\rho\s}-f_\m^{\hph{\m}\s}\wtH_{\n\rho\s}\rs,\non\\
\nabla_yH_{\m\n\rho} &=& -\frac{3}{2}\whn_{[\m}\vp\wtf_{\n\rho]}+\frac{3}{2}e^\vp\ls 2a_{[\m}f_\n^{\hph{\n}\s}\wtf_{\rho]\s}+\whn^\s\vp\, a_{[\m}\wtH_{\n\rho]\s}-f_{[\m}^{\hph{[\m}\s}\wtH_{\n\rho]\s}\rs,\\
\nabla_yH_{\m\n y} &=& \hlf e^\vp\ls 2f_{[\m}^{\hph{[\m}\rho}\wtf_{\n]\rho}+\whn^\rho\vp\wtH_{\m\n\rho}\rs.\non
\eea

Finally, we must identify how these reduced fields behave under the orientifold projection.  These follow easily from the behavior of the ten-dimensional fields.  For $\whn^n\Phi$, $\whn^n\vp$, and $\whn^n\widehat{R}$, we must have an even number of normal indices to survive the projection, while $\whn^n\wtH$ must have an odd number of normal indices.  For the pair of vectors, there are two cases; either the involution acts on the circle fiber, or it leaves it invariant.  In the former case the O-plane is transverse to the circle, and $\whn^nf$ should have an odd number of normal indices, while $\whn^n\wtf$ should have an even number.  In the latter case, with the O-plane parallel to the circle direction, it is reversed - $\whn^nf$ should have an even number of normal indices, while $\whn^n\wtf$ should have an odd number.

\subsection{Trivial product}
\label{subsec:TrivialProduct}

As our first example of a simplified background to consider, we will take the case of a product space $\mathcal{B}\times S^1$, where the $S^1$ is constant radius ($\vp$ is constant).  We allow an arbitrary metric $\widehat{g}$ and $B$-field $\widehat{B}$ on $\mathcal{B}$, and a dilaton which depends on the coordinates $x^\m$ of $\mathcal{B}$, but we allow no cross-terms in the metric or $B$-field (so $a_\m=b_\m=0$).

For this background, the reduction of the couplings is very simple - we just replace each $R$ by $\widehat{R}$, each $\nabla$ by $\widehat{\nabla}$, and each $H$ by $\wtH$.  The general coupling can be put in the form
\be
\sqrt{-g}f(\Phi)\mathcal{L}[\nabla\Phi,R,H,\nabla].
\ee
In the case that the O-plane wraps the circle fiber, this reduces to the following coupling on the base,
\be
\label{eq:TrivialParallel}
\sqrt{-g}f(\Phi)\mathcal{L}[\nabla\Phi,R,H,\nabla]=_\parallel\sqrt{-\widehat{g}}e^{\hlf\vp}f(\Phi)\mathcal{L}[\whn\Phi,\widehat{R},\wtH,\whn],
\ee
while for the case that the circle fiber is normal to the O-plane we have
\be
\sqrt{-g}f(\Phi)\mathcal{L}[\nabla\Phi,R,H,\nabla]=_\perp\sqrt{-\widehat{g}}f(\Phi)\mathcal{L}[\whn\Phi,\widehat{R},\wtH,\whn].
\ee
Finally, under T-duality, the latter couplings map as
\be
\label{eq:TrivialTDual}
\sqrt{-\widehat{g}}f(\Phi)\mathcal{L}[\whn\Phi,\widehat{R},\wtH,\whn]\quad\longrightarrow\quad\sqrt{-\widehat{g}}f(\Phi-\hlf\vp)\mathcal{L}[\whn\Phi,\widehat{R},\wtH,\whn].
\ee

Comparing (\ref{eq:TrivialTDual}) with (\ref{eq:TrivialParallel}), we conclude that T-duality requires
\be
e^{\hlf\vp}f(\Phi)=f(\Phi-\hlf\vp).
\ee
Since this should hold for all $\Phi$ and constant $\vp$, we conclude that
\be
f(\Phi)=ce^{-\Phi},
\ee
for some constant $c$.

There are of course other means we could have used to fix the dilaton dependence of these couplings, but it is somewhat gratifying to see that it in our formalism it follows simply from consistency with T-duality, without adding any extra assumptions.  In the rest of the paper, we will assume that the coupling functions $f(\Phi)$ all have this form.

\section{Warped product}
\label{sec:Warped}

The next class of backgrounds we will consider are warped products of flat space with a circle.  We take $g_{\m\n}=\eta_{\m\n}$, $g_{\m y}=0$, $g_{yy}=e^\vp$, $B_{\m y}=0$, and $B_{\m\n}$, $\Phi$, and $\vp$ are arbitrary functions of the base coordinates $x^\m$.  The only nonvanishing Christoffel symbols for this metric are
\be
\G^\m_{yy}=-\hlf e^\vp\p^\m\vp,\qquad\G^y_{\m y}=\hlf\p_\m\vp,
\ee
and this gives us the following expressions
\be
\nabla_\m\Phi=\p_\m\Phi,\quad\nabla_y\Phi=0,\qquad\nabla_{\m\n}\Phi=\p_{\m\n}\Phi,\quad\nabla_{\m y}\Phi=0,\quad\nabla_{yy}\Phi=\hlf e^\vp\p^\m\Phi\p_\m\vp,
\ee
\be
R_{\m\n\rho\s}=0,\qquad R_{\m\n\rho y}=0,\qquad R_{\m y\n y}=\frac{1}{4}e^\vp\ls -\p_\m\vp\p_\n\vp-2\p_{\m\n}\vp\rs,
\ee
\be
\nabla_\m H_{\n\rho\s}=\p_\m H_{\n\rho\s},\qquad\nabla_\m H_{\n\rho y}=0,\qquad\nabla_yH_{\m\n\rho}=0,\qquad\nabla_yH_{\m\n y}=\hlf e^\vp\p^\rho\vp H_{\m\n\rho}.
\ee
Of course, when pulled back to the orientifold-plane we need to impose various projections on the fields as well.  Note that in the absence of $a_\m$ and $b_\m$ there is no distinction between $H_{\m\n\rho}$ and $\wtH_{\m\n\rho}$, so we use the former to save on tildes.

For each coupling, we want to reduce in the case that the circle is parallel to the O-plane and in the case that the circle is perpendicular to the O-plane, and in the latter case we also want to apply T-duality, using the Buscher rules
\be
\vp\rightarrow -\vp,\qquad\Phi\rightarrow\Phi-\hlf\vp.
\ee

Let's illustrate this in the case of our four two-derivative couplings.
\bea
c_1\sqrt{-g}e^{-\Phi}\nabla^a\Phi\nabla_a\Phi &=_\parallel & c_1e^{-\Phi+\hlf\vp}\p^a\Phi\p_a\Phi\non\\
& =_\perp& c_1e^{-\Phi}\p^a\Phi\p_a\Phi\non\\
&\rr& \frac{c_1}{4}e^{-\Phi+\hlf\vp}\ls 4\p^a\Phi\p_a\Phi-4\p^a\Phi\p_a\vp+\p^a\vp\p_a\vp\rs,\non
\eea
\bea
c_2\sqrt{-g}e^{-\Phi}H^{abi}H_{abi} &=_\parallel& c_2e^{-\Phi+\hlf\vp}H^{abi}H_{abi}\non\\
&=_\perp& c_2e^{-\Phi}H^{abi}H_{abi}\non\\
&\rr& c_2e^{-\Phi+\hlf\vp}H^{abi}H_{abi},\non
\eea
\bea
c_3\sqrt{-g}e^{-\Phi}H^{ijk}H_{ijk} &=_\parallel& c_3e^{-\Phi+\hlf\vp}H^{ijk}H_{ijk}\non\\
&=_\perp& c_3e^{-\Phi}H^{ijk}H_{ijk}\non\\
&\rr& c_3e^{-\Phi+\vp}H^{ijk}H_{ijk},\non
\eea
and
\bea
c_4\sqrt{-g}e^{-\Phi}R^{ab}_{\hph{ab}ab} &=_\parallel& \frac{c_4}{2}e^{-\Phi+\hlf\vp}\ls -\p^a\vp\p_a\vp-2\p^a_{\hph{a}a}\vp\rs\non\\
&=_\perp& 0\non\\
&\rr& 0.\non
\eea
The corresponding computations for the forty-eight four-derivative couplings are presented in appendix \ref{app:Warped}.  In either case, the parallel reduction must be equal to the T-dual of the perpendicular reduction, up to terms that re total derivatives or are proportional to Bianchi identities.

Among the reduced fields $\Phi$, $\vp$, and $H$, the only Bianchi identity we have is
\be
\p_{[\m}H_{\n\rho\s]}=0.
\ee
Since this is already two-derivative order, there's no way to get a two derivative coupling by contracting this with other fields.  For four derivative couplings, there are ten possible couplings which are enumerated in appendix \ref{app:Warped}.

At two derivative order, the only total derivatives constructed using the reduced fields are
\bea
y_1\p^a\lp e^{-\Phi+\hlf\vp}\p_a\Phi\rp &=& \frac{y_1}{2}e^{-\Phi+\hlf\vp}\ls -2\p^a\Phi\p_a\Phi+\p^a\Phi\p_a\vp+2\p^a_{\hph{a}a}\Phi\rs,\non
\eea
and
\bea
y_2\p^a\lp e^{-\Phi+\hlf\vp}\p_a\vp\rp &=& \frac{y_2}{2}e^{-\Phi+\hlf\vp}\ls -2\p^a\Phi\p_a\vp+\p^a\vp\p_a\vp+2\p^a_{\hph{a}a}\vp\rs.
\eea
At four derivatives there are twenty-eight couplings, listed in appendix \ref{app:Warped}.

Thus, subtracting the T-dual of the perpendicular couplings from the parallel couplings, and adding in an arbitrary multiple of the total derivatives, we find at two derivatives
\begin{multline}
0=e^{-\Phi+\hlf\vp}\left\{ -y_1\p^a\Phi\p_a\Phi+\hlf\lp 2c_1+y_1-2y_2\rp\p^a\Phi\p_a\vp+\frac{1}{4}\lp -c_1-2c_4+2y_2\rp\p^a\vp\p_a\vp\right.\\
\left.\vphantom{\frac{1}{4}}+y_1\p^a_{\hph{a}a}\Phi+\lp -c_4+y_2\rp\p^a_{\hph{a}a}\vp\right\}.
\end{multline}
This gives five linear equations for the $c_i$ and $y_i$.  In this case the only solution is that $c_1=c_4=y_1=y_2=0$.  The $H^2$ couplings $c_2$ and $c_3$ do not appear in this system, and remain unconstrained.

The same procedure is carried out for the four-derivative couplings in appendix \ref{app:Warped}.  The resulting system leaves twenty-four of our forty-eight couplings unconstrained (ten of these are the $H^4$ couplings), fixing the remaining twenty-four in terms of them.  The end result is
\begin{multline}
\label{eq:WarpedFinalResult}
\mathcal{L}=\int d^{p+1}x\sqrt{-g}e^{-2\Phi}\left\{c_1\ls\nabla^a\Phi\nabla_a\Phi\nabla^b\Phi\nabla_b\Phi-3\nabla^a\Phi\nabla_a\Phi\nabla^b_{\hph{b}b}\Phi+2\nabla^a_{\hph{a}a}\Phi\nabla^b_{\hph{b}b}\Phi\right.\right.\\
\left.\left.-2\nabla^{ij}\Phi\nabla_{ij}\Phi-2\nabla^a\Phi\nabla_a\Phi R^{bc}_{\hph{bc}bc}+2\nabla^a\Phi\nabla^b\Phi R_{a\hph{c}bc}^{\hph{a}c}+2\nabla^a_{\hph{a}a}\Phi R^{bc}_{\hph{bc}bc}-4\nabla^{ij}\Phi R^a_{\hph{a}iaj}+2R^{ab\hph{a}c}_{\hph{ab}a}R_{b\hph{d}cd}^{\hph{b}d}\right.\right.\\
\left.\left.-R^{abcd}R_{abcd}-2R^{ai\hph{a}j}_{\hph{ai}a}R^b_{\hph{b}ibj}\rs+c_4\ls\nabla^a\Phi\nabla^b\Phi H_a^{\hph{a}ci}H_{bci}+2\nabla^a\Phi H_a^{\hph{a}bi}\nabla^cH_{bci}-\nabla^aH_a^{\hph{a}bi}\nabla^cH_{bci}\rs\right.\\
\left.+c_{18}\ls\nabla^{ab}\Phi H_a^{\hph{a}ci}H_{bci}+R^{ab\hph{a}c}_{\hph{ab}a}H_b^{\hph{b}di}H_{cdi}-\hlf\nabla^iH^{ajk}\nabla_iH_{ajk}\rs\right.\\
\left.+c_{19}\ls\nabla^{ij}\Phi H^{ab}_{\hph{ab}i}H_{abj}+R^{ai\hph{a}j}_{\hph{ai}a}H^{bc}_{\hph{bc}i}H_{bcj}\rs+c_{20}\ls\nabla^{ij}\Phi H_i^{\hph{i}k\ell}H_{jk\ell}+R^{ai\hph{a}j}_{\hph{ai}a}H_i^{\hph{i}k\ell}H_{jk\ell}\rs\right.\\
\left.+c_{28}R^{abcd}H_{ab}^{\hph{ab}i}H_{cdi}+c_{29}R^{abij}H_{ab}^{\hph{ab}k}H_{ijk}+c_{30}R^{abij}H_{a\hph{c}i}^{\hph{a}c}H_{bcj}+c_{33}R^{aibj}H_{a\hph{c}i}^{\hph{a}c}H_{bcj}\right.\\
\left.+c_{34}R^{ijk\ell}H_{ij}^{\hph{ij}m}H_{k\ell m}+c_{37}\ls R^{ab}_{\hph{ab}ab}R^{cd}_{\hph{cd}cd}-4R^{ab\hph{a}c}_{\hph{ab}a}R_{b\hph{d}cd}^{\hph{b}d}+R^{abcd}R_{abcd}\rs+c_{40}R^{abij}R_{abij}\right.\\
\left.+c_{46}\ls\nabla^aH^{bci}\nabla_aH_{bci}+\nabla^iH^{ajk}\nabla_iH_{ajk}\rs+c_{47}\nabla^aH^{ijk}\nabla_aH_{ijk}+(H^4)\right\}.
\end{multline}

We note with satisfaction that the $R^2$ and $(\nabla H)^2$ terms are consistent with known results~\cite{Bachas:1999um}, \cite{Garousi:2009dj} (we would require $c_{37}=0$ and $c_{40}=c_1$ to match the $R^2$ terms, and $c_4=-2c_1$, $c_{18}=0$, $c_{46}=c_1/2$, and $c_{47}=c_1/6$ to match the $(\nabla H)^2$ terms\footnote{To carry out the match with~\cite{Garousi:2009dj}, we need to rewrite the coupling $\nabla^iH^{abc}\nabla_iH_{abc}$ in our basis, using the null vectors from appendix \ref{app:Classification},
\be
\int\sqrt{-g}e^{-\Phi}\nabla^iH^{abc}\nabla_iH_{abc}=\int\sqrt{-g}e^{-\Phi}\ls 3\nabla^aH^{bci}\nabla_aH_{bci}+6\nabla^aH_a^{\hph{a}bi}\nabla^cH_{bci}+\lp RH^2\rp+\lp\nabla\Phi H\nabla H\rp\rs.
\ee}), but many other coefficients have now been fixed.  It is interesting to also note that $c_{37}$ multiplies the Gauss-Bonnet combination for the pulled back metric $g_{ab}$, which can be argued to vanish (in our basis of couplings), but which can not be checked directly by two-point amplitudes.

To fix more coefficients, we need to consider a different class of backgrounds.

\section{Twisted product}
\label{sec:Twisted}

In the twisted product, we set $g_{\m\n}=\eta_{\m\n}+e^\vp a_\m a_\n$, $g_{\m y}=e^\vp a_\m$, $g_{yy}=e^\vp$, $B_{\m\n}=\whB_{\m\n}-\hlf a_\m b_\n+\hlf a_\n b_\m$, and $B_{\m y}=b_\m$, with $\vp$ and $\Phi$ constant, and with $a_\m$ and $b_\m$ being arbitrary functions of the base coordinates $x^\m$.  We define field strengths
\be
f_{\m\n}=2\p_{[\m}a_{\n]},\qquad \wtf_{\m\n}=2\p_{[\m}b_{\n]}.
\ee
As described in section \ref{subsec:Generalities}, it is also useful to define
\be
\wtH_{\m\n\rho}=H_{\m\n\rho}-3a_{[\m}\wtf_{\n\rho]}=3\p_{[\m}\whB_{\n\rho]}-\frac{3}{2}f_{[\m\n}b_{\rho]}-\frac{3}{2}a_{[\m}\wtf_{\n\rho]}.
\ee
The Buscher rules will act by
\be
\vp\rr -\vp,\qquad\Phi\rr\Phi-\hlf\vp,\qquad a_\m\leftrightarrow b_\m,\quad f_{\m\n}\leftrightarrow\wtf_{\m\n},
\ee
and $\wtH_{\m\n\rho}$ is invariant.

The Christoffel symbols are given by
\be
\G^\rho_{\m\n}=\hlf e^\vp\lp a_\m f_\n^{\hph{\n}\rho}+a_\n f_\m^{\hph{\m}\rho}\rp,\qquad \G^\rho_{\m y}=\hlf e^\vp f_\m^{\hph{\m}\rho},\qquad \G^\rho_{yy}=0,
\ee
\be
\G^y_{\m\n}=\hlf\lp\p_\m a_\n+\p_\n a_\m\rp-\hlf e^\vp\lp a_\m a^\rho f_{\n\rho}+a_\n a^\rho f_{\m\rho}\rp,\qquad\G^y_{\m y}=-\hlf e^\vp a^\n f_{\m\n},\qquad\G^y_{yy}=0.\non
\ee
This gives
\begin{multline}
R_{\m\n\rho\s}=\hlf e^\vp\ls 2a_{[\m}\p_{\n]}f_{\rho\s}+a_\rho\p_\s f_{\m\n}-a_\s\p_\rho f_{\m\n}-f_{\m\n}f_{\rho\s}-f_{[\m|\rho|}f_{\n]\s}\rs\\
+\hlf e^{2\vp}\ls a_{[\m}a_{|\rho|}f_{\n]}^{\hph{\n]}\tau}f_{\s\tau}-a_{[\m}a_{|\s|}f_{\n]}^{\hph{\n]}\tau}f_{\rho\tau}\rs,\non
\end{multline}
\be 
\label{eq:TwistedR}
R_{\m\n\rho y}=-\hlf e^\vp\p_\rho f_{\m\n}-\hlf e^{2\vp}a_{[\m}f_{\n]}^{\hph{\n]}\s}f_{\rho\s},\qquad R_{\m y\n y}=\frac{1}{4} e^{2\vp}f_\m^{\hph{\m}\rho}f_{\n\rho},
\ee
\be
\label{eq:TwistedH}
H_{\n\rho\s}=\wtH_{\m\n\rho}+3a_{[\m}\wtf_{\n\rho]},\qquad H_{\m\n y}=\wtf_{\m\n},
\ee
\begin{multline}
\nabla_\m H_{\n\rho\s}=\frac{3}{2}f_{\m[\n}\wtf_{\rho\s]}+3a_{[\n}\p_{|\m|}\wtf_{\rho\s]}+\p_\m\wtH_{\n\rho\s}\\
+\frac{3}{2}e^\vp\ls 2a_\m a_{[\n}f_\rho^{\hph{\rho}\tau}\wtf_{\s]\tau}-a_\m f_{[\n}^{\hph{[\n}\tau}\wtH_{\rho\s]\tau}-a_{[\n}f_{|\m|}^{\hph{|\m|}\tau}\wtH_{\rho\s]\tau}\rs,\non
\end{multline}
\be
\label{eq:TwistedDH}
\nabla_\m H_{\n\rho y}=\p_\m\wtf_{\n\rho}+\hlf e^\vp\ls 2a_\m f_{[\n}^{\hph{[\n}\s}\wtf_{\rho]\s}-f_\m^{\hph{\m}\s}\wtH_{\n\rho\s}\rs,
\ee
\be
\nabla_yH_{\m\n\rho}=\frac{3}{2}e^\vp\ls 2a_{[\m}f_\n^{\hph{\n}\s}\wtf_{\rho]\s}-f_{[\m}^{\hph{[\m}\s}\wtH_{\n\rho]\s}\rs,\qquad\nabla_yH_{\m\n y}=e^\vp f_{[\m}^{\hph{[\m}\rho}\wtf_{\n]\rho}.\non
\ee

Again, we will illustrate the procedure with the two-derivative couplings.
\bea
c_1\sqrt{-g}e^{-\Phi}\nabla^a\Phi\nabla_a\Phi &=_\parallel& 0\non\\
&=_\perp& 0\non\\
&\rr& 0,\non
\eea
\bea
c_2\sqrt{-g}e^{-\Phi}H^{abi}H_{abi} &=_\parallel& c_2e^{-\Phi+\hlf\vp}\ls\wtH^{abi}\wtH_{abi}+2e^{-\vp}\wtf^{ai}\wtf_{ai}\rs\non\\
&=_\perp& c_2e^{-\Phi}\ls\wtH^{abi}\wtH_{abi}+e^{-\vp}\wtf^{ab}\wtf_{ab}\rs\non\\
&\rr& c_2e^{-\Phi+\hlf\vp}\ls\wtH^{abi}\wtH_{abi}+e^\vp f^{ab}f_{ab}\rs,\non
\eea
\bea
c_3\sqrt{-g}e^{-\Phi}H^{ijk}H_{ijk} &=_\parallel& c_3e^{-\Phi+\hlf\vp}\wtH^{ijk}\wtH_{ijk}\non\\
&=_\perp& c_3e^{-\Phi}\ls\wtH^{ijk}\wtH_{ijk}+3e^{-\vp}\wtf^{ij}\wtf_{ij}\rs\non\\
&\rr& c_3e^{-\Phi+\hlf\vp}\ls\wtH^{ijk}\wtH_{ijk}+3e^\vp f^{ij}f_{ij}\rs,\non
\eea
\bea
c_4\sqrt{-g}e^{-\Phi}R^{ab}_{\hph{ab}ab} &=_\parallel& c_4e^{-\Phi+\hlf\vp}\ls -\frac{1}{4}e^\vp f^{ab}f_{ab}\rs\non\\
&=_\perp& 0\non\\
&\rr& 0.\non
\eea
The corresponding computations for the four-derivative couplings are presented in appendix \ref{app:Twisted}.  Note that we have some useful selection rules.  The total number of $\wtH$ and $\wtf$ which appear must be even, since this just counts the number of $H$ fields in the original covariant coupling.  Furthermore, the total number of $f$ and $\wtf$ fields which appear must also be even.  To see this, we note that in the expansions (\ref{eq:TwistedR})-(\ref{eq:TwistedH}), the parity of the number of $a_\m$ and $b_\m$ is equal to the parity of the number of $y$ indices.  Since the total coupling has no free indices, there must be an even total number of $f$ and $\wtf$.

There are three Bianchi identities that can be relevant for the reduced fields $a$, $b$, and $\wtH$,
\bea
\p_{[\m}f_{\n\rho]} &=& 0,\non\\
\p_{[\m}\wtf_{\n\rho]} &=& 0,\\
\p_{[\m}\wtH_{\n\rho\s]}+\frac{3}{2}f_{[\m\n}\wtf_{\rho\s]} &=& 0.\non
\eea
These all have two derivatives already, so won't play a role in constraining the two-derivative couplings.  For the four-derivative couplings, there are twenty-three combinations one can write down (in the parallel case), and they are listed in appendix \ref{app:Twisted}.

In the two-derivative case, our selection rules prevent us from writing any total derivative terms either (we assume that parity-odd terms, which would be dimension dependent, are not allowed).  At four derivatives, there are twenty possible total derivative combinations that we can construct which are consistent with the selection rules, and they are enumerated in appendix \ref{app:Twisted}.

So for the two-derivative case, demanding that the parallel reductions minus the T-duals of the perpendicular reductions vanish up to Bianchi identities and total derivatives, we find
\be
0=e^{-\Phi+\hlf\vp}\left\{ 2c_2e^{-\vp}\wtf^{ai}\wtf_{ai}+\frac{1}{4}\lp -4c_2-c_4\rp e^\vp f^{ab}f_{ab}-3c_3e^\vp f^{ij}f_{ij}\right\}.
\ee
This imposes three linear equations which force $c_2=c_3=c_4=0$.  Only the $(\nabla\Phi)^2$ coupling $c_1$ is unfixed by this result; the other couplings are forced to vanish.  Note that this is consistent with our warped product analysis which left $c_2$ and $c_3$ unfixed and forced $c_1=c_4=0$.  Combining the two results, we learn that $c_1=c_2=c_3=c_4=0$.  In other words, there is no two-derivative NS-NS sector action which we can write an on O-plane which is compatible with T-duality!

In appendix \ref{app:Twisted}, the analogous procedure is carried out for the forty-eight four-derivative couplings.  Seventeen couplings involve derivatives of the dilaton, and these cannot be directly fixed by considering the twisted product backgrounds.  Of the remaining thirty-one couplings, we find that there is only one free parameter, $c_7$, and all of the other thirty couplings are fixed in terms of that one (in fact we find that fourteen of them must vanish).  The resulting action is
\begin{multline}
\label{eq:TwistedFinalResult}
\mathcal{L}=\sqrt{-g}e^{-\Phi}\left\{ c_7\ls H^{abi}H_{ab}^{\hph{ab}j}H^{cd}_{\hph{cd}i}H_{cdj}-H^{abi}H_{ab}^{\hph{ab}j}H_i^{\hph{i}k\ell}H_{jk\ell}-\frac{2}{3}H^{abi}H_a^{\hph{a}cj}H_{bc}^{\hph{bc}k}H_{ijk}\right.\right.\\
\left.\left.+\hlf H^{abi}H_a^{\hph{a}cj}H_{b\hph{d}j}^{\hph{b}d}H_{cdi}+\frac{1}{6}H^{ijk}H_i^{\hph{i}\ell m}H_{j\ell}^{\hph{j\ell}n}H_{kmn}-4R^{ab\hph{a}c}_{\hph{ab}a}H_b^{\hph{b}di}H_{cdi}+4R^{abij}H_{ab}^{\hph{ab}k}H_{ijk}\right.\right.\\
\left.\left.+8R^{abij}H_{a\hph{c}i}^{\hph{a}c}H_{bci}-6R^{ai\hph{a}j}_{\hph{ai}a}H^{bc}_{\hph{bc}i}H_{bcj}+2R^{ai\hph{a}j}_{\hph{ai}a}H_i^{\hph{i}k\ell}H_{jk\ell}-8R^{ab\hph{a}c}_{\hph{ab}a}R_{b\hph{d}cd}^{\hph{b}d}+4R^{abcd}R_{abcd}\right.\right.\\
\left.\left.-4R^{abij}R_{abij}+8R^{ai\hph{a}j}_{\hph{ai}a}R^b_{\hph{b}ibj}-8\nabla^aH_a^{\hph{a}bi}\nabla^cH_{bci}-2\nabla^aH^{bci}\nabla_aH_{bci}-\frac{2}{3}\nabla^aH^{ijk}\nabla_aH_{ijk}\rs\right.\\
\left.+(\Phi\mathrm{\ terms})\right\}.
\end{multline}

Note that this is completely consistent with (\ref{eq:WarpedFinalResult}) and with previously known $R^2$ and $(\nabla H)^2$ couplings.

\section{Combined results}
\label{sec:Combined}

\subsection{Final Result}

As mentioned above, for the two-derivative couplings the twisted product analysis fixed all the couplings which did not involve derivatives of the dilaton to vanish.  Meanwhile, the warped product analysis showed that all couplings which weren't purely $H^2$ must vanish.  Between these two results, we see that the entire two-derivative action is fixed to vanish.

At four derivatives, we found that the twisted product again fixed every coupling that did not involve derivatives of the dilaton, up to one overall parameter.  And by examining (\ref{eq:WarpedFinalResult}), we see that the warped product analysis in turn relates every dilaton coupling to a coupling that does not involve the dilaton.  Thus, by combining the two analyses, we fix the entire four-derivative action up to one overall constant,
\begin{multline}
\label{eq:MainResult}
\mathcal{L}=-c_1\sqrt{-g}e^{-\Phi}\left\{ -\nabla^a\Phi\nabla_a\Phi\nabla^b\Phi\nabla_b\Phi+2\nabla^a\Phi\nabla^b\Phi H_a^{\hph{a}ci}H_{bci}+\frac{1}{4}H^{abi}H_{ab}^{\hph{ab}j}H^{cd}_{\hph{cd}i}H_{cdj}\right.\\
\left.-\frac{1}{4}H^{abi}H_{ab}^{\hph{ab}j}H_i^{\hph{i}k\ell}H_{jk\ell}-\frac{1}{6}H^{abi}H_a^{\hph{a}cj}H_{bc}^{\hph{bc}k}H_{ijk}+\frac{1}{8}H^{abi}H_a^{\hph{a}cj}H_{b\hph{d}j}^{\hph{b}d}H_{cdi}+\frac{1}{24}H^{ijk}H_i^{\hph{i}\ell m}H_{j\ell}^{\hph{j\ell}n}H_{kmn}\right.\\
\left.+3\nabla^a\Phi\nabla_a\Phi\nabla^b_{\hph{b}b}\Phi-\nabla^{ab}\Phi H_a^{\hph{a}ci}H_{bci}-\frac{3}{2}\nabla^{ij}\Phi H^{ab}_{\hph{ab}i}H_{abj}+\hlf\nabla^{ij}\Phi H_i^{\hph{i}k\ell}H_{jk\ell}-2\nabla^a_{\hph{a}a}\Phi\nabla^b_{\hph{b}b}\Phi\right.\\
\left.+2\nabla^{ij}\Phi\nabla_{ij}\Phi+2\nabla^a\Phi\nabla_a\Phi R^{bc}_{\hph{bc}bc}-2\nabla^a\Phi\nabla^b\Phi R_{a\hph{c}bc}^{\hph{a}c}-R^{ab\hph{a}c}_{\hph{ab}a}H_b^{\hph{b}di}H_{cdi}+R^{abij}H_{ab}^{\hph{ab}k}H_{ijk}\right.\\
\left.+2R^{abij}H_{a\hph{c}i}^{\hph{a}c}H_{bcj}-\frac{3}{2}R^{ai\hph{a}j}_{\hph{ai}a}H^{bc}_{\hph{bc}i}H_{bcj}+\hlf R^{ai\hph{a}j}_{\hph{ai}a}H_i^{\hph{i}k\ell}H_{jk\ell}-2\nabla^a_{\hph{a}a}\Phi R^{bc}_{\hph{bc}bc}+4\nabla^{ij}\Phi R^a_{\hph{a}iaj}\right.\\
\left.-2R^{ab\hph{a}c}_{\hph{ab}a}R_{b\hph{d}cd}^{\hph{b}d}+R^{abcd}R_{abcd}-R^{abij}R_{abij}+2R^{ai\hph{a}j}_{\hph{ai}a}R^b_{\hph{b}ibj}+4\nabla^a\Phi H_a^{\hph{a}bi}\nabla^cH_{bci}-2\nabla^aH_a^{\hph{a}bi}\nabla^cH_{bci}\right.\\
\left.-\hlf\nabla^aH^{bci}\nabla_aH_{bci}-\frac{1}{6}\nabla^aH^{ijk}\nabla_aH_{ijk}\right\}.
\end{multline}

By comparing with~\cite{Garousi:2009dj}, we can fix the constant as well to be
\be
c_1=-T_p'\frac{\pi^2\lp\al'\rp^2}{96},
\ee
where $T_p'=2^{p-5}T_p$ is the (absolute value of the) O-plane tension, i.e.\ the zero-derivative action is $S_0=T_p'\int d^{p+1}xe^{-\Phi}\sqrt{-g}$, and
\be
T_p=\frac{2\pi}{g_s}\lp 4\pi^2\al'\rp^{-\frac{p+1}{2}},
\ee
is the D$p$-brane tension.

For comparison, the action on a D$p$-brane is
\begin{multline}
S_{Dp}=-T_p\int d^{p+1}xe^{-\Phi}\sqrt{-\det{g+B+2\pi\al'F}}\\
+T_p\frac{\pi^2\lp\al'\rp^2}{48}\int d^{p+1}e^{-\Phi}\sqrt{-g}\lp R^{abcd}R_{abcd}+\cdots\rp.
\end{multline}
The index structure of the $R^2$ squared terms~\cite{Bachas:1999um} and $(\nabla H)^2$ term~\cite{Garousi:2009dj} have the same structure as they do for D-branes, essentially because the two-point $\R\P^2$ amplitude of NS-NS vertex operators is related to the disc amplitude simply by a kinematic factor.

\subsection{A dilaton-free rewriting}

The method we have used relied on the fact that we could consistently use the bulk equations of motion to eliminate terms in which two normal indices were contracted inside the same field, and that this elimination didn't mix with our other classes of null vectors, in particular those coming from total derivatives.  However, now that we have our final result in hand, we are free to switch to a different basis of couplings.  One interesting choice, inspired by the structures which actually appear when computing amplitudes~\cite{Garousi:1996ad,Garousi:2009dj,Becker:2010ij,Becker:2011bw}, is to instead only keep self-contractions built with the matrix
\be
D^{\m\n}=\lp\begin{matrix}\d^{ab} & 0 \\ 0 & -\d^{ij}\end{matrix}\rp.
\ee

To this end we can define quantities
\bea
\lp D\nabla^2\Phi\rp &=& D^{\m\n}\nabla_\m\nabla_\n\Phi=\nabla^a_{\hph{a}a}\Phi-\nabla^i_{\hph{i}i}\Phi,\non\\
\lp DR\rp_{\m\n} &=& D^{\rho\s}R_{\m\rho\n\s}=R_{\m\hph{a}\n a}^{\hph{\m}a}-R_{\m\hph{i}\n i}^{\hph{\m}i},\\
\lp D\nabla H\rp_{ai} &=& D^{\m\n}\nabla_\m H_{ai\n}=-\nabla^bH_{abi}-\nabla^jH_{aij}.\non
\eea

We can then use the equations of motion and heavy use of null vectors (i.e.\ much use of Bianchi identities, equations of motion, and integrations by parts) to rewrite our action above as
\begin{multline}
\label{eq:DilatonFreeResult}
\mathcal{L}=-c_1\sqrt{-g}e^{-\Phi}\left\{\frac{3}{32}H^{abi}H_{ab}^{\hph{ab}j}H^{cd}_{\hph{cd}i}H_{cdj}-\frac{5}{16}H^{abi}H_{ab}^{\hph{ab}j}H_i^{\hph{i}k\ell}H_{jk\ell}-\frac{3}{8}H^{abi}H_{a\hph{c}i}^{\hph{a}c}H_b^{\hph{b}dj}H_{cdj}\right.\\
\left.-\frac{1}{6}H^{abi}H_a^{\hph{a}cj}H_{bc}^{\hph{bc}k}H_{ijk}+\frac{1}{8}H^{abi}H_a^{\hph{a}cj}H_{b\hph{d}j}^{\hph{b}d}H_{cdi}+\frac{3}{32}H^{ijk}H_{ij}^{\hph{ij}\ell}H_k^{\hph{k}mn}H_{\ell mn}\right.\\
\left.+\frac{1}{24}H^{ijk}H_i^{\hph{i}\ell m}H_{j\ell}^{\hph{j\ell}n}H_{kmn}-\lp DR\rp^{ab}H_a^{\hph{a}ci}H_{bci}+R^{abij}H_{ab}^{\hph{ab}k}H_{ijk}+2R^{abij}H_{a\hph{c}i}^{\hph{a}c}H_{bcj}\right.\\
\left.-\hlf\lp DR\rp^{ij}H^{ab}_{\hph{ab}i}H_{abj}+\hlf\lp DR\rp^{ij}H_i^{\hph{i}k\ell}H_{jk\ell}-\hlf\lp DR\rp^{ab}\lp DR\rp_{ab}+R^{abcd}R_{abcd}-R^{abij}R_{abij}\right.\\
\left.+\hlf\lp DR\rp^{ij}\lp DR\rp_{ij}+\hlf\lp D\nabla H\rp^{ai}\lp D\nabla H\rp_{ai}-\hlf\nabla^aH^{bci}\nabla_aH_{bci}-\frac{1}{6}\nabla^aH^{ijk}\nabla_aH_{ijk}\right\}.
\end{multline}
Writing things in this way, the dilaton dependence has entirely vanished except for the overall factor of $e^{-\Phi}$.  It would be very interesting to understand why this situation arises.  Note also that the connection with computation of $\R\P^2$ amplitudes is not very direct, since to compare with the string scattering calculations we must first convert to Einstein frame, which will make the dilaton couplings reappear.

\acknowledgments

The authors would like to thank K.~Becker, R.~Minasian, and S.~Sethi for useful discussions and comments.  D.~R. would like to thank the Mitchell Institute for hospitality and support while this work was being completed.  D.~R. is supported by funding from the European Research Council, ERC grant agreement no.\ 268088-EMERGRAV.  Z.~W. is supported by NSF grants PHY-0906222 and PHY-1001140, Texas A{\&}M University, and the Mitchell Institute for Fundamental Physics and Astronomy.


\appendix

\section{Classification of covariant couplings}
\label{app:Classification}

As in the rest of the paper, we use indices from the beginning of the alphabet, $a$, $b$, etc.\ for directions along the O-plane world-volume, and indices from the middle of the alphabet, $i$, $j$, etc.\ for normal directions.  Occasionally Greek letters will appear, $\m$, $\n$, etc.; in this section these represent all ten directions, tangent and normal.

Each coupling below will be accompanied in the Lagrangian by a factor of $\sqrt{-g}$ as well as a function of the dilaton $f(\Phi)$.  In section \ref{subsec:TrivialProduct} it is shown that T-duality requires the function to be of the form $f(\Phi)=ce^{-\Phi}$ for some constant $c$.  At any rate, to save space these factors will be omitted from the couplings below.  It should be understood that in the action, each coupling will appear with integration and measure $\int\sqrt{-g}e^{-\Phi}[\cdots]$.

As discussed in section \ref{subsec:Lexicography}, terms are built out of letters which consist of symmetrized covariant derivatives acting on covariant fields ($\Phi$, $R_{\m\n\rho\s}$, or $H_{\m\n\rho}$).  The orientifold projection demands that the number of normal indices on a letter built from $\Phi$ or $R$ must be even, while on a letter built from $H$ there must be an odd number of normal indices.  We also require an even number of $H$-letters in each term (note that at even derivative order, we need an even number of $H$ fields in order to have an even total number of indices).  Moreover, as in section \ref{subsec:Redundancies} we can always use the leading order bulk equations of motion to remove any term that includes a contraction of normal indices within a given letter.

As described in \ref{subsec:Lexicography}, we order letters by complexity.  The ones that will appear here, in order, are
\be
\left\{\nabla\Phi, H, \nabla^2\Phi, R, \nabla H, \nabla^3\Phi, \nabla R, \nabla^2H, \nabla^4\Phi, \nabla^2R\right\}.
\ee
We then order terms by comparing their most complex letter, then moving to their next most complex letter, and so on.  Within a term we order the letters by starting with all the $\Phi$-letters, with increasing numbers of derivatives, then the $R$-letters, and finally the $H$-letters.  Finally, terms that differ only in their index structure are ordered by the minimal lexicographic order of their indices, read from left to right, using the rules we have outlined and the basic symmetries of the letters (i.e.\ that all derivatives are symmetrized, that $H_{\m\n\rho}$ is antisymmetric, and that the Riemann tensor satisfies $R_{\m\n\rho\s}=R_{\rho\s\m\n}=-R_{\n\m\rho\s}$) and exchanges of identical letters..

The list of allowed terms, where we do not yet worry about Bianchi identities or integration by parts\footnote{It would not be difficult to skip ahead and take account of Bianchi identites and total derivatives by hand.  However, we are trying to proceed in the most systematic possible manner, both to allay any doubts about our procedure, and also because we are in the process of computerizing this approach to work in some more general contexts.} is,
\begin{itemize}
\item $\nabla^a\Phi\nabla_a\Phi\nabla^b\Phi\nabla_b\Phi$,
\item $\nabla^a\Phi\nabla_a\Phi H^{bci}H_{bci}$, $\nabla^a\Phi\nabla_a\Phi H^{ijk}H_{ijk}$, $\nabla^a\Phi\nabla^b\Phi H_a^{\hph{a}ci}H_{bci}$,
\item $H^{abi}H_{abi}H^{cdj}H_{cdj}$, $H^{abi}H_{abi}H^{jk\ell}H_{jk\ell}$, $H^{abi}H_{ab}^{\hph{ab}j}H^{cd}_{\hph{cd}i}H_{cdj}$, $H^{abi}H_{ab}^{\hph{ab}j}H_i^{\hph{i}k\ell}H_{jk\ell}$,\\ $H^{abi}H_{a\hph{c}i}^{\hph{a}c}H_b^{\hph{b}dj}H_{cdj}$, $H^{abi}H_a^{\hph{a}cj}H_{bc}^{\hph{bc}k}H_{ijk}$, $H^{abi}H_a^{\hph{a}cj}H_{b\hph{d}j}H_{cdi}$, $H^{ijk}H_{ijk}H^{\ell mn}H_{\ell mn}$,\\ $H^{ijk}H_{ij}^{\hph{ij}\ell}H_k^{\hph{k}mn}H_{\ell mn}$, $H^{ijk}H_i^{\hph{i}\ell m}H_{j\ell}^{\hph{j\ell}n}H_{kmn}$,
\item $\nabla^a\Phi\nabla_a\Phi\nabla^b_{\hph{b}b}\Phi$, $\nabla^a\Phi\nabla^b\Phi\nabla_{ab}\Phi$,
\item $\nabla^a_{\hph{a}a}\Phi H^{bci}H_{bci}$, $\nabla^a_{\hph{a}a}\Phi H^{ijk}H_{ijk}$, $\nabla^{ab}\Phi H_a^{\hph{a}ci}H_{bci}$, $\nabla^{ij}\Phi H^{ab}_{\hph{ab}i}H_{abj}$, $\nabla^{ij}\Phi H_i^{\hph{i}k\ell}H_{jk\ell}$,
\item $\nabla^a_{\hph{a}a}\Phi\nabla^b_{\hph{b}b}\Phi$, $\nabla^{ab}\Phi\nabla_{ab}\Phi$, $\nabla^{ij}\Phi\nabla_{ij}\Phi$,
\item $\nabla^a\Phi\nabla_a\Phi R^{bc}_{\hph{bc}bc}$, $\nabla^a\Phi\nabla^b\Phi R_{a\hph{c}bc}^{\hph{a}c}$,
\item $R^{ab}_{\hph{ab}ab}H^{cdi}H_{cdi}$, $R^{ab}_{\hph{ab}ab}H^{ijk}H_{ijk}$, $R^{ab\hph{a}c}_{\hph{ab}a}H_b^{\hph{b}di}H_{cdi}$, $R^{abcd}H_{ab}^{\hph{ab}i}H_{cdi}$, $R^{abcd}H_{ac}^{\hph{ac}i}H_{bdi}$,\\ $R^{abij}H_{ab}^{\hph{ab}k}H_{ijk}$, $R^{abij}H_{a\hph{c}i}^{\hph{a}c}H_{bcj}$, $R^{ai\hph{a}j}_{\hph{ai}a}H^{bc}_{\hph{bc}i}H_{bcj}$, $R^{ai\hph{a}j}_{\hph{ai}a}H_i^{\hph{i}k\ell}H_{jk\ell}$, $R^{aibj}H_{ab}^{\hph{ab}k}H_{ijk}$,\\ $R^{aibj}H_{a\hph{c}i}^{\hph{a}c}H_{bcj}$, $R^{aibj}H_{a\hph{c}j}H_{bci}$, $R^{ijk\ell}H_{ij}^{\hph{ij}m}H_{k\ell m}$, $R^{ijk\ell}H_{ik}^{\hph{ik}m}H_{j\ell m}$,
\item $\nabla^a_{\hph{a}a}\Phi R^{bc}_{\hph{bc}bc}$, $\nabla^{ab}\Phi R_{a\hph{c}bc}^{\hph{a}c}$, $\nabla^{ij}\Phi R^a_{\hph{a}iaj}$,
\item $R^{ab}_{\hph{ab}ab}R^{cd}_{\hph{cd}cd}$, $R^{ab\hph{a}c}_{\hph{ab}a}R_{b\hph{d}cd}^{\hph{b}d}$, $R^{abcd}R_{abcd}$, $R^{abcd}R_{acbd}$, $R^{abij}R_{abij}$, $R^{abij}R_{aibj}$, $R^{ai\hph{a}j}_{\hph{ai}a}R^b_{\hph{b}ibj}$, $R^{aibj}R_{aibj}$, $R^{aibj}R_{ajbi}$, $R^{ijk\ell}R_{ijk\ell}$, $R^{ijk\ell}R_{ikj\ell}$,
\item $\nabla^a\Phi H_a^{\hph{a}bi}\nabla^cH_{bci}$, $\nabla^a\Phi H^{bci}\nabla_aH_{bci}$, $\nabla^a\Phi H^{bci}\nabla_bH_{aci}$, $\nabla^a\Phi H^{bci}\nabla_iH_{abc}$, $\nabla^a\Phi H^{ijk}\nabla_aH_{ijk}$, $\nabla^a\Phi H^{ijk}\nabla_iH_{ajk}$,
\item $\nabla^aH_a^{\hph{a}bi}\nabla^cH_{bci}$, $\nabla^aH^{bci}\nabla_aH_{bci}$, $\nabla^aH^{bci}\nabla_bH_{aci}$, $\nabla^aH^{bci}\nabla_iH_{abc}$, $\nabla^aH^{ijk}\nabla_aH_{ijk}$,\\ $\nabla^aH^{ijk}\nabla_iH_{ajk}$, $\nabla^iH^{abc}\nabla_iH_{abc}$, $\nabla^iH^{ajk}\nabla_iH_{ajk}$, $\nabla^iH^{ajk}\nabla_jH_{aik}$,
\item $\nabla^a\Phi\nabla_{a\hph{b}b}^{\hph{a}b}\Phi$,
\item $\nabla^a\Phi\nabla_aR^{bc}_{\hph{bc}bc}$, $\nabla^a\Phi\nabla^bR_{a\hph{c}bc}^{\hph{a}c}$,
\item $H^{abi}\nabla_a^{\hph{a}c}H_{bci}$, $H^{abi}\nabla^c_{\hph{c}c}H_{abi}$, $H^{abi}\nabla^c_{\hph{c}i}H_{abc}$, $H^{ijk}\nabla^a_{\hph{a}a}H_{ijk}$, $H^{ijk}\nabla^a_{\hph{a}i}H_{ajk}$,
\item $\nabla^{a\hph{a}b}_{\hph{a}a\hph{b}b}\Phi$,
\item $\nabla^a_{\hph{a}a}R^{bc}_{\hph{bc}bc}$, $\nabla^{ab}R_{a\hph{c}bc}^{\hph{a}c}$.
\end{itemize}

We should think about these terms as spanning an $80$-dimensional vector space of couplings.  However, many of the vectors in this space are actually zero in the physical action, either because they are proportional to a Bianchi identity, or because they correspond to total derivatives on the world-volume.  The physical space of couplings will correspond to the quotient of the full space by this subspace of null couplings.  Our objective is to find a (lexicographically earliest) subset of the couplings above whose images under projection to the quotient space form a basis of the quotient space.

To accomplish this we now list all terms which are zero by virtue of Bianchi identities or total derivatives.  First the Bianchi identities.  There are three basic ones to consider,
\be
R_{[\m\n\rho]\s}=0,\qquad\nabla_{[\m}H_{\n\rho\s]}=0,\qquad\nabla_{[\m}R_{\n\rho]\s\tau}=0.
\ee
Any term that is built by acting on these with covariant derivatives or multiplying them with other letters should be zero.  Occasionally we will omit the details of some terms which are obtained by replacing commutators of covariant derivatives with Riemann tensors, since these will inevitably involve only terms which are earlier in our ordering than the other terms in a given vector, and they will not matter when we are deciding which couplings can be eliminated using these null vectors.
\bea
3H^{abi}H_a^{\hph{a}cj}R_{[bci]j} &=& R^{abij}H_{a\hph{c}i}^{\hph{a}c}H_{bcj}-R^{aibj}H_{a\hph{c}i}^{\hph{a}c}H_{bcj}+R^{aibj}H_{a\hph{c}j}^{\hph{a}c}H_{bci},\non\\
3H^{abi}H_a^{\hph{a}cj}R_{[bij]c} &=& R^{abij}H_{a\hph{c}i}^{\hph{a}c}H_{bcj}-R^{aibj}H_{a\hph{c}i}^{\hph{a}c}H_{bcj}+R^{aibj}H_{a\hph{c}j}^{\hph{a}c}H_{bci},\non\\
3H^{abi}H^{cd}_{\hph{cd}i}R_{[abc]d} &=& R^{abcd}H_{ab}^{\hph{ab}i}H_{cdi}-2R^{abcd}H_{ac}^{\hph{ac}i}H_{bdi},\non\\
3H^{abi}H_i^{\hph{i}jk}R_{[abj]k} &=& R^{abij}H_{ab}^{\hph{ab}k}H_{ijk}-2R^{aibj}H_{ab}^{\hph{ab}k}H_{ijk},\non\\
3H^{abi}H_i^{\hph{i}jk}R_{[ajk]b} &=& R^{abij}H_{ab}^{\hph{ab}k}H_{ijk}-2R^{aibj}H_{ab}^{\hph{ab}k}H_{ijk},\non\\
3H^{ijk}H_i^{\hph{i}\ell m}R_{[jk\ell]m} &=& R^{ijk\ell}H_{ij}^{\hph{ij}m}H_{k\ell m}-2R^{ijk\ell}H_{ik}^{\hph{ik}m}H_{j\ell m},\non
\eea
\bea
3R^{abcd}R_{[abc]d} &=& R^{abcd}R_{abcd}-2R^{abcd}R_{acbd},\non\\
3R^{abij}R_{[abi]j} &=& R^{abij}R_{abij}-2R^{abij}R_{aibj},\non\\
3R^{abij}R_{[aij]b} &=& R^{abij}R_{abij}-2R^{abij}R_{aibj},\non\\
3R^{aibj}R_{[abi]j} &=& R^{abij}R_{aibj}-R^{aibj}R_{aibj}+R^{aibj}R_{ajbi},\non\\
3R^{aibj}R_{[aij]b} &=& R^{abij}R_{aibj}-R^{aibj}R_{aibj}+R^{aibj}R_{ajbi},\non\\
3R^{ijk\ell}R_{[ijk]\ell} &=& R^{ijk\ell}R_{ijk\ell}-2R^{ijk\ell}R_{ikj\ell},\non
\eea
\bea
4\nabla^a\Phi H^{bci}\nabla_{[a}H_{bci]} &=& \nabla^a\Phi H^{bci}\nabla_aH_{bci}-2\nabla^a\Phi H^{bci}\nabla_bH_{aci}-\nabla^a\Phi H^{bci}\nabla_iH_{abc},\non\\
4\nabla^a\Phi H^{ijk}\nabla_{[a}H_{ijk]} &=& \nabla^a\Phi H^{ijk}\nabla_aH_{ijk}-3\nabla^a\Phi H^{ijk}\nabla_iH_{ajk},\non
\eea
\bea
4\nabla^aH^{bci}\nabla_{[a}H_{bci]} &=& \nabla^aH^{bci}\nabla_aH_{bci}-2\nabla^aH^{bci}\nabla_bH_{aci}-\nabla^aH^{bci}\nabla_iH_{abc},\non\\
4\nabla^aH^{ijk}\nabla_{[a}H_{ijk]} &=& \nabla^aH^{ijk}\nabla_aH_{ijk}-3\nabla^aH^{ijk}\nabla_iH_{ajk},\non\\
4\nabla^iH^{abc}\nabla_{[a}H_{bci]} &=& 3\nabla^aH^{bci}\nabla_iH_{abc}-\nabla^iH^{abc}\nabla_iH_{abc},\non\\
4\nabla^iH^{ajk}\nabla_{[a}H_{ijk]} &=& \nabla^aH^{ijk}\nabla_iH_{ajk}-\nabla^iH^{ajk}\nabla_iH_{ajk}+2\nabla^iH^{ajk}\nabla_jH_{aik},\non
\eea
\bea
4H^{abi}\nabla^c\nabla_{[a}H_{bci]} &=& 2H^{abi}\nabla_a^{\hph{a}c}H_{bci}+H^{abi}\nabla^c_{\hph{c}c}H_{abi}-H^{abi}\nabla^c_{\hph{c}i}H_{abc}+\lp RH^2\rp,\non\\
4H^{ijk}\nabla^a\nabla_{[a}H_{ijk]} &=& H^{ijk}\nabla^a_{\hph{a}a}H_{ijk}-3H^{ijk}\nabla^a_{\hph{a}i}H_{ajk}+\lp RH^2\rp,\non
\eea
\bea
3\nabla^a\Phi\nabla_{[a}R_{bc]}^{\hph{bc]}bc} &=& \nabla^a\Phi\nabla_aR^{bc}_{\hph{bc}bc}-2\nabla^a\Phi\nabla^bR_{a\hph{c}bc}^{\hph{a}c},\non\\
3\nabla^a\nabla_{[a}R_{bc]}^{\hph{bc]}bc} &=& \nabla^a_{\hph{a}a}R^{bc}_{\hph{bc}bc}-2\nabla^{ab}R_{a\hph{c}bc}^{\hph{a}c}
.\non
\eea
Note that this collection of null vectors is not linearly independent.

Similarly, we can find all the total derivatives\footnote{Actually, because of the factor of $e^{-\Phi}$ which multiplies all of these couplings in the action, these total derivatives are not truly null; integration by parts would replace the total derivative $\nabla^a$ by a factor of $\nabla^a\Phi$.  The resulting terms are always lower in the lexicographic ordering however, and do not affect the determination of which couplings can be eliminated.},
\be
\nabla^a\lp\nabla_a\Phi\nabla^b\Phi\nabla_b\Phi\rp=\nabla^a\Phi\nabla_a\Phi\nabla^b_{\hph{b}b}\Phi+2\nabla^a\Phi\nabla^b\Phi\nabla_{ab}\Phi,\non
\ee
\bea
\nabla^a\lp\nabla_a\Phi H^{bci}H_{bci}\rp &=& \nabla^a_{\hph{a}a}\Phi H^{bci}H_{bci}+2\nabla^a\Phi H^{bci}\nabla_aH_{bci},\non\\
\nabla^a\lp\nabla_a\Phi H^{ijk}H_{ijk}\rp &=& \nabla^a_{\hph{a}a}\Phi H^{ijk}H_{ijk}+2\nabla^a\Phi H^{ijk}\nabla_aH_{ijk},\non\\
\nabla^a\lp\nabla^b\Phi H_a^{\hph{a}ci}H_{bci}\rp &=& \nabla^{ab}\Phi H_a^{\hph{a}ci}H_{bci}-\nabla^a\Phi H_a^{\hph{a}bi}\nabla^cH_{bci}+\nabla^a\Phi H^{bci}\nabla_bH_{aci},\non
\eea
\bea
\nabla^a\lp\nabla_a\Phi\nabla^b_{\hph{b}b}\Phi\rp &=& \nabla^a_{\hph{a}a}\Phi\nabla^b_{\hph{b}b}\Phi+\nabla^a\Phi\nabla_{a\hph{b}b}^{\hph{a}b}\Phi+\lp\lp\nabla\Phi\rp^2R\rp,\non\\
\nabla^a\lp\nabla^b\Phi\nabla_{ab}\Phi\rp &=& \nabla^{ab}\Phi\nabla_{ab}\Phi+\nabla^a\Phi\nabla_{a\hph{b}b}^{\hph{a}b}\Phi+\lp\lp\nabla\Phi\rp^2R\rp,\non
\eea
\bea
\nabla^a\lp\nabla_a\Phi R^{bc}_{\hph{bc}bc}\rp &=& \nabla^a_{\hph{a}a}\Phi R^{bc}_{\hph{bc}bc}+\nabla^a\Phi\nabla_aR^{bc}_{\hph{bc}bc},\non\\
\nabla^a\lp\nabla^b\Phi R_{a\hph{c}bc}^{\hph{a}c}\rp &=& \nabla^{ab}\Phi R_{a\hph{c}bc}^{\hph{a}c}+\nabla^a\Phi\nabla^bR_{a\hph{c}bc}^{\hph{a}c},\non
\eea
\bea
\nabla^a\lp H_a^{\hph{a}bi}\nabla^cH_{bci}\rp &=& \nabla^aH_a^{\hph{a}bi}\nabla^cH_{bci}+H^{abi}\nabla_a^{\hph{a}c}H_{bci}+\lp RH^2\rp,\non\\
\nabla^a\lp H^{bci}\nabla_aH_{bci}\rp &=& \nabla^aH^{bci}\nabla_aH_{bci}+H^{abi}\nabla^c_{\hph{c}c}H_{abi},\non\\
\nabla^a\lp H^{bci}\nabla_bH_{aci}\rp &=& \nabla^aH^{bci}\nabla_bH_{aci}-H^{abi}\nabla_a^{\hph{a}c}H_{bci}+\lp RH^2\rp,\non\\
\nabla^a\lp H^{bci}\nabla_iH_{abc}\rp &=& \nabla^aH^{bci}\nabla_iH_{abc}+H^{abi}\nabla^c_{\hph{c}i}H_{abc}+\lp RH^2\rp,\non\\
\nabla^a\lp H^{ijk}\nabla_aH_{ijk}\rp &=& \nabla^aH^{ijk}\nabla_aH_{ijk}+H^{ijk}\nabla^a_{\hph{a}a}H_{ijk},\non\\
\nabla^a\lp H^{ijk}\nabla_iH_{ajk}\rp &=& \nabla^aH^{ijk}\nabla_iH_{ajk}+H^{ijk}\nabla^a_{\hph{a}i}H_{ajk}+\lp RH^2\rp,\non
\eea
\be
\nabla^a\lp\nabla_{a\hph{b}b}^{\hph{a}b}\Phi\rp=\nabla^{a\hph{a}b}_{\hph{a}a\hph{b}b}\Phi+\lp\nabla^2\Phi R,\nabla\Phi\nabla R\rp,\non
\ee
\bea
\nabla^a\lp\nabla_aR^{bc}_{\hph{bc}bc}\rp &=& \nabla^a_{\hph{a}a}R^{bc}_{\hph{bc}bc},\non\\
\nabla^a\lp\nabla^bR_{a\hph{c}bc}^{\hph{a}c}\rp &=& \nabla^{ab}R_{a\hph{c}bc}^{\hph{a}c},\non
\eea

At this point the remaining work is only linear algebra.  It can be checked that the Bianchi identities and total derivatives span a $32$-dimensional subspace of our $80$-dimensional space of couplings, leaving a $48$-dimensional quotient space representing physical couplings.  A basis for these physical couplings is given in section \ref{subsec:List}.

\section{Reduction and duality of couplings in the warped product}
\label{app:Warped}

For each of the forty-eight possible four-derivative couplings, we must reduce them in the case that $y$ is parallel to the brane, $y$ is perpendicular to the brane, and then compute the T-duality of the latter.  Below we omit the $H^4$ couplings $c_5,\cdots, c_{14}$ because all three expressions (parallel, perpendicular, and T-dual to perpendicular) are trivially equal, and we don't get any constraints on these coefficients.  There are also some other coefficients (certain $RH^2$, $R^2$, and $(\nabla H)^2$ terms) which will not be constrained, but we include them below for completeness.

\bea
c_1\sqrt{-g}e^{-\Phi}\nabla^a\Phi\nabla_a\Phi\nabla^b\Phi\nabla_b\Phi &=_\parallel& c_1e^{-\Phi+\hlf\vp}\p^a\Phi\p_a\Phi\p^b\Phi\p_b\Phi\non\\
&=_\perp& c_1e^{-\Phi}\p^a\Phi\p_a\Phi\p^b\Phi\p_b\Phi\non\\
&\rr& \frac{c_1}{16}e^{-\Phi+\hlf\vp}\ls16\p^a\Phi\p_a\Phi\p^b\Phi\p_b\Phi-32\p^a\Phi\p_a\Phi\p^b\Phi\p_b\vp\right.\non\\
&& \quad\left.+8\p^a\Phi\p_a\Phi\p^b\vp\p_b\vp+16\p^a\Phi\p^b\Phi\p_a\vp\p_b\vp\right.\non\\
&& \quad\left.-8\p^a\Phi\p_a\vp\p^b\vp\p_b\vp+\p^a\vp\p_a\vp\p^b\vp\p_b\vp\rs,\non
\eea
\bea
c_2\sqrt{-g}e^{-\Phi}\nabla^a\Phi\nabla_a\Phi H^{bci}H_{bci} &=_\parallel& c_2e^{-\Phi+\hlf\vp}\p^a\Phi\p_a\Phi H^{bci}H_{bci}\non\\
&=_\perp& c_2e^{-\Phi}\p^a\Phi\p_a\Phi H^{bci}H_{bci}\non\\
&\rr& \frac{c_2}{4}e^{-\Phi+\hlf\vp}\ls 4\p^a\Phi\p_a\Phi H^{bci}H_{bci}-4\p^a\Phi\p_a\vp H^{bci}H_{bci}\right.\non\\
&& \quad\left.+\p^a\vp\p_a\vp H^{bci}H_{bci}\rs,\non
\eea
\bea
c_3\sqrt{-g}e^{-\Phi}\nabla^a\Phi\nabla_a\Phi H^{ijk}H_{ijk} &=_\parallel& c_3e^{-\Phi+\hlf\vp}\p^a\Phi\p_a\Phi H^{ijk}H_{ijk}\non\\
&=_\perp& c_3e^{-\Phi}\p^a\Phi\p_a\Phi H^{ijk}H_{ijk}\non\\
&\rr& \frac{c_3}{4}e^{-\Phi+\hlf\vp}\ls 4\p^a\Phi\p_a\Phi H^{ijk}H_{ijk}-4\p^a\Phi\p_a\vp H^{ijk}H_{ijk}\right.\non\\
&& \quad\left.+\p^a\vp\p_a\vp H^{ijk}H_{ijk}\rs.\non
\eea
\bea
c_4\sqrt{-g}e^{-\Phi}\nabla^a\Phi\nabla^b\Phi H_a^{\hph{a}ci}H_{bci} &=_\parallel& c_4e^{-\Phi+\hlf\vp}\p^a\Phi\p^b\Phi H_a^{\hph{a}ci}H_{bci}\non\\
&=_\perp& c_4e^{-\Phi}\p^a\Phi\p^b\Phi H_a^{\hph{a}ci}H_{bci}\non\\
&\rr& \frac{c_4}{4}e^{-\Phi+\hlf\vp}\ls 4\p^a\Phi\p^b\Phi H_a^{\hph{a}ci}H_{bci}-4\p^a\Phi\p^b\vp H_a^{\hph{a}ci}H_{bci}\right.\non\\
&& \quad\left.+\p^a\vp\p^b\vp H_a^{\hph{a}ci}H_{bci}\rs,\non
\eea
\bea
c_{15}\sqrt{-g}e^{-\Phi}\nabla^a\Phi\nabla_a\Phi\nabla^b_{\hph{b}b}\Phi &=_\parallel& \frac{c_{15}}{2}e^{-\Phi+\hlf\vp}\ls\p^a\Phi\p_a\Phi\p^b\Phi\p_b\vp+2\p^a\Phi\p_a\Phi\p^b_{\hph{b}b}\Phi\rs\non\\
&=_\perp& c_{15}e^{-\Phi}\p^a\Phi\p_a\Phi\p^b_{\hph{b}b}\Phi\non\\
&\rr& \frac{c_{15}}{8}e^{-\Phi+\hlf\vp}\ls 8\p^a\Phi\p_a\Phi\p^b_{\hph{b}b}\Phi-8\p^a\Phi\p^b_{\hph{b}b}\Phi\p_a\vp+2\p^a_{\hph{a}a}\Phi\p^b\vp\p_b\vp\right.\non\\
&& \quad\left.-4\p^a\Phi\p_a\Phi\p^b_{\hph{b}b}\vp+4\p^a\Phi\p_a\vp\p^b_{\hph{b}b}\vp-\p^a\vp\p_a\vp\p^b_{\hph{b}b}\vp\rs,\non
\eea
\bea
c_{16}\sqrt{-g}e^{-\Phi}\nabla^a_{\hph{a}a}\Phi H^{bci}H_{bci} &=_\parallel& \frac{c_{16}}{2}e^{-\Phi+\hlf\vp}\ls\p^a\Phi\p_a\vp H^{bci}H_{bci}+2\p^a_{\hph{a}a}\Phi H^{bci}H_{bci}\rs\non\\
&=_\perp& c_{16}e^{-\Phi}\p^a_{\hph{a}a}\Phi H^{bci}H_{bci}\non\\
&\rr& \frac{c_{16}}{2}e^{-\Phi+\hlf\vp}\ls 2\p^a_{\hph{a}a}\Phi H^{bci}H_{bci}-\p^a_{\hph{a}a}\vp H^{bci}H_{bci}\rs,\non
\eea
\bea
c_{17}\sqrt{-g}e^{-\Phi}\nabla^a_{\hph{a}a}\Phi H^{ijk}H_{ijk} &=_\parallel& \frac{c_{17}}{2}e^{-\Phi+\hlf\vp}\ls\p^a\Phi\p_a\vp H^{ijk}H_{ijk}+2\p^a_{\hph{a}a}\Phi H^{ijk}H_{ijk}\rs\non\\
&=_\perp& c_{17}e^{-\Phi}\p^a_{\hph{a}a}\Phi H^{ijk}H_{ijk}\non\\
&\rr& \frac{c_{17}}{2}e^{-\Phi+\hlf\vp}\ls 2\p^a_{\hph{a}a}\Phi H^{ijk}H_{ijk}-\p^a_{\hph{a}a}\vp H^{ijk}H_{ijk}\rs,\non
\eea
\bea
c_{18}\sqrt{-g}e^{-\Phi}\nabla^{ab}\Phi H_a^{\hph{a}ci}H_{bci} &=_\parallel& c_{18}e^{-\Phi+\hlf\vp}\p^{ab}\Phi H_a^{\hph{a}ci}H_{bci}\non\\
&=_\perp& c_{18}e^{-\Phi}\p^{ab}\Phi H_a^{\hph{a}ci}H_{bci}\non\\
&\rr& \frac{c_{18}}{2}e^{-\Phi+\hlf\vp}\ls 2\p^{ab}\Phi H_a^{\hph{a}ci}H_{bci}-\p^{ab}\vp H_a^{\hph{a}ci}H_{bci}\rs,\non
\eea
\bea
c_{19}\sqrt{-g}e^{-\Phi}\nabla^{ij}\Phi H^{ab}_{\hph{ab}i}H_{abj} &=_\parallel& c_{19}e^{-\Phi+\hlf\vp}\p^{ij}\Phi H^{ab}_{\hph{ab}i}H_{abj}\non\\
&=_\perp& c_{19}e^{-\Phi}\p^{ij}\Phi H^{ab}_{\hph{ab}i}H_{abj}\non\\
&\rr& \frac{c_{19}}{2}e^{-\Phi+\hlf\vp}\ls 2\p^{ij}\Phi H^{ab}_{\hph{ab}i}H_{abj}-\p^{ij}\vp H^{ab}_{\hph{ab}i}H_{abj}\rs\non
\eea
\bea
c_{20}\sqrt{-g}e^{-\Phi}\nabla^{ij}\Phi H_i^{\hph{i}k\ell}H_{jk\ell} &=_\parallel& c_{20}e^{-\Phi+\hlf\vp}\p^{ij}\Phi H_i^{\hph{i}k\ell}H_{jk\ell}\non\\
&=_\perp& c_{20}e^{-\Phi}\p^{ij}\Phi H_i^{\hph{i}k\ell}H_{jk\ell}\non\\
&\rr& \frac{c_{20}}{2}e^{-\Phi+\hlf\vp}\ls 2\p^{ij}\Phi H_i^{\hph{i}k\ell}H_{jk\ell}-\p^{ij}\vp H_i^{\hph{i}k\ell}H_{jk\ell}\rs,\non
\eea
\bea
c_{21}\sqrt{-g}e^{-\Phi}\nabla^a_{\hph{a}a}\Phi\nabla^b_{\hph{b}b}\Phi &=_\parallel& \frac{c_{21}}{4}e^{-\Phi+\hlf\vp}\ls\p^a\Phi\p^b\Phi\p_a\vp\p_b\vp+4\p^a\Phi\p^b_{\hph{b}b}\Phi\p_a\vp+4\p^a_{\hph{a}a}\Phi\p^b_{\hph{b}b}\Phi\rs\non\\
&=_\perp& c_{21}e^{-\Phi}\p^a_{\hph{a}a}\Phi\p^b_{\hph{b}b}\Phi\non\\
&\rr& \frac{c_{21}}{4}e^{-\Phi+\hlf\vp}\ls 4\p^a_{\hph{a}a}\Phi\p^b_{\hph{b}b}\Phi-4\p^a_{\hph{a}a}\Phi\p^b_{\hph{b}b}\vp+\p^a_{\hph{a}a}\vp\p^b_{\hph{b}b}\vp\rs,\non
\eea
\bea
c_{22}\sqrt{-g}e^{-\Phi}\nabla^{ij}\Phi\nabla_{ij}\Phi &=_\parallel& c_{22}e^{-\Phi+\hlf\vp}\p^{ij}\Phi\p_{ij}\Phi\non\\
&=_\perp& \frac{c_{22}}{4}e^{-\Phi}\ls\p^a\Phi\p^b\Phi\p_a\vp\p_b\vp+4\p^{ij}\Phi\p_{ij}\Phi\rs\non\\
&\rr& \frac{c_{22}}{16}e^{-\Phi+\hlf\vp}\ls 4\p^a\Phi\p^b\Phi\p_a\vp\p_b\vp-4\p^a\Phi\p_a\vp\p^b\vp\p_b\vp\right.\non\\
&& \quad\left.+\p^a\vp\p_a\vp\p^b\vp\p_b\vp+16\p^{ij}\Phi\p_{ij}\Phi-16\p^{ij}\Phi\p_{ij}\vp+4\p^{ij}\vp\p_{ij}\vp\rs,\non
\eea
\bea
c_{23}\sqrt{-g}e^{-\Phi}\nabla^a\Phi\nabla_a\Phi R^{bc}_{\hph{bc}bc} &=_\parallel& \frac{c_{23}}{2}e^{-\Phi+\hlf\vp}\ls -\p^a\Phi\p_a\Phi\p^b\vp\p_b\vp-2\p^a\Phi\p_a\Phi\p^b_{\hph{b}b}\vp\rs\non\\
&=_\perp& 0\non\\
&\rr& 0,\non
\eea
\bea
c_{24}\sqrt{-g}e^{-\Phi}\nabla^a\Phi\nabla^b\Phi R_{a\hph{c}bc}^{\hph{a}c} &=_\parallel& \frac{c_{24}}{4}e^{-\Phi+\hlf\vp}\ls -\p^a\Phi\p^b\Phi\p_a\vp\p_b\vp-2\p^a\Phi\p^b\Phi\p_{ab}\vp\rs\non\\
&=_\perp& 0\non\\
&\rr& 0,\non
\eea
\bea
c_{25}\sqrt{-g}e^{-\Phi}R^{ab}_{\hph{ab}ab}H^{cdi}H_{cdi} &=_\parallel& \frac{c_{25}}{2}e^{-\Phi+\hlf\vp}\ls -\p^a\vp\p_a\vp H^{bci}H_{bci}-2\p^a_{\hph{a}a}\vp H^{bci}H_{bci}\rs\non\\
&=_\perp& 0\non\\
&\rr& 0,\non
\eea
\bea
c_{26}\sqrt{-g}e^{-\Phi}R^{ab}_{\hph{ab}ab}H^{ijk}H_{ijk} &=_\parallel& \frac{c_{26}}{2}e^{-\Phi+\hlf\vp}\ls -\p^a\vp\p_a\vp H^{ijk}H_{ijk}-2\p^a_{\hph{a}a}\vp H^{ijk}H_{ijk}\rs,\non\\
&=_\perp& 0\non\\
&\rr& 0,\non
\eea
\bea
c_{27}\sqrt{-g}e^{-\Phi}R^{ab\hph{a}c}_{\hph{ab}a}H_b^{\hph{b}di}H_{cdi} &=_\parallel& \frac{c_{27}}{4}e^{-\Phi+\hlf\vp}\ls -\p^a\vp\p^b\vp H_a^{\hph{a}ci}H_{bci}-2\p^{ab}\vp H_a^{\hph{a}ci}H_{bci}\rs\non\\
&=_\perp& 0\non\\
&\rr& 0,\non
\eea
\bea
c_{28}\sqrt{-g}e^{-\Phi}R^{abcd}H_{ab}^{\hph{ab}i}H_{cdi} &=_\parallel& 0\non\\
&=_\perp& 0\non\\
&\rr& 0,\non
\eea
\bea
c_{29}\sqrt{-g}e^{-\Phi}R^{abij}H_{ab}^{\hph{ab}k}H_{ijk} &=_\parallel& 0\non\\
&=_\perp& 0\non\\
&\rr& 0,\non
\eea
\bea
c_{30}\sqrt{-g}e^{-\Phi}R^{abij}H_{a\hph{c}i}^{\hph{a}c}H_{bcj} &=_\parallel& 0\non\\
&=_\perp& 0\non\\
&\rr& 0,\non
\eea
\bea
c_{31}\sqrt{-g}e^{-\Phi}R^{ai\hph{a}j}_{\hph{ai}a}H^{bc}_{\hph{bc}i}H_{bcj} &=_\parallel& -\frac{c_{31}}{2}e^{-\Phi+\hlf\vp}\p^{ij}\vp H^{ab}_{\hph{bc}i}H_{abj}\non\\
&=_\perp& 0\non\\
&\rr& 0,\non
\eea
\bea
c_{32}\sqrt{-g}e^{-\Phi}R^{ai\hph{a}j}_{\hph{ai}a}H_i^{\hph{i}k\ell}H_{jk\ell} &=_\parallel& -\frac{c_{32}}{2}e^{-\Phi+\hlf\vp}\p^{ij}\vp H_i^{\hph{i}k\ell}H_{jk\ell}\non\\
&=_\perp& 0\non\\
&\rr& 0,\non
\eea
\bea
c_{33}\sqrt{-g}e^{-\Phi}R^{aibj}H_{a\hph{c}i}^{\hph{a}c}H_{bcj} &=_\parallel& 0\non\\
&=_\perp& 0\non\\
&\rr& 0,\non
\eea
\bea
c_{34}\sqrt{-g}e^{-\Phi}R^{ijk\ell}H_{ij}^{\hph{ij}m}H_{k\ell m} &=_\parallel& 0\non\\
&=_\perp& 0\non\\
&\rr& 0,\non
\eea
\bea
c_{35}\sqrt{-g}e^{-\Phi}\nabla^a_{\hph{a}a}\Phi R^{bc}_{\hph{bc}bc} &=_\parallel& \frac{c_{35}}{4}e^{-\Phi+\hlf\vp}\ls -\p^a\Phi\p_a\vp\p^b\vp\p_b\vp-2\p^a_{\hph{a}a}\Phi\p^b\vp\p_b\vp-2\p^a\Phi\p_a\vp\p^b_{\hph{b}b}\vp\right.\non\\
&& \quad\left.-4\p^a_{\hph{a}a}\Phi\p^b_{\hph{b}b}\vp\rs\non\\
&=_\perp& 0\non\\
&\rr& 0,\non
\eea
\bea
c_{36}\sqrt{-g}e^{-\Phi}\nabla^{ij}\Phi R^a_{\hph{a}iaj} &=_\parallel& -\frac{c_{36}}{2}e^{-\Phi+\hlf\vp}\p^{ij}\Phi\p_{ij}\vp\non\\
&=_\perp& \frac{c_{36}}{8}e^{-\Phi}\ls -\p^a\Phi\p_a\vp\p^b\vp\p_b\vp-2\p^a\Phi\p_a\vp\p^b_{\hph{b}b}\vp\rs\non\\
&\rr& \frac{c_{36}}{16}e^{-\Phi+\hlf\vp}\ls 2\p^a\Phi\p_a\vp\p^b\vp\p_b\vp-\p^a\vp\p_a\vp\p^b\vp\p_b\vp-4\p^a\Phi\p_a\vp\p^b_{\hph{b}b}\vp\right.\non\\
&& \quad\left.+2\p^a\vp\p_a\vp\p^b_{\hph{b}b}\vp\rs\non
\eea
\bea
c_{37}\sqrt{-g}e^{-\Phi}R^{ab}_{\hph{ab}ab}R^{cd}_{\hph{cd}cd} &=_\parallel& \frac{c_{37}}{4}e^{-\Phi+\hlf\vp}\ls\p^a\vp\p_a\vp\p^b\vp\p_b\vp+4\p^a\vp\p_a\vp\p^b_{\hph{b}b}\vp+4\p^a_{\hph{a}a}\vp\p^b_{\hph{b}b}\vp\rs\non\\
&=_\perp& 0\non\\
&\rr& 0,\non
\eea
\bea
c_{38}\sqrt{-g}e^{-\Phi}R^{ab\hph{a}c}_{\hph{ab}a}R_{b\hph{d}cd}^{\hph{b}d} &=\parallel& \frac{c_{38}}{8}e^{-\Phi+\hlf\vp}\ls\p^a\vp\p_a\vp\p^b\vp\p_b\vp+2\p^a\vp\p_a\vp\p^b_{\hph{b}b}\vp+2\p^a\vp\p^b\vp\p_{ab}\vp\right.\non\\
&& \quad\left.+2\p^a_{\hph{a}a}\vp\p^b_{\hph{b}b}\vp+2\p^{ab}\vp\p_{ab}\vp\rs\non\\
&=_\perp& 0\non\\
&\rr& 0,\non
\eea
\bea
c_{39}\sqrt{-g}e^{-\Phi}R^{abcd}R_{abcd} &=_\parallel& \frac{c_{39}}{4}e^{-\Phi+\hlf\vp}\ls\p^a\vp\p_a\vp\p^b\vp\p_b\vp+4\p^a\vp\p^b\vp\p_{ab}\vp+4\p^{ab}\vp\p_{ab}\vp\rs\non\\
&=_\perp& 0\non\\
&\rr& 0,\non
\eea
\bea
c_{40}\sqrt{-g}e^{-\Phi}R^{abij}R_{abij} &=_\parallel& 0\non\\
&=_\perp& 0\non\\
&\rr& 0,\non
\eea
\bea
c_{41}\sqrt{-g}e^{-\Phi}R^{ai\hph{a}j}_{\hph{ai}a}R^b_{\hph{b}ibj} &=_\parallel& \frac{c_{41}}{4}e^{-\Phi+\hlf\vp}\p^{ij}\vp\p_{ij}\vp \non\\
&=_\perp& \frac{c_{41}}{16}e^{-\Phi}\ls\p^a\vp\p_a\vp\p^b\vp\p_b\vp+4\p^a\vp\p_a\vp\p^b_{\hph{b}b}\vp+4\p^a_{\hph{a}a}\vp\p^b_{\hph{b}b}\vp\rs\non\\
&\rr& \frac{c_{41}}{16}e^{-\Phi+\hlf\vp}\ls\p^a\vp\p_a\vp\p^b\vp\p_b\vp-4\p^a\vp\p_a\vp\p^b_{\hph{b}b}\vp+4\p^a_{\hph{a}a}\vp\p^b_{\hph{b}b}\vp\rs,\non
\eea
\bea
c_{42}\sqrt{-g}e^{-\Phi}R^{aibj}R_{aibj} &=_\parallel& \frac{c_{42}}{4}e^{-\Phi+\hlf\vp}\p^{ij}\vp\p_{ij}\vp\non\\
&=_\perp& \frac{c_{42}}{16}e^{-\Phi}\ls\p^a\vp\p_a\vp\p^b\vp\p_b\vp+4\p^a\vp\p^b\vp\p_{ab}\vp+4\p^{ab}\vp\p_{ab}\vp\rs\non\\
&\rr& \frac{c_{42}}{16}e^{-\Phi+\hlf\vp}\ls\p^a\vp\p_a\vp\p^b\vp\p_b\vp-4\p^a\vp\p^b\vp\p_{ab}\vp+4\p^{ab}\vp\p_{ab}\vp\rs,\non
\eea
\bea
c_{43}\sqrt{-g}e^{-\Phi}R^{ijk\ell}R_{ijk\ell} &=_\parallel& 0\non\\
&=_\perp& c_{43}e^{-\Phi}\p^{ij}\vp\p_{ij}\vp\non\\
&\rr& c_{43}e^{-\Phi+\hlf\vp}\p^{ij}\vp\p_{ij}\vp,\non
\eea
\bea
c_{44}\sqrt{-g}e^{-\Phi}\nabla^a\Phi H_a^{\hph{a}bi}\nabla^cH_{bci} &=_\parallel& \frac{c_{44}}{2}e^{-\Phi+\hlf\vp}\ls -\p^a\Phi\p^b\vp H_a^{\hph{a}ci}H_{bci}+2\p^a\Phi H_a^{\hph{a}bi}\p^cH_{bci}\rs\non\\
&=_\perp& c_{44}e^{-\Phi}\p^a\Phi H_a^{\hph{a}bi}\p^cH_{bci}\non\\
&\rr& \frac{c_{44}}{2}e^{-\Phi+\hlf\vp}\ls 2\p^a\Phi H_a^{\hph{a}bi}\p^cH_{bci}-\p^a\vp H_a^{\hph{a}bi}\p^cH_{bci}\rs,\non
\eea
\bea
c_{45}\sqrt{-g}e^{-\Phi}\nabla^aH_a^{\hph{a}bi}\nabla^cH_{bci} &=_\parallel& \frac{c_{45}}{4}e^{-\Phi+\hlf\vp}\ls -\p^a\vp\p^b\vp H_a^{\hph{a}ci}H_{bci}+4\p^a\vp H_a^{\hph{a}bi}\p^cH_{bci}\right.\non\\
&& \quad\left.+4\p^aH_a^{\hph{a}bi}\p^cH_{bci}\rs\non\\
&=_\perp& c_{45}e^{-\Phi}\p^aH_a^{\hph{a}bi}\p^cH_{bci}\non\\
&\rr& c_{45}e^{-\Phi+\hlf\vp}\p^aH_a^{\hph{a}bi}\p^cH_{bci},\non
\eea
\bea
c_{46}\sqrt{-g}e^{-\Phi}\nabla^aH^{bci}\nabla_aH_{bci} &=_\parallel& \frac{c_{46}}{2}e^{-\Phi+\hlf\vp}\ls\p^a\vp\p^b\vp H_a^{\hph{a}ci}H_{bci}+2\p^aH^{bci}\p_aH_{bci}\rs\non\\
&=_\perp& c_{46}e^{-\Phi}\p^aH^{bci}\p_aH_{bci}\non\\
&\rr& c_{46}e^{-\Phi+\hlf\vp}\p^aH^{bci}\p_aH_{bci},\non
\eea
\bea
c_{47}\sqrt{-g}e^{-\Phi}\nabla^aH^{ijk}\nabla_aH_{ijk} &=_\parallel& c_{47}e^{-\Phi+\hlf\vp}\p^aH^{ijk}\p_aH_{ijk}\non\\
&=_\perp& c_{47}e^{-\Phi}\p^aH^{ijk}\p_aH_{ijk}\non\\
&\rr& c_{47}e^{-\Phi+\hlf\vp}\p^aH^{ijk}\p_aH_{ijk},\non
\eea
\bea
c_{48}\sqrt{-g}e^{-\Phi}\nabla^iH^{ajk}\nabla_iH_{ajk} &=_\parallel& c_{48}e^{-\Phi+\hlf\vp}\p^iH^{ajk}\p_iH_{ajk}\non\\
&=_\perp& \frac{c_{48}}{2}e^{-\Phi}\ls\p^a\vp\p^b\vp H_a^{\hph{a}ci}H_{bci}+2\p^iH^{ajk}\p_iH_{ajk}\rs\non\\
&\rr& \frac{c_{48}}{2}e^{-\Phi+\hlf\vp}\ls\p^a\vp\p^b\vp H_a^{\hph{a}ci}H_{bci}+2\p^iH^{ajk}\p_iH_{ajk}\rs.\non
\eea

We will also need the warped product combinations (with $y$ parallel to the O-plane) which are zero either because of a Bianchi identity or because they are a total derivative on the O-plane.  Proceeding as in section \ref{sec:Classifying}, we list the Bianchi combinations, this time built only from $\Phi$, $\vp$, and $H_{\m\n\rho}$:
\bea
4x_1\p^a\Phi H^{bci}\p_{[a}H_{bci]} &=& x_1\ls\p^a\Phi H^{bci}\p_aH_{bci}-2\p^a\Phi H^{bci}\p_bH_{aci}-\p^a\Phi H^{bci}\p_iH_{abc}\rs,\non\\
4x_2\p^a\Phi H^{ijk}\p_{[a}H_{ijk]} &=& x_2\ls\p^a\Phi H^{ijk}\p_aH_{ijk}-3\p^a\Phi H^{ijk}\p_iH_{ajk},\rs\non\\
4x_3\p^a\vp H^{bci}\p_{[a}H_{bci]} &=& x_3\ls\p^a\vp H^{bci}\p_aH_{bci}-2\p^a\vp H^{bci}\p_bH_{aci}-\p^a\vp H^{bci}\p_iH_{abc}\rs,\non\\
4x_4\p^a\vp H^{ijk}\p_{[a}H_{ijk]} &=& x_4\ls\p^a\vp H^{ijk}\p_aH_{ijk}-3\p^a\vp H^{ijk}\p_iH_{ajk}\rs,\non
\eea
\bea
4x_5\p^aH^{bci}\p_{[a}H_{bci]} &=& x_5\ls\p^aH^{bci}\p_aH_{bci}-2\p^aH^{bci}\p_bH_{aci}-\p^aH^{bci}\p_iH_{abc}\rs,\non\\
4x_6\p^aH^{ijk}\p_{[a}H_{ijk]} &=& x_6\ls\p^aH^{ijk}\p_aH_{ijk}-3\p^aH^{ijk}\p_iH_{ajk}\rs,\non\\
4x_7\p^iH^{abc}\p_{[a}H_{bci]} &=& x_7\ls 3\p^aH^{bci}\p_iH_{abc}-\p^iH^{abc}\p_iH_{abc}\rs,\non\\
4x_8\p^iH^{ajk}\p_{[a}H_{ijk]} &=& x_8\ls\p^aH^{ijk}\p_iH_{ajk}-\p^iH^{ajk}\p_iH_{ajk}+2\p^iH^{ajk}\p_jH_{aik}\rs,\non
\eea
\bea
4x_9H^{abi}\p^c\p_{[a}H_{bci]} &=& x_9\ls 2H^{abi}\p_a^{\hph{a}c}H_{bci}+H^{abi}\p^c_{\hph{c}c}H_{abi}-H^{abi}\p^c_{\hph{c}i}H_{abc}\rs,\non\\
4x_{10}H^{ijk}\p^a\p_{[a}H_{ijk]} &=& x_{10}\ls H^{ijk}\p^a_{\hph{a}a}H_{ijk}-3H^{ijk}\p^a_{\hph{a}i}H_{ajk}\rs,\non
\eea
and the total derivatives,
\bea
y_1\p^a\lp e^{-\Phi+\hlf\vp}\p_a\Phi\p^b\Phi\p_b\Phi\rp &=& \frac{y_1}{2}e^{-\Phi+\hlf\vp}\ls -2\p^a\Phi\p_a\Phi\p^b\Phi\p_b\Phi+\p^a\Phi\p_a\Phi\p^b\Phi\p_b\vp\right.\non\\
&& \quad\left.+2\p^a\Phi\p_a\Phi\p^b_{\hph{b}b}\Phi+4\p^a\Phi\p^b\Phi\p_{ab}\Phi\rs,\non
\eea
\bea
y_2\p^a\lp e^{-\Phi+\hlf\vp}\p_a\Phi\p^b\Phi\p_b\vp\rp &=& \frac{y_2}{2}e^{-\Phi+\hlf\vp}\ls -2\p^a\Phi\p_a\Phi\p^b\Phi\p_b\vp+\p^a\Phi\p^b\Phi\p_a\vp\p_b\vp\right.\non\\
&& \quad\left.+2\p^a\Phi\p_a^{\hph{a}b}\Phi\p_b\vp+2\p^a\Phi\p^b_{\hph{b}b}\Phi\p_a\vp+2\p^a\Phi\p^b\Phi\p_{ab}\vp\rs,\non\\
y_3\p^a\lp e^{-\Phi+\hlf\vp}\p^b\Phi\p_b\Phi\p_a\vp\rp &=& \frac{y_3}{2}e^{-\Phi+\hlf\vp}\ls -2\p^a\Phi\p_a\Phi\p^b\Phi\p_b\vp+\p^a\Phi\p_a\Phi\p^b\vp\p_b\vp\right.\non\\
&& \quad\left.+4\p^a\Phi\p_a^{\hph{a}b}\Phi\p_b\vp+2\p^a\Phi\p_a\Phi\p^b_{\hph{b}b}\vp\rs,\non
\eea
\bea
y_4\p^a\lp e^{-\Phi+\hlf\vp}\p_a\Phi\p^b\vp\p_b\vp\rp &=& \frac{y_4}{2}e^{-\Phi+\hlf\vp}\ls -2\p^a\Phi\p_a\Phi\p^b\vp\p_b\vp+\p^a\Phi\p_a\vp\p^b\vp\p_b\vp\right.\non\\
&& \quad\left.+2\p^a_{\hph{a}a}\Phi\p^b\vp\p_b\vp+4\p^a\Phi\p^b\vp\p_{ab}\vp\rs,\non\\
y_5\p^a\lp e^{-\Phi+\hlf\vp}\p^b\Phi\p_a\vp\p_b\vp\rp &=& \frac{y_5}{2}e^{-\Phi+\hlf\vp}\ls -2\p^a\Phi\p^b\Phi\p_a\vp\p_b\vp+\p^a\Phi\p_a\vp\p^b\vp\p_b\vp\right.\non\\
&& \quad\left.+2\p^{ab}\Phi\p_a\vp\p_b\vp+2\p^a\Phi\p_a\vp\p^b_{\hph{b}b}\vp+2\p^a\Phi\p^b\vp\p_{ab}\vp\rs,\non
\eea
\bea
y_6\p^a\lp e^{-\Phi+\hlf\vp}\p_a\vp\p^b\vp\p_b\vp\rp &=& \frac{y_6}{2}e^{-\Phi+\hlf\vp}\ls -2\p^a\Phi\p_a\vp\p^b\vp\p_b\vp+\p^a\vp\p_a\vp\p^b\vp\p_b\vp\right.\non\\
&& \quad\left.+2\p^a\vp\p_a\vp\p^b_{\hph{b}b}\vp+4\p^a\vp\p^b\vp\p_{ab}\vp\rs,\non
\eea
\bea
y_7\p^a\lp e^{-\Phi+\hlf\vp}\p_a\Phi H^{bci}H_{bci}\rp &=& \frac{y_7}{2}e^{-\Phi+\hlf\vp}\ls -2\p^a\Phi\p_a\Phi H^{bci}H_{bci}+\p^a\Phi\p_a\vp H^{bci}H_{bci}\right.\non\\
&& \quad\left. +2\p^a_{\hph{a}a}\Phi H^{bci}H_{bci}+4\p^a\Phi H^{bci}\p_aH_{bci}\rs,\non\\
y_8\p^a\lp e^{-\Phi+\hlf\vp}\p^a\Phi H^{ijk}H_{ijk}\rp &=& \frac{y_8}{2}e^{-\Phi+\hlf\vp}\ls -2\p^a\Phi\p_a\Phi H^{ijk}H_{ijk}+\p^a\Phi\p_a\vp H^{ijk}H_{ijk}\right.\non\\
&& \quad\left. +2\p^a_{\hph{a}a}\Phi H^{ijk}H_{ijk}+4\p^a\Phi H^{ijk}\p_aH_{ijk}\rs,\non\\
y_9\p^a\lp e^{-\Phi+\hlf\vp}\p^b\Phi H_a^{\hph{a}ci}H_{bci}\rp &=& \frac{y_9}{2}e^{-\Phi+\hlf\vp}\ls -2\p^a\Phi\p^b\Phi H_a^{\hph{a}ci}H_{bci}+\p^a\Phi\p^b\vp H_a^{\hph{a}ci}H_{bci}\right.\non\\
&& \quad\left. +2\p^{ab}\Phi H_a^{\hph{a}ci}H_{bci}-2\p^a\Phi H_a^{\hph{a}bi}\p^cH_{bci}+2\p^a\Phi H^{bci}\p_bH_{aci}\rs,\non
\eea
\bea
y_{10}\p^a\lp e^{-\Phi+\hlf\vp}\p_a\vp H^{bci}H_{bci}\rp &=& \frac{y_{10}}{2}e^{-\Phi+\hlf\vp}\ls -2\p^a\Phi\p_a\vp H^{bci}H_{bci}+\p^a\vp\p_a\vp H^{bci}H_{bci}\right.\non\\
&& \quad\left. +2\p^a_{\hph{a}a}\vp H^{bci}H_{bci}+4\p^a\vp H^{bci}\p_aH_{bci}\rs,\non\\
y_{11}\p^a\lp e^{-\Phi+\hlf\vp}\p^a\vp H^{ijk}H_{ijk}\rp &=& \frac{y_{11}}{2}e^{-\Phi+\hlf\vp}\ls -2\p^a\Phi\p_a\vp H^{ijk}H_{ijk}+\p^a\vp\p_a\vp H^{ijk}H_{ijk}\right.\non\\
&& \quad\left. +2\p^a_{\hph{a}a}\vp H^{ijk}H_{ijk}+4\p^a\vp H^{ijk}\p_aH_{ijk}\rs,\non\\
y_{12}\p^a\lp e^{-\Phi+\hlf\vp}\p^b\vp H_a^{\hph{a}ci}H_{bci}\rp &=& \frac{y_{12}}{2}e^{-\Phi+\hlf\vp}\ls -2\p^a\Phi\p^b\vp H_a^{\hph{a}ci}H_{bci}+\p^a\vp\p^b\vp H_a^{\hph{a}ci}H_{bci}\right.\non\\
&& \quad\left. +2\p^{ab}\vp H_a^{\hph{a}ci}H_{bci}-2\p^a\vp H_a^{\hph{a}bi}\p^cH_{bci}+2\p^a\vp H^{bci}\p_bH_{aci}\rs,\non
\eea
\bea
y_{13}\p^a\lp e^{-\Phi+\hlf\vp}\p_a\Phi\p^b_{\hph{b}b}\Phi\rp &=& \frac{y_{13}}{2}e^{-\Phi+\hlf\vp}\ls -2\p^a\Phi\p_a\Phi\p^b_{\hph{b}b}\Phi+\p^a\Phi\p^b_{\hph{b}b}\Phi\p_a\vp+2\p^a_{\hph{a}a}\Phi\p^b_{\hph{b}b}\Phi\right.\non\\
&& \quad\left. +2\p^a\Phi\p_{a\hph{b}b}^{\hph{a}b}\Phi\rs,\non\\
y_{14}\p^a\lp e^{-\Phi+\hlf\vp}\p^b\Phi\p_{ab}\Phi\rp &=& \frac{y_{14}}{2}e^{-\Phi+\hlf\vp}\ls -2\p^a\Phi\p^b\Phi\p_{ab}\Phi+\p^a\Phi\p_a^{\hph{a}b}\Phi\p_b\vp+2\p^{ab}\Phi\p_{ab}\Phi\right.\non\\
&& \quad\left. +2\p^a\Phi\p_{a\hph{b}b}^{\hph{a}b}\Phi\rs,\non
\eea
\bea
y_{15}\p^a\lp e^{-\Phi+\hlf\vp}\p_a^{\hph{a}b}\Phi\p_b\vp\rp &=& \frac{y_{15}}{2}e^{-\Phi+\hlf\vp}\ls -2\p^a\Phi\p_a^{\hph{a}b}\Phi\p_b\vp+\p^{ab}\Phi\p_a\vp\p_b\vp+2\p^{ab}\Phi\p_{ab}\vp\right.\non\\
&& \quad\left. +2\p^{a\hph{a}b}_{\hph{a}a}\Phi\p_b\vp\rs,\non\\
y_{16}\p^a\lp e^{-\Phi+\hlf\vp}\p^b_{\hph{b}b}\Phi\p_a\vp\rp &=& \frac{y_{16}}{2}e^{-\Phi+\hlf\vp}\ls -2\p^a\Phi\p^b_{\hph{b}b}\Phi\p_a\vp+\p^a_{\hph{a}a}\Phi\p^b\vp\p_b\vp+2\p^a_{\hph{a}a}\Phi\p^b_{\hph{b}b}\vp\right.\non\\
&& \quad\left. +2\p^{a\hph{a}b}_{\hph{a}a}\Phi\p_b\vp\rs,\non
\eea
\bea
y_{17}\p^a\lp e^{-\Phi+\hlf\vp}\p_a\Phi\p^b_{\hph{b}b}\vp\rp &=& \frac{y_{17}}{2}e^{-\Phi+\hlf\vp}\ls -2\p^a\Phi\p_a\Phi\p^b_{\hph{b}b}\vp+\p^a\Phi\p_a\vp\p^b_{\hph{b}b}\vp+2\p^a_{\hph{a}a}\Phi\p^b_{\hph{b}b}\vp\right.\non\\
&& \quad\left. +2\p^a\Phi\p_{a\hph{b}b}^{\hph{a}b}\vp\rs,\non\\
y_{18}\p^a\lp e^{-\Phi+\hlf\vp}\p^b\Phi\p_{ab}\vp\rp &=& \frac{y_{18}}{2}e^{-\Phi+\hlf\vp}\ls -2\p^a\Phi\p^b\Phi\p_{ab}\vp+\p^a\Phi\p^b\vp\p_{ab}\vp+2\p^{ab}\Phi\p_{ab}\vp\right.\non\\
&& \quad\left. +2\p^a\Phi\p_{a\hph{b}b}^{\hph{a}b}\vp\rs,\non
\eea
\bea
y_{19}\p^a\lp e^{-\Phi+\hlf\vp}\p_a\vp\p^b_{\hph{b}b}\vp\rp &=& \frac{y_{19}}{2}e^{-\Phi+\hlf\vp}\ls -2\p^a\Phi\p_a\vp\p^b_{\hph{b}b}\vp+\p^a\vp\p_a\vp\p^b_{\hph{b}b}\vp+2\p^a_{\hph{a}a}\vp\p^b_{\hph{b}b}\vp\right.\non\\
&& \quad\left. +2\p^a\vp\p_{a\hph{b}b}^{\hph{a}b}\vp\rs,\non\\
y_{20}\p^a\lp e^{-\Phi+\hlf\vp}\p^b\vp\p_{ab}\vp\rp &=& \frac{y_{20}}{2}e^{-\Phi+\hlf\vp}\ls -2\p^a\Phi\p^b\vp\p_{ab}\vp+\p^a\vp\p^b\vp\p_{ab}\vp+2\p^{ab}\vp\p_{ab}\vp\right.\non\\
&& \quad\left. +2\p^a\vp\p_{a\hph{b}b}^{\hph{a}b}\vp\rs,\non
\eea
\bea
y_{21}\p^a\lp e^{-\Phi+\hlf\vp}H_a^{\hph{a}bi}\p^cH_{bci}\rp &=& \frac{y_{21}}{2}e^{-\Phi+\hlf\vp}\ls -2\p^a\Phi H_a^{\hph{a}bi}\p^cH_{bci}+\p^a\vp H_a^{\hph{a}bi}\p^cH_{bci}\right.\non\\
&& \quad\left. +2\p^aH_a^{\hph{a}bi}\p^cH_{bci}+2H^{abi}\p_a^{\hph{a}c}H_{bci}\rs,\non\\
y_{22}\p^a\lp e^{-\Phi+\hlf\vp}H^{bci}\p_aH_{bci}\rp &=& \frac{y_{22}}{2}e^{-\Phi+\hlf\vp}\ls -2\p^a\Phi H^{bci}\p_aH_{bci}+\p^a\vp H^{bci}\p_aH_{bci}\right.\non\\
&& \quad\left. +2\p^aH^{bci}\p_aH_{bci}+2H^{abi}\p^c_{\hph{c}c}H_{abi}\rs,\non\\
y_{23}\p^a\lp e^{-\Phi+\hlf\vp}H^{bci}\p_bH_{aci}\rp &=& \frac{y_{23}}{2}e^{-\Phi+\hlf\vp}\ls -2\p^a\Phi H^{bci}\p_bH_{aci}+\p^a\vp H^{bci}\p_bH_{aci}\right.\non\\
&& \quad\left. +2\p^aH^{bci}\p_bH_{aci}-2H^{abi}\p_a^{\hph{a}c}H_{bci}\rs,\non\\
y_{24}\p^a\lp e^{-\Phi+\hlf\vp}H^{bci}\p_iH_{abc}\rp &=& \frac{y_{24}}{2}e^{-\Phi+\hlf\vp}\ls -2\p^a\Phi H^{bci}\p_iH_{abc}+\p^a\vp H^{bci}\p_iH_{abc}\right.\non\\
&& \quad\left. +2\p^aH^{bci}\p_iH_{abc}+2H^{abi}\p^c_{\hph{c}i}H_{abc}\rs,\non\\
y_{25}\p^a\lp e^{-\Phi+\hlf\vp}H^{ijk}\p_aH_{ijk}\rp &=& \frac{y_{25}}{2}e^{-\Phi+\hlf\vp}\ls -2\p^a\Phi H^{ijk}\p_aH_{ijk}+\p^a\vp H^{ijk}\p_aH_{ijk}\right.\non\\
&& \quad\left. +2\p^aH^{ijk}\p_aH_{ijk}+2H^{ijk}\p^a_{\hph{a}a}H_{ijk}\rs,\non\\
y_{26}\p^a\lp e^{-\Phi+\hlf\vp}H^{ijk}\p_iH_{ajk}\rp &=& \frac{y_{26}}{2}e^{-\Phi+\hlf\vp}\ls -2\p^a\Phi H^{ijk}\p_iH_{ajk}+\p^a\vp H^{ijk}\p_iH_{ajk}\right.\non\\
&& \quad\left. +2\p^aH^{ijk}\p_iH_{ajk}+2H^{ijk}\p^a_{\hph{a}i}H_{ajk}\rs,\non
\eea
\bea
y_{27}\p^a\lp e^{-\Phi+\hlf\vp}\p_{a\hph{b}b}^{\hph{a}b}\Phi\rp &=& \frac{y_{27}}{2}e^{-\Phi+\hlf\vp}\ls -2\p^a\Phi\p_{a\hph{b}b}^{\hph{a}b}\Phi+\p^{a\hph{a}b}_{\hph{a}a}\Phi\p_b\vp+2\p^{a\hph{a}b}_{\hph{a}a\hph{b}b}\Phi\rs,\non
\eea
\bea
y_{28}\p^a\lp e^{-\Phi+\hlf\vp}\p_{a\hph{b}b}^{\hph{a}b}\vp\rp &=& \frac{y_{28}}{2}e^{-\Phi+\hlf\vp}\ls -2\p^a\Phi\p_{a\hph{b}b}^{\hph{a}b}\vp+\p^a\vp\p_{a\hph{b}b}^{\hph{a}b}\vp+2\p^{a\hph{a}b}_{\hph{a}a\hph{b}b}\vp\rs.\non
\eea

Finally, enforcing that the parallel action minus the T-dual of the perpendicular action is zero, up to Bianchi identities and total derivatives, leads to
\bea
0 &=& e^{-\Phi+\hlf\vp}\left\{
\lp -y_1\rp\p^a\Phi\p_a\Phi\p^b\Phi\p_b\Phi
+\frac{1}{2}\lp 4c_1+c_{15}-2y_2-2y_3\rp\p^a\Phi\p_a\Phi\p^b\Phi\p_b\vp\right.\non\\
&& \left.
+\frac{1}{2}\lp -c_1-c_{23}+y_3-2y_4\rp\p^a\Phi\p_a\Phi\p^b\vp\p_b\vp\right.\non\\
&& \left.
+\frac{1}{4}\lp -4c_1+c_{21}-c_{22}-c_{24}+2y_2-4y_5\rp\p^a\Phi\p^b\Phi\p_a\vp\p_b\vp\right.\non\\
&& \left.
+\frac{1}{8}\lp 4c_1+2c_{22}-2c_{35}-c_{36}+4y_1+4y_4+4y_5-8y_6\rp\p^a\Phi\p_a\vp\p^b\vp\p_b\vp\right.\non\\
&& \left.
+\frac{1}{16}\lp -c_1-c_{22}+c_{36}+4c_{37}+2c_{38}+4c_{39}-c_{41}-c_{42}+8y_6\rp\p^a\vp\p_a\vp\p^b\vp\p_b\vp\right.\non\\
&& \left.
+\lp -y_7\rp\p^a\Phi\p_a\Phi H^{bci}H_{bci}
+\lp -y_8\rp\p^a\Phi\p_a\Phi H^{ijk}H_{ijk}
+\lp -y_9\rp\p^a\Phi\p^b\Phi H_a^{\hph{a}ci}H_{bci}\right.\non\\
&& \left.
+\frac{1}{2}\lp 2c_2+c_{16}+y_7-2y_{10}\rp\p^a\Phi\p_a\vp H^{bci}H_{bci}\right.\non\\
&& \left.
+\frac{1}{2}\lp 2c_3+c_{17}+y_8-2y_{11}\rp\p^a\Phi\p_a\vp H^{ijk}H_{ijk}\right.\non\\
&& \left.
+\frac{1}{2}\lp 2c_4-c_{44}+y_9-2y_{12}\rp\p^a\Phi\p^b\vp H_a^{\hph{a}ci}H_{bci}
+\frac{1}{4}\lp -c_2-2c_{25}+2y_{10}\rp\p^a\vp\p_a\vp H^{bci}H_{bci}\right.\non
\eea
\bea
&& \left.
+\frac{1}{4}\lp -c_3-2c_{26}+2y_{11}\rp\p^a\vp\p_a\vp H^{ijk}H_{ijk}\right.\non\\
&& \left.
+\frac{1}{4}\lp -c_4-c_{27}-c_{45}+2c_{46}-2c_{48}+2y_{12}\rp\p^a\vp\p^b\vp H_a^{\hph{a}ci}H_{bci}
+\lp y_1-y_{13}\rp\p^a\Phi\p_a\Phi\p^b_{\hph{b}b}\Phi\right.\non\\
&& \left.
+\lp 2y_1-y_{14}\rp\p^a\Phi\p^b\Phi\p_{ab}\Phi
+\frac{1}{2}\lp 2y_2+4y_3+y_{14}-2y_{15}\rp\p^a\Phi\p_a^{\hph{a}b}\Phi\p_b\vp\right.\non\\
&& \left.
+\frac{1}{2}\lp 2c_{15}+2c_{21}+2y_2+y_{13}-2y_{16}\rp\p^a\Phi\p^b_{\hph{b}b}\Phi\p_a\vp\right.\non\\
&& \left.
+\frac{1}{4}\lp -c_{15}-2c_{35}+4y_4+2y_{16}\rp\p^a_{\hph{a}a}\Phi\p^b\vp\p_b\vp
+\frac{1}{2}\lp 2y_5+y_{15}\rp\p^{ab}\Phi\p_a\vp\p_b\vp\right.\non\\
&& \left.
+\lp y_7\rp\p^a_{\hph{a}a}\Phi H^{bci}H_{bci}
+\lp y_8\rp\p^a_{\hph{a}a}\Phi H^{ijk}H_{ijk}
+\lp y_9\rp\p^{ab}\Phi H_a^{\hph{a}ci}H_{bci}
+\lp y_{13}\rp\p^a_{\hph{a}a}\Phi\p^b_{\hph{b}b}\Phi\right.\non\\
&& \left.
+\lp y_{14}\rp\p^{ab}\Phi\p_{ab}\Phi
+\frac{1}{2}\lp c_{15}-2c_{23}+2y_3-2y_{17}\rp\p^a\Phi\p_a\Phi\p^b_{\hph{b}b}\vp\right.\non\\
&& \left.
+\frac{1}{2}\lp -c_{24}+2y_2-2y_{18}\rp\p^a\Phi\p^b\Phi\p_{ab}\vp\right.\non\\
&& \left.
+\frac{1}{4}\lp -2c_{15}-2c_{35}+c_{36}+4y_5+2y_{17}-4y_{19}\rp\p^a\Phi\p_a\vp\p^b_{\hph{b}b}\vp\right.\non\\
&& \left.
+\frac{1}{2}\lp 4y_4+2y_5+y_{18}-2y_{20}\rp\p^a\Phi\p^b\vp\p_{ab}\vp\right.\non\\
&& \left.
+\frac{1}{8}\lp c_{15}-c_{36}+8c_{37}+2c_{38}+2c_{41}+8y_6+4y_{19}\rp\p^a\vp\p_a\vp\p^b_{\hph{b}b}\vp\right.\non\\
&& \left.
+\frac{1}{4}\lp c_{38}+4c_{39}+c_{42}+8y_6+2y_{20}\rp\p^a\vp\p^b\vp\p_{ab}\vp
+\frac{1}{2}\lp c_{16}-2c_{25}+2y_{10}\rp\p^a_{\hph{a}a}\vp H^{bci}H_{bci}\right.\non\\
&& \left.
+\frac{1}{2}\lp c_{17}-2c_{26}+2y_{11}\rp\p^a_{\hph{a}a}\vp H^{ijk}H_{ijk}
+\frac{1}{2}\lp c_{18}-c_{27}+2y_{12}\rp\p^{ab}\vp H_a^{\hph{a}ci}H_{bci}\right.\non\\
&& \left.
+\frac{1}{2}\lp c_{19}-c_{31}\rp\p^{ij}\vp H^{ab}_{\hph{ab}i}H_{abj}
+\frac{1}{2}\lp c_{20}-c_{32}\rp\p^{ij}\vp H_i^{\hph{i}k\ell}H_{jk\ell}\right.\non\\
&& \left.
+\lp c_{21}-c_{35}+y_{16}+y_{17}\rp\p^a_{\hph{a}a}\Phi\p^b_{\hph{b}b}\vp
+\lp y_{15}+y_{18}\rp\p^{ab}\Phi\p_{ab}\vp
+\frac{1}{2}\lp 2c_{22}-c_{36}\rp\p^{ij}\Phi\p_{ij}\vp\right.\non\\
&& \left.
+\frac{1}{4}\lp -c_{21}+4c_{37}+c_{38}-c_{41}+4y_{19}\rp\p^a_{\hph{a}a}\vp\p^b_{\hph{b}b}\vp
+\frac{1}{4}\lp c_{38}+4c_{39}-c_{42}+4y_{20}\rp\p^{ab}\vp\p_{ab}\vp\right.\non\\
&& \left.
+\frac{1}{4}\lp -c_{22}+c_{41}+c_{42}-4c_{43}\rp\p^{ij}\vp\p_{ij}\vp
+\lp -y_9-y_{21}\rp\p^a\Phi H_a^{\hph{a}bi}\p^cH_{bci}\right.\non\\
&& \left.
+\lp x_1+2y_7-y_{22}\rp\p^a\Phi H^{bci}\p_aH_{bci}
+\lp -2x_1+y_9-y_{23}\rp\p^a\Phi H^{bci}\p_bH_{aci}\right.\non\\
&& \left.
+\lp -x_1-y_{24}\rp\p^a\Phi H^{bci}\p_iH_{abc}
+\lp x_2+2y_8-y_{25}\rp\p^a\Phi H^{ijk}\p_aH_{ijk}\right.\non\\
&& \left.
+\lp -3x_2-y_{26}\rp\p^a\Phi H^{ijk}\p_iH_{ajk}
+\frac{1}{2}\lp c_{44}+2c_{45}-2y_{12}+y_{21}\rp\p^a\vp H_a^{\hph{a}bi}\p^cH_{bci}\right.\non\\
&& \left.
+\frac{1}{2}\lp 2x_3+4y_{10}+y_{22}\rp\p^a\vp H^{bci}\p_aH_{bci}
+\frac{1}{2}\lp -4x_3+2y_{12}+y_{23}\rp\p^a\vp H^{bci}\p_bH_{aci}\right.\non\\
&& \left.
+\frac{1}{2}\lp -2x_3+y_{24}\rp\p^a\vp H^{bci}\p_iH_{abc}
+\frac{1}{2}\lp 2x_4+4y_{11}+y_{25}\rp\p^a\vp H^{ijk}\p_aH_{ijk}\right.\non
\eea
\bea
&& \left.
+\frac{1}{2}\lp -6x_4+y_{26}\rp\p^a\vp H^{ijk}\p_iH_{ajk}
+\lp y_{21}\rp\p^aH_a^{\hph{a}bi}\p^cH_{bci}
+\lp x_5+y_{22}\rp\p^aH^{bci}\p_aH_{bci}\right.\non\\
&& \left.
+\lp -2x_5+y_{23}\rp\p^aH^{bci}\p_bH_{aci}
+\lp -x_5+3x_7+y_{24}\rp\p^aH^{bci}\p_iH_{abc}\right.\non\\
&& \left.
+\lp x_6+y_{25}\rp\p^aH^{ijk}\p_aH_{ijk}
+\lp -3x_6+x_8+y_{26}\rp\p^aH^{ijk}\p_iH_{ajk}
+\lp -x_7\rp\p^iH^{abc}\p_iH_{abc}\right.\non\\
&& \left.
+\lp -x_8\rp\p^iH^{ajk}\p_iH_{ajk}
+\lp 2x_8\rp\p^iH^{ajk}\p_jH_{aik}
+\lp y_{13}+y_{14}-y_{27}\rp\p^a\Phi\p_{a\hph{b}b}^{\hph{a}b}\Phi\right.\non\\
&& \left.
+\frac{1}{2}\lp 2y_{15}+2y_{16}+y_{27}\rp\p^{a\hph{a}b}_{\hph{a}a}\Phi\p_b\vp
+\lp y_{17}+y_{18}-y_{28}\rp\p^a\Phi\p_{a\hph{b}b}^{\hph{a}b}\vp\right.\non\\
&& \left.
+\frac{1}{2}\lp 2y_{19}+2y_{20}+y_{28}\rp\p^a\vp\p_{a\hph{b}b}^{\hph{a}b}\vp
+\lp 2x_9+y_{21}-y_{23}\rp H^{abi}\p_a^{\hph{a}c}H_{bci}\right.\non\\
&& \left.
+\lp x_9+y_{22}\rp H^{abi}\p^c_{\hph{c}c}H_{abi}
+\lp -x_9+y_{24}\rp H^{abi}\p^c_{\hph{c}i}H_{abc}
+\lp x_{10}+y_{25}\rp H^{ijk}\p^a_{\hph{a}a}H_{ijk}\right.\non\\
&& \left.
+\lp -3x_{10}+y_{26}\rp H^{ijk}\p^a_{\hph{a}i}H_{ajk}
+\lp y_{27}\rp\p^{a\hph{a}b}_{\hph{a}a\hph{b}b}\Phi
+\lp y_{28}\rp\p^{a\hph{a}b}_{\hph{a}a\hph{b}b}\vp
\vphantom{\hlf}\right\}.\non
\eea

Setting this lengthy expression to zero just gives a large number of linear equations for the coefficients $c_i$, $x_i$, and $y_i$.  Terms with different numbers of $H$ fields don't mix in the warped product.  Turning first to the terms with no $H$ fields, the solution to this linear system is given by
\be
c_{15}=-3c_1,\quad c_{21}=2c_1,\quad c_{22}=-2c_1,\quad c_{23}=-2c_1,\quad c_{24}=2c_1,\quad c_{35}=2c_1\non
\ee
\be
c_{36}=-4c_1,\quad c_{38}=2c_1-4c_{37},\quad c_{39}=-c_1+c_{37},\quad c_{41}=-2c_1,\quad c_{42}=c_{43}=0,
\ee
\be
y_1=0,\quad y_3=\hlf c_1-y_2,\quad y_4=\frac{3}{4}c_1-\hlf y_2,\quad y_5=-\hlf c_1+\hlf y_2,\quad y_6=\frac{1}{8}c_1,\non
\ee
\be
y_{13}=y_{14}=0,\quad y_{15}=c_1-y_2,\quad y_{16}=-c_1+y_2,\quad y_{17}=c_1-y_2,\quad y_{18}=-c_1+y_2,\non
\ee
\be
y_{19}=-\hlf c_1,\quad y_{20}=\hlf c_1,\quad y_{27}=y_{28}=0.\non
\ee
The coefficients $c_1$, $c_{37}$, $c_{40}$, and $y_2$ are arbitrary.

For the coefficients involving two $H$ fields, we find relations
\be
c_2=c_3=c_{16}=c_{17}=c_{25}=c_{26}=x_7=x_8=y_7=y_8=y_9=y_{10}=y_{11}=y_{12}=y_{21}=0,\non
\ee
\be
c_{27}=c_{18},\quad c_{31}=c_{19},\quad c_{32}=c_{20},\non
\ee
\be
c_{44}=2c_4,\quad c_{45}=-c_4,\quad c_{48}=-\hlf c_{18}+c_{46},
\ee
\be
x_3=-\hlf x_1,\quad x_5=-x_1,\quad x_9=-x_1,\quad y_{22}=x_1,\quad y_{23}=-2x_1,\quad y_{24}=-x_1,\non
\ee
\be
x_4=-\hlf x_2,\quad x_6=-x_2,\quad x_{10}=-x_2,\quad y_{25}=x_2,\quad y_{26}=-3x_2,\non
\ee
with $c_4$, $c_{18}$, $c_{19}$, $c_{20}$, $c_{28}$, $c_{29}$, $c_{30}$, $c_{33}$, $c_{34}$, $c_{46}$, $c_{47}$, $x_1$, and $x_2$ unconstrained. 

As mentioned before, the terms with four $H$ fields are all unconstrained.

\section{Reduction and duality of couplings in the twisted product}
\label{app:Twisted}

Now for each coupling that does not involve derivatives of the dilaton, we will reduce the couplings in the case that the circle is parallel to the O-plane, and then subtract the T-dual of the reduction when the circle is perpendicular.  Computing,
\begin{multline}
c_5\sqrt{-g}e^{-\Phi}H^{abi}H_{abi}H^{cdj}H_{cdj}\longrightarrow c_5e^{-\Phi+\hlf\vp}\ls 4e^{-2\vp}\wtf^{ai}\wtf_{ai}\wtf^{bj}\wtf_{bj}+4e^{-\vp}\wtf^{ai}\wtf_{ai}\wtH^{bcj}\wtH_{bcj}\right.\\
\left. -2e^\vp f^{ab}f_{ab}\wtH^{cdi}\wtH_{cdi}-e^{2\vp}f^{ab}f_{ab}f^{cd}f_{cd}\rs,\non
\end{multline}
\begin{multline}
c_6\sqrt{-g}e^{-\Phi}H^{abi}H_{abi}H^{jk\ell}H_{jk\ell}\longrightarrow c_6e^{-\Phi+\hlf\vp}\ls 2e^{-\vp}\wtf^{ai}\wtf_{ai}\wtH^{jk\ell}\wtH_{jk\ell}\right.\\
\left. +e^\vp\lp -f^{ab}f_{ab}\wtH^{ijk}\wtH_{ijk}-3f^{ij}f_{ij}\wtH^{abk}\wtH_{abk}\rp-3e^{2\vp}f^{ab}f_{ab}f^{ij}f_{ij}\rs,\non
\end{multline}
\begin{multline}
c_7\sqrt{-g}e^{-\Phi}H^{abi}H_{ab}^{\hph{ab}j}H^{cd}_{\hph{cd}i}H_{cdj}\longrightarrow c_7e^{-\Phi+\hlf\vp}\ls 4e^{-2\vp}\wtf^{ai}\wtf_a^{\hph{a}j}\wtf^b_{\hph{b}i}\wtf_{bj}+4e^{-\vp}\wtf^{ai}\wtf_a^{\hph{a}j}\wtH^{bc}_{\hph{bc}i}\wtH_{bcj}\right.\\
\left. -2e^\vp f^{ab}f^{cd}\wtH_{ab}^{\hph{ab}i}\wtH_{cdi}-e^{2\vp}f^{ab}f_{ab}f^{cd}f_{cd}\rs,\non
\end{multline}
\begin{multline}
c_8\sqrt{-g}e^{-\Phi}H^{abi}H_{ab}^{\hph{ab}j}H_i^{\hph{i}k\ell}H_{jk\ell}\longrightarrow c_8e^{-\Phi+\hlf\vp}\ls 2e^{-\vp}\wtf^{ai}\wtf_a^{\hph{a}j}\wtH_i^{\hph{i}k\ell}\wtH_{jk\ell}\right.\\
\left. +e^\vp\lp -2f^{ab}f^{ij}\wtH_{ab}^{\hph{ab}k}\wtH_{ijk}-2f^{ij}f_i^{\hph{i}k}\wtH^{ab}_{\hph{ab}j}\wtH_{abk}\rp-e^{2\vp}f^{ab}f_{ab}f^{ij}f_{ij}\rs,\non
\end{multline}
\begin{multline}
c_9\sqrt{-g}e^{-\Phi}H^{abi}H_{a\hph{c}i}^{\hph{a}c}H_b^{\hph{b}dj}H_{cdj}\longrightarrow c_9e^{-\Phi+\hlf\vp}\ls e^{-2\vp}\lp\wtf^{ai}\wtf_{ai}\wtf^{bj}\wtf_{bj}+\wtf^{ai}\wtf_a^{\hph{a}j}\wtf^b_{\hph{b}i}\wtf_{bj}\rp\right.\\
\left. +e^{-\vp}\lp 2\wtf^{ai}\wtf^b_{\hph{b}i}\wtH_a^{\hph{a}cj}\wtH_{bcj}+2\wtf^{ai}\wtf^{bj}\wtH_{a\hph{c}i}^{\hph{a}c}\wtH_{bcj}\rp-2e^\vp f^{ab}f_a^{\hph{a}c}\wtH_b^{\hph{b}di}\wtH_{cdi}-e^{2\vp}f^{ab}f_a^{\hph{a}c}f_b^{\hph{b}d}f_{cd}\rs,\non
\end{multline}
\begin{multline}
c_{10}\sqrt{-g}e^{-\Phi}H^{abi}H_a^{\hph{a}cj}H_{bc}^{\hph{bc}k}H_{ijk}\longrightarrow c_{10}e^{-\Phi+\hlf\vp}\ls  3e^{-\vp}\wtf^{ai}\wtf^{bj}\wtH_{ab}^{\hph{ab}k}\wtH_{ijk}-3e^\vp f^{ab}f^{ij}\wtH_{a\hph{c}i}^{\hph{a}c}\wtH_{bcj}\rs,\non
\end{multline}
\begin{multline}
c_{11}\sqrt{-g}e^{-\Phi}H^{abi}H_a^{\hph{a}cj}H_{b\hph{d}j}^{\hph{b}d}H_{cdi}\longrightarrow c_{11}e^{-\Phi+\hlf\vp}\ls 2e^{-2\vp}\wtf^{ai}\wtf_a^{\hph{a}j}\wtf^b_{\hph{b}i}\wtf_{bj}+4e^{-\vp}\wtf^{ai}\wtf^{bj}\wtH_{a\hph{c}j}^{\hph{a}c}\wtH_{bci}\right.\\
\left. -2e^\vp f^{ab}f^{cd}\wtH_{ac}^{\hph{ac}i}\wtH_{bdi}-e^{2\vp}f^{ab}f_a^{\hph{a}c}f_b^{\hph{b}d}f_{cd}\rs,\non
\end{multline}
\begin{multline}
c_{12}\sqrt{-g}e^{-\Phi}H^{ijk}H_{ijk}H^{\ell mn}H_{\ell mn}\longrightarrow c_{12}e^{-\Phi+\hlf\vp}\ls -6e^\vp f^{ij}f_{ij}\wtH^{k\ell m}\wtH_{k\ell m}-9e^{2\vp}f^{ij}f_{ij}f^{k\ell}f_{k\ell}\rs,\non
\end{multline}
\begin{multline}
c_{13}\sqrt{-g}e^{-\Phi}H^{ijk}H_{ij}^{\hph{ij}\ell}H_k^{\hph{k}mn}H_{\ell mn}\\
\longrightarrow c_{13}e^{-\Phi+\hlf\vp}\ls e^\vp\lp -4f^{ij}f_i^{\hph{i}k}\wtH_j^{\hph{j}\ell m}\wtH_{k\ell m}-2f^{ij}f^{k\ell}\wtH_{ij}^{\hph{ij}m}\wtH_{k\ell m}\rp\right.\\
\left. +e^{2\vp}\lp -f^{ij}f_{ij}f^{k\ell}f_{k\ell}-4f^{ij}f_i^{\hph{i}k}f_j^{\hph{j}\ell}f_{k\ell}\rp\rs,\non
\end{multline}
\begin{multline}
c_{14}\sqrt{-g}e^{-\Phi}H^{ijk}H_i^{\hph{i}\ell m}H_{j\ell}^{\hph{j\ell}n}H_{kmn}\\
\longrightarrow c_{14}e^{-\Phi+\hlf\vp}\ls -6e^\vp f^{ij}f^{k\ell}\wtH_{ik}^{\hph{ik}m}\wtH_{j\ell m}-3e^{2\vp}f^{ij}f_i^{\hph{i}k}f_j^{\hph{j}\ell}f_{k\ell}\rs,\non
\end{multline}
\begin{multline}
c_{25}\sqrt{-g}e^{-\Phi}R^{ab}_{\hph{ab}ab}H^{cdi}H_{cdi}\longrightarrow c_{25}e^{-\Phi+\hlf\vp}\ls -\hlf f^{ab}f_{ab}\wtf^{ci}\wtf_{ci}-\frac{1}{4}e^\vp f^{ab}f_{ab}\wtH^{cdi}\wtH_{cdi}\rs,\non
\end{multline}
\begin{multline}
c_{26}\sqrt{-g}e^{-\Phi}R^{ab}_{\hph{ab}ab}H^{ijk}H_{ijk}\longrightarrow c_{26}e^{-\Phi+\hlf\vp}\ls -\frac{1}{4}e^\vp f^{ab}f_{ab}\wtH^{ijk}\wtH_{ijk}\rs,\non
\end{multline}
\begin{multline}
c_{27}\sqrt{-g}e^{-\Phi}R^{ab\hph{a}c}_{\hph{ab}a}H_b^{\hph{b}di}H_{cdi}\\
\longrightarrow c_{27}e^{-\Phi+\hlf\vp}\ls\frac{1}{4}\lp f^{ab}f_{ab}\wtf^{ci}\wtf_{ci}-2f^{ab}f_a^{\hph{a}c}\wtf_b^{\hph{b}i}\wtf_{ci}-4\p^af_a^{\hph{a}b}\wtf^{ci}\wtH_{bci}\rp-\hlf e^\vp f^{ab}f_a^{\hph{a}c}\wtH_b^{\hph{b}di}\wtH_{cdi}\rs,\non
\end{multline}
\begin{multline}
c_{28}\sqrt{-g}e^{-\Phi}R^{abcd}H_{ab}^{\hph{ab}i}H_{cdi}\longrightarrow c_{28}e^{-\Phi+\hlf\vp}\ls f^{ab}f_a^{\hph{a}c}\wtf_b^{\hph{b}i}\wtf_{ci}+2\p^af^{bc}\wtf_a^{\hph{a}i}\wtH_{bci}\right.\\
\left. +\hlf e^\vp\lp -f^{ab}f^{cd}\wtH_{ab}^{\hph{ab}i}\wtH_{cdi}-f^{ab}f^{cd}\wtH_{ac}^{\hph{ac}i}\wtH_{bdi}\rp\rs,\non
\end{multline}
\begin{multline}
c_{29}\sqrt{-g}e^{-\Phi}R^{abij}H_{ab}^{\hph{ab}k}H_{ijk}\longrightarrow c_{29}e^{-\Phi+\hlf\vp}\ls\hlf e^{-\vp}\wtf^{ai}\wtf^{bj}\wtH_{ab}^{\hph{ab}k}\wtH_{ijk}\right.\\
\left. +\hlf\lp f^{ab}f^{ij}\wtf_{ai}\wtf_{bj}+2\p^af^{ij}\wtf_a^{\hph{a}k}\wtH_{ijk}-2f^{ij}\p_i\wtf^{ab}\wtH_{abj}\rp-\hlf e^\vp f^{ab}f^{ij}\wtH_{ab}^{\hph{ab}k}\wtH_{ijk}\rs,\non
\end{multline}
\begin{multline}
c_{30}\sqrt{-g}e^{-\Phi}R^{abij}H_{a\hph{c}i}^{\hph{a}c}H_{bcj}\longrightarrow c_{30}e^{-\Phi+\hlf\vp}\ls\frac{1}{4}e^{-\vp}\lp\wtf^{ai}\wtf^{bj}\wtH_{a\hph{c}i}^{\hph{a}c}\wtH_{bcj}-\wtf^{ai}\wtf^{bj}\wtH_{a\hph{c}j}^{\hph{a}c}\wtH_{bci}\rp\right.\\
\left. +\hlf\lp -f^{ab}f^{ij}\wtf_{ai}\wtf_{bj}+2\p^af^{ij}\wtf^b_{\hph{b}i}\wtH_{abj}+2f^{ab}\p^i\wtf_a^{\hph{a}c}\wtH_{bci}\rp-\hlf e^\vp f^{ab}f^{ij}\wtH_{a\hph{c}i}^{\hph{a}c}\wtH_{bcj}\rs,\non
\end{multline}
\begin{multline}
c_{31}\sqrt{-g}e^{-\Phi}R^{ai\hph{a}j}_{\hph{ai}a}H^{bc}_{\hph{bc}i}H_{bcj}\longrightarrow c_{31}e^{-\Phi+\hlf\vp}\ls\frac{3}{4}e^{-\vp}\wtf^{ai}\wtf_a^{\hph{a}j}\wtH^{bc}_{\hph{bc}i}\wtH_{bcj}\right.\\
\left. +\frac{1}{4}\lp -f^{ab}f_{ab}\wtf^{ci}\wtf_{ci}+2f^{ij}f_i^{\hph{i}k}\wtf^a_{\hph{a}j}\wtf_{ak}+4f^{ab}\p^c\wtf_c^{\hph{c}i}\wtH_{abi}\rp+\frac{1}{4}e^\vp f^{ij}f_i^{\hph{i}k}\wtH^{ab}_{\hph{ab}j}\wtH_{abk}\rs,\non
\end{multline}
\begin{multline}
c_{32}\sqrt{-g}e^{-\Phi}R^{ai\hph{a}j}_{\hph{ai}a}H_i^{\hph{i}k\ell}H_{jk\ell}\longrightarrow c_{32}e^{-\Phi+\hlf\vp}\ls\frac{3}{4}e^{-\vp}\wtf^{ai}\wtf_a^{\hph{a}j}\wtH_i^{\hph{i}k\ell}H_{jk\ell}\right.\\
\left. +\frac{1}{4}\lp -f^{ij}f_{ij}\wtf^{ak}\wtf_{ak}+6f^{ij}f_i^{\hph{i}k}\wtf^a_{\hph{a}j}\wtf_{ak}+4f^{ij}\p^a\wtf_a^{\hph{a}k}\wtH_{ijk}\rp+\frac{1}{4}e^\vp f^{ij}f_i^{\hph{i}k}\wtH_j^{\hph{j}\ell m}\wtH_{k\ell m}\rs,\non
\end{multline}
\begin{multline}
c_{33}\sqrt{-g}e^{-\Phi}R^{aibj}H_{a\hph{c}i}^{\hph{a}c}H_{bcj}\longrightarrow c_{33}e^{-\Phi+\hlf\vp}\ls\frac{1}{4}e^{-\vp}\lp 2\wtf^{ai}\wtf^{bj}\wtH_{a\hph{c}i}^{\hph{a}c}\wtH_{bcj}+\wtf^{ai}\wtf^{bj}\wtH_{a\hph{c}j}^{\hph{a}c}\wtH_{bci}\rp\right.\\
\left. +\frac{1}{4}\lp -f^{ab}f_a^{\hph{a}c}\wtf_b^{\hph{b}i}\wtf_{ci}-f^{ab}f^{ij}\wtf_{ai}\wtf_{bj}+f^{ij}f_i^{\hph{i}k}\wtf^a_{\hph{a}j}\wtf_{ak}+4\p^if^{aj}\wtf^b_{\hph{b}i}\wtH_{abj}-4f^{ab}\p_a\wtf^{ci}\wtH_{bci}\rp\right.\\
\left. -\frac{1}{4}e^\vp f^{ab}f^{ij}\wtH_{a\hph{c}i}^{\hph{a}c}\wtH_{bcj}\rs,\non
\end{multline}
\begin{multline}
c_{34}\sqrt{-g}e^{-\Phi}R^{ijk\ell}H_{ij}^{\hph{ij}m}H_{k\ell m}\longrightarrow c_{34}e^{-\Phi+\hlf\vp}\ls -f^{ij}f_i^{\hph{i}k}\wtf^a_{\hph{a}j}\wtf_{ak}-2f^{ij}\p_i\wtf^{k\ell}\wtH_{jk\ell}\right.\\
\left. +\hlf e^\vp\lp -f^{ij}f^{k\ell}\wtH_{ij}^{\hph{ij}m}\wtH_{k\ell m}-f^{ij}f^{k\ell}\wtH_{ik}^{\hph{ik}m}\wtH_{j\ell m}\rp\rs,\non
\end{multline}
\begin{multline}
c_{37}\sqrt{-g}e^{-\Phi}R^{ab}_{\hph{ab}ab}R^{cd}_{\hph{cd}cd}\longrightarrow c_{37}e^{-\Phi+\hlf\vp}\ls\frac{1}{16}e^{2\vp}f^{ab}f_{ab}f^{cd}f_{cd}\rs,\non
\end{multline}
\begin{multline}
c_{38}\sqrt{-g}e^{-\Phi}R^{ab\hph{a}c}_{\hph{ab}a}R_{b\hph{d}cd}^{\hph{b}d}\longrightarrow c_{38}e^{-\Phi+\hlf\vp}\ls -\hlf e^\vp\p^af_a^{\hph{a}b}\p^cf_{bc}\right.\\
\left. +\frac{1}{16}e^{2\vp}\lp f^{ab}f_{ab}f^{cd}f_{cd}+4f^{ab}f_a^{\hph{a}c}f_b^{\hph{b}d}f_{cd}\rp\rs,\non
\end{multline}
\begin{multline}
c_{39}\sqrt{-g}e^{-\Phi}R^{abcd}R_{abcd}\longrightarrow c_{39}e^{-\Phi+\hlf\vp}\ls e^\vp\p^af^{bc}\p_af_{bc}\right.\\
\left. +\frac{1}{8}e^{2\vp}\lp 3f^{ab}f_{ab}f^{cd}f_{cd}+5f^{ab}f_a^{\hph{a}c}f_b^{\hph{b}d}f_{cd}\rp\rs,\non
\end{multline}
\begin{multline}
c_{40}\sqrt{-g}e^{-\Phi}R^{abij}R_{abij}\longrightarrow c_{40}e^{-\Phi+\hlf\vp}\ls\frac{1}{8}e^{-2\vp}\lp -\wtf^{ai}\wtf_{ai}\wtf^{bj}\wtf_{bj}+\wtf^{ai}\wtf_a^{\hph{a}j}\wtf^b_{\hph{b}i}\wtf_{bj}\rp\right.\\
\left. -\hlf e^{-\vp}\p^i\wtf^{ab}\p_i\wtf_{ab}+\hlf e^\vp\p^af^{ij}\p_af_{ij}+\frac{1}{4}e^{2\vp}f^{ab}f_{ab}f^{ij}f_{ij}\rs,\non
\end{multline}
\begin{multline}
c_{41}\sqrt{-g}e^{-\Phi}R^{ai\hph{a}j}_{\hph{ai}a}R^b_{\hph{b}ibj}\longrightarrow c_{41}e^{-\Phi+\hlf\vp}\ls\frac{1}{16}e^{-2\vp}\lp -\wtf^{ai}\wtf_{ai}\wtf^{bj}\wtf_{bj}-9\wtf^{ai}\wtf_a^{\hph{a}j}\wtf^b_{\hph{b}i}\wtf_{bj}\rp\right.\\
\left. -\hlf e^{-\vp}\p^a\wtf_a^{\hph{a}i}\p^b\wtf_{bi}+\frac{1}{16}e^{2\vp}f^{ij}f_i^{\hph{i}k}f_j^{\hph{j}\ell}f_{k\ell}\rs,\non
\end{multline}
\begin{multline}
c_{42}\sqrt{-g}e^{-\Phi}R^{aibj}R_{aibj}\longrightarrow c_{42}e^{-\Phi+\hlf\vp}\ls\frac{1}{16}e^{-2\vp}\lp -5\wtf^{ai}\wtf_{ai}\wtf^{bj}\wtf_{bj}-5\wtf^{ai}\wtf_a^{\hph{a}j}\wtf^b_{\hph{b}i}\wtf_{bj}\rp\right.\\
\left. -\hlf e^{-\vp}\p^a\wtf^{bi}\p_a\wtf_{bi}+\hlf e^\vp\p^if^{aj}\p_if_{aj}+\frac{1}{16}e^{2\vp}\lp f^{ab}f_{ab}f^{ij}f_{ij}+f^{ij}f_i^{\hph{i}k}f_j^{\hph{j}\ell}f_{k\ell}\rp\rs,\non
\end{multline}
\begin{multline}
c_{43}\sqrt{-g}e^{-\Phi}R^{ijk\ell}R_{ijk\ell}\longrightarrow c_{43}e^{-\Phi+\hlf\vp}\ls -\frac{1}{4}e^{-2\vp}\wtf^{ai}\wtf_a^{\hph{a}j}\wtf^b_{\hph{b}i}\wtf_{bj}-e^{-\vp}\p^i\wtf^{jk}\p_i\wtf_{jk}\right.\\
\left. +\frac{1}{8}e^{2\vp}\lp 3f^{ij}f_{ij}f^{k\ell}f_{k\ell}+3f^{ij}f_i^{\hph{i}k}f_j^{\hph{j}\ell}f_{k\ell}\rp\rs,\non
\end{multline}
\begin{multline}
c_{45}\sqrt{-g}e^{-\Phi}\nabla^aH_a^{\hph{a}bi}\nabla^cH_{bci}\longrightarrow c_{45}e^{-\Phi+\hlf\vp}\ls\frac{1}{4}e^{-\vp}\lp\wtf^{ai}\wtf^{bj}\wtH_{a\hph{c}i}^{\hph{a}c}\wtH_{bcj}-4\p^a\wtf_a^{\hph{a}i}\p^b\wtf_{bi}\rp\right.\\
\left. +\frac{1}{4}\lp f^{ab}f_a^{\hph{a}c}\wtf_b^{\hph{b}i}\wtf_{ci}-f^{ij}f_i^{\hph{i}k}\wtf^a_{\hph{a}j}\wtf_{ak}+4\p^af_a^{\hph{a}b}\wtf^{ci}\wtH_{bci}-4f^{ab}\p^c\wtf_c^{\hph{c}i}\wtH_{abi}-4f^{ab}\wtf_a^{\hph{a}i}\p^c\wtH_{bci}\right.\right.\\
\left.\left. +4f^{ij}\wtf^a_{\hph{a}i}\p^b\wtH_{abj}\rp+\frac{1}{4}e^\vp\lp -f^{ab}f^{cd}\wtH_{ab}^{\hph{ab}i}\wtH_{cdi}-4\p^af_a^{\hph{a}b}\p^cf_{bc}\rp\rs,\non
\end{multline}
\begin{multline}
c_{46}\sqrt{-g}e^{-\Phi}\nabla^aH^{bci}\nabla_aH_{bci}\longrightarrow c_{46}e^{-\Phi+\hlf\vp}\ls\frac{1}{4}e^{-\vp}\lp -\wtf^{ai}\wtf_a^{\hph{a}j}\wtH^{bc}_{\hph{bc}i}\wtH_{bcj}+8\p^a\wtf^{bi}\p_a\wtf_{bi}\rp\right.\\
\left. +\frac{1}{4}\lp f^{ab}f_{ab}\wtf^{ci}\wtf_{ci}-4f^{ab}f^{ij}\wtf_{ai}\wtf_{bj}+2f^{ij}f_i^{\hph{i}k}\wtf^a_{\hph{a}j}\wtf_{ak}+4\p^af^{bc}\wtf_a^{\hph{a}i}\wtH_{bci}-8f^{ab}\p_a\wtf^{ci}\wtH_{bci}\right.\right.\\
\left.\left. +8f^{ab}\wtf^{ci}\p_a\wtH_{bci}-4f^{ab}\wtf^{ci}\p_c\wtH_{abi}\rp+\frac{1}{4}e^\vp\lp 4f^{ab}f_a^{\hph{a}c}\wtH_b^{\hph{b}di}\wtH_{cdi}-2f^{ab}f^{cd}\wtH_{ac}^{\hph{ac}i}\wtH_{bdi}\right.\right.\\
\left.\left. -4f^{ab}f^{ij}\wtH_{a\hph{c}i}^{\hph{a}c}\wtH_{bcj}+f^{ij}f_i^{\hph{i}k}\wtH^{ab}_{\hph{ab}j}\wtH_{abk}-4\p^af^{bc}\p_af_{bc}\rp\rs,\non
\end{multline}
\begin{multline}
c_{47}\sqrt{-g}e^{-\Phi}\nabla^aH^{ijk}\nabla_aH_{ijk}\longrightarrow c_{47}e^{-\Phi+\hlf\vp}\ls -\frac{3}{4}e^{-\vp}\wtf^{ai}\wtf_a^{\hph{a}j}\wtH_i^{\hph{i}k\ell}\wtH_{jk\ell}\right.\\
\left. +\frac{1}{4}\lp -3f^{ij}f_{ij}\wtf^{ak}\wtf_{ak}+6f^{ij}f_i^{\hph{i}k}\wtf^a_{\hph{a}j}\wtf_{ak}+12\p^af^{ij}\wtf_a^{\hph{a}k}\wtH_{ijk}-12f^{ij}\wtf^{ak}\p_a\wtH_{ijk}\rp\right.\\
\left. +\frac{1}{4}e^\vp\lp 3f^{ij}f_i^{\hph{i}k}\wtH_j^{\hph{j}\ell m}\wtH_{k\ell m}-6f^{ij}f^{k\ell}\wtH_{ik}^{\hph{ik}m}\wtH_{j\ell m}-12\p^af^{ij}\p_af_{ij}\rp\rs,\non
\end{multline}
\begin{multline}
c_{48}\sqrt{-g}e^{-\Phi}\nabla^iH^{ajk}\nabla_iH_{ajk}\longrightarrow c_{48}e^{-\Phi+\hlf\vp}\ls\frac{1}{4}e^{-\vp}\lp -\wtf^{ai}\wtf_a^{\hph{a}j}\wtH_i^{\hph{i}k\ell}\wtH_{jk\ell}-4\wtf^{ai}\wtf^b_{\hph{b}i}\wtH_a^{\hph{a}cj}\wtH_{bcj}\right.\right.\\
\left.\left. +4\wtf^{ai}\wtf^{bj}\wtH_{ab}^{\hph{ab}k}\wtH_{ijk}+2\wtf^{ai}\wtf^{bj}\wtH_{a\hph{c}j}^{\hph{a}c}\wtH_{bci}+4\p^i\wtf^{jk}\p_i\wtf_{jk}\rp+\frac{1}{4}\lp -2f^{ab}f_a^{\hph{a}c}\wtf_b^{\hph{b}i}\wtf_{ci}+4f^{ab}f^{ij}\wtf_{ai}\wtf_{bj}\right.\right.\\
\left.\left. +f^{ij}f_{ij}\wtf^{ak}\wtf_{ak}-4f^{ij}f_i^{\hph{i}k}\wtf^a_{\hph{a}j}\wtf_{ak}+8\p^if^{aj}\wtf^b_{\hph{b}i}\wtH_{abj}-4f^{ij}\p_i\wtf^{k\ell}\wtH_{jk\ell}-8f^{ij}\wtf^{ak}\p_i\wtH_{ajk}\right.\right.\\
\left.\left. +4f^{ij}\wtf^{ak}\p_k\wtH_{aij}\rp+\frac{1}{4}e^\vp\lp f^{ij}f_i^{\hph{i}k}\wtH_j^{\hph{j}\ell m}\wtH_{k\ell m}-8\p^if^{aj}\p_if_{aj}\rp\rs,\non
\end{multline}

The relevant Bianchi identities are (we omit the constant prefactors of $e^{-\Phi}$ and powers of $e^\vp$)
\bea
3x_1'\wtf^{ai}\wtH^{bc}_{\hph{bc}i}\p_{[a}f_{bc]} &=& x_1'\ls\p^af^{bc}\wtf_a^{\hph{a}i}\wtH_{bci}-2\p^af^{bc}\wtf_b^{\hph{b}i}\wtH_{aci}\rs,\non\\
3x_2'\wtf^{ak}\wtH^{ij}_{\hph{ij}k}\p_{[a}f_{ij]} &=& x_2'\ls\p^af^{ij}\wtf_a^{\hph{a}k}\wtH_{ijk}-2\p^if^{aj}\wtf_a^{\hph{a}k}\wtH_{ijk}\rs,\non\\
3x_3'\wtf^{bi}\wtH^{a\hph{b}j}_{\hph{a}b}\p_{[a}f_{ij]} &=& x_3'\ls\p^af^{ij}\wtf^b_{\hph{b}i}\wtH_{abj}-\p^if^{aj}\wtf^b_{\hph{b}i}\wtH_{abj}+\p^if^{aj}\wtf^b_{\hph{b}j}\wtH_{abi}\rs,\non
\eea
\bea
3x_4'\p^af^{bc}\p_{[a}f_{bc]} &=& x_4'\ls\p^af^{bc}\p_af_{bc}-2\p^af^{bc}\p_bf_{ac}\rs,\non\\
3x_5'\p^af^{ij}\p_{[a}f_{ij]} &=& x_5'\ls\p^af^{ij}\p_af_{ij}-2\p^af^{ij}\p_if_{aj}\rs,\non\\
3x_6'\p^if^{aj}\p_{[a}f_{ij]} &=& x_6'\ls\p^af^{ij}\p_if_{aj}-\p^if^{aj}\p_if_{aj}+\p^if^{aj}\p_jf_{ai}\rs,\non
\eea
\bea
3x_7'f^{ab}\p^c\p_{[a}f_{bc]} &=& x_7'\ls 2f^{ab}\p_a^{\hph{a}c}f_{bc}+f^{ab}\p^c_{\hph{c}c}f_{ab}\rs,\non\\
3x_8'f^{ij}\p^a\p_{[a}f_{ij]} &=& x_8'\ls f^{ij}\p^a_{\hph{a}a}f_{ij}-2f^{ij}\p^a_{\hph{a}i}f_{aj}\rs,\non
\eea
\bea
3x_9'f^{ac}\wtH^{b\hph{c}i}_{\hph{b}c}\p_{[a}\wtf_{bi]} &=& x_9'\ls -f^{ab}\p_a\wtf^{ci}\wtH_{bci}+f^{ab}\p^c\wtf_a^{\hph{a}i}\wtH_{bci}-f^{ab}\p^i\wtf_a^{\hph{a}c}\wtH_{bci}\rs,\non\\
3x_{10}'f^{ij}\wtH^{ab}_{\hph{ab}j}\p_{[a}\wtf_{bi]} &=& x_{10}'\ls 2f^{ij}\p^a\wtf^b_{\hph{b}i}\wtH_{abj}+f^{ij}\p_i\wtf^{ab}\wtH_{abj}\rs,\non\\
3x_{11}'f^{i\ell}\wtH^{jk}_{\hph{jk}\ell}\p_{[i}\wtf_{jk]} &=& x_{11}'\ls f^{ij}\p_i\wtf^{k\ell}\wtH_{jk\ell}-2f^{ij}\p^k\wtf_i^{\hph{i}\ell}\wtH_{jk\ell}\rs,\non
\eea
\bea
3x_{12}'\p^a\wtf^{bi}\p_{[a}\wtf_{bi]} &=& x_{12}'\ls \p^a\wtf^{bi}\p_a\wtf_{bi}-\p^a\wtf^{bi}\p_b\wtf_{ai}+\p^a\wtf^{bi}\p_i\wtf_{ab}\rs,\non\\
3x_{13}'\p^i\wtf^{ab}\p_{[a}\wtf_{bi]} &=& x_{13}'\ls 2\p^a\wtf^{bi}\p_i\wtf_{ab}+\p^i\wtf^{ab}\p_i\wtf_{ab}\rs,\non\\
3x_{14}'\p^i\wtf^{jk}\p_{[i}\wtf_{jk]} &=& x_{14}'\ls\p^i\wtf^{jk}\p_i\wtf_{jk}-2\p^i\wtf^{jk}\p_j\wtf_{ik}\rs,\non
\eea
\bea
3x_{15}'\wtf^{ai}\p^b\p_{[a}\wtf_{bi]} &=& x_{15}'\ls\wtf^{ai}\p_a^{\hph{a}b}\wtf_{bi}-\wtf^{ai}\p^b_{\hph{b}b}\wtf_{ai}+\wtf^{ai}\p^b_{\hph{b}i}\wtf_{ab}\rs.\non
\eea
The Bianchi identities involving $\wtH$ are slightly more complicated as discussed in section \ref{subsec:Redundancies}.
\bea
x_{16}'f^{ab}\wtf^{ci}\lp 4\p_{[a}\wtH_{bci]}+6f_{[ab}\wtf_{ci]}\rp &=& x_{16}'\ls f^{ab}f_{ab}\wtf^{ci}\wtf_{ci}-2f^{ab}f_a^{\hph{a}c}\wtf_b^{\hph{b}i}\wtf_{ci}+2f^{ab}\wtf^{ci}\p_a\wtH_{bci}\right.\non\\
&& \quad\left. +f^{ab}\wtf^{ci}\p_c\wtH_{abi}-f^{ab}\wtf^{ci}\p_i\wtH_{abc}\rs,\non\\
x_{17}'f^{ij}\wtf^{ak}\lp 4\p_{[a}\wtH_{ijk]}+6f_{[ai}\wtf_{jk]}\rp &=& x_{17}'\ls f^{ij}f_{ij}\wtf^{ak}\wtf_{ak}-2f^{ij}f_i^{\hph{i}k}\wtf^a_{\hph{a}j}\wtf_{ak}+f^{ij}\wtf^{ak}\p_a\wtH_{ijk}\right.\non\\
&& \quad\left. -2f^{ij}\wtf^{ak}\p_i\wtH_{ajk}-f^{ij}\wtf^{ak}\p_k\wtH_{aij}\rs,\non
\eea
\bea
x_{18}'\p^a\wtH^{bci}\lp 4\p_{[a}\wtH_{bci]}+6f_{[ab}\wtf_{ci]}\rp &=& x_{18}'\ls 2f^{ab}\wtf^{ci}\p_a\wtH_{bci}+f^{ab}\wtf^{ci}\p_c\wtH_{abi}+\p^a\wtH^{bci}\p_a\wtH_{bci}\right.\non\\
&& \quad\left. -2\p^a\wtH^{bci}\p_b\wtH_{aci}-\p^a\wtH^{bci}\p_i\wtH_{abc}\rs,\non\\
x_{19}'\p^a\wtH^{ijk}\lp 4\p_{[a}\wtH_{ijk]}+6f_{[ai}\wtf_{jk]}\rp &=& x_{19}'\ls 3f^{ij}\wtf^{ak}\p_a\wtH_{ijk}+\p^a\wtH^{ijk}\p_a\wtH_{ijk}-3\p^a\wtH^{ijk}\p_i\wtH_{ajk}\rs,\non\\
x_{20}'\p^i\wtH^{abc}\lp 4\p_{[a}\wtH_{bci]}+6f_{[ab}\wtf_{ci]}\rp &=& x_{20}'\ls 3f^{ab}\wtf^{ci}\p_i\wtH_{abc}+3\p^a\wtH^{bci}\p_i\wtH_{abc}-\p^i\wtH^{abc}\p_i\wtH_{abc}\rs,\non\\
x_{21}'\p^i\wtH^{ajk}\lp 4\p_{[a}\wtH_{ijk]}+6f_{[ai}\wtf_{jk]}\rp &=& x_{21}'\ls 2f^{ij}\wtf^{ak}\p_i\wtH_{ajk}+f^{ij}\wtf^{ak}\p_k\wtH_{aij}+\p^a\wtH^{ijk}\p_i\wtH_{ajk}\right.\non\\
&& \quad\left. -\p^i\wtH_{ajk}\p_i\wtH_{ajk}+2\p^i\wtH^{ajk}\p_j\wtH_{aik}\rs,\non
\eea
\bea
x_{22}'\wtH^{abi}\p^c\lp 4\p_{[a}\wtH_{bci]}+6f_{[ab}\wtf_{ci]}\rp &=& x_{22}'\ls 2\p^af_a^{\hph{a}b}\wtf^{ci}\wtH_{bci}+\p^af^{bc}\wtf_a^{\hph{a}i}\wtH_{bci}+2f^{ab}\p_a\wtf^{ci}\wtH_{bci}\right.\non\\
&& \quad\left. +f^{ab}\p^c\wtf_c^{\hph{c}i}\wtH_{abi}+2\wtH^{abi}\p_a^{\hph{a}c}\wtH_{bci}+\wtH^{abi}\p^c_{\hph{c}c}\wtH_{abi}\right.\non\\
&& \quad\left. -\wtH^{abi}\p^c_{\hph{c}i}\wtH_{abc}\rs,\non\\
x_{23}'\wtH^{ijk}\p^a\lp 4\p_{[a}\wtH_{ijk]}+6f_{[ai}\wtf_{jk]}\rp &=& x_{23}'\ls 3\p^af^{ij}\wtf_a^{\hph{a}k}\wtH_{ijk}+3f^{ij}\p^a\wtf_a^{\hph{a}k}\wtH_{ijk}+\wtH^{ijk}\p^a_{\hph{a}a}\wtH_{ijk}\right.\non\\
&& \quad\left. -3\wtH^{ijk}\p^a_{\hph{a}i}\wtH_{ajk}\rs,\non
\eea

Finally we also need the total derivatives,
\bea
y_1'\p^a\lp f_a^{\hph{a}b}\wtf^{ci}\wtH_{bci}\rp &=& y_1'\ls\p^af_a^{\hph{a}b}\wtf^{ci}\wtH_{bci}+f^{ab}\p_a\wtf^{ci}\wtH_{bci}+f^{ab}\wtf^{ci}\p_a\wtH_{bci}\rs,\non\\
y_2'\p^a\lp f^{bc}\wtf_a^{\hph{a}i}\wtH_{bci}\rp &=& y_2'\ls\p^af^{bc}\wtf_a^{\hph{a}i}\wtH_{bci}+f^{ab}\p^c\wtf_c^{\hph{c}i}\wtH_{abi}+f^{ab}\wtf^{ci}\p_c\wtH_{abi}\rs,\non\\
y_3'\p^a\lp f^{bc}\wtf_b^{\hph{b}i}\wtH_{aci}\rp &=& y_3'\ls\p^af^{bc}\wtf_b^{\hph{b}i}\wtH_{aci}-f^{ab}\p^c\wtf_a^{\hph{a}i}\wtH_{bci}-f^{ab}\wtf_a^{\hph{a}i}\p^c\wtH_{bci}\rs,\non\\
y_4'\p^a\lp f^{ij}\wtf_a^{\hph{a}k}\wtH_{ijk}\rp &=& y_4'\ls\p^af^{ij}\wtf_a^{\hph{a}k}\wtH_{ijk}+f^{ij}\p^a\wtf_a^{\hph{a}k}\wtH_{ijk}+f^{ij}\wtf^{ak}\p_a\wtH_{ijk}\rs,\non\\
y_5'\p^a\lp f^{ij}\wtf^b_{\hph{b}i}\wtH_{abj}\rp &=& y_5'\ls\p^af^{ij}\wtf^b_{\hph{b}i}\wtH_{abj}+f^{ij}\p^a\wtf^b_{\hph{b}i}\wtH_{abj}-f^{ij}\wtf^a_{\hph{a}i}\p^b\wtH_{abj}\rs,\non
\eea
\bea
y_6'\p^a\lp f_a^{\hph{a}b}\p^cf_{bc}\rp &=& y_6'\ls\p^af_a^{\hph{a}b}\p^cf_{bc}+f^{ab}\p_a^{\hph{a}c}f_{bc}\rs,\non\\
y_7'\p^a\lp f^{bc}\p_af_{bc}\rp &=& y_7'\ls\p^af^{bc}\p_af_{bc}+f^{ab}\p^c_{\hph{c}c}f_{ab}\rs,\non\\
y_8'\p^a\lp f^{bc}\p_bf_{ac}\rp &=& y_8'\ls\p^af^{bc}\p_bf_{ac}-f^{ab}\p_a^{\hph{a}c}f_{bc}\rs,\non\\
y_9'\p^a\lp f^{ij}\p_af_{ij}\rp &=& y_9'\ls\p^af^{ij}\p_af_{ij}+f^{ij}\p^a_{\hph{a}a}f_{ij}\rs,\non\\
y_{10}'\p^a\lp f^{ij}\p_if_{aj}\rp &=& y_{10}'\ls\p^af^{ij}\p_if_{aj}+f^{ij}\p^a_{\hph{a}i}f_{aj}\rs,\non
\eea
\bea
y_{11}'\p^a\lp\wtf_a^{\hph{a}i}\p^b\wtf_{bi}\rp &=& y_{11}'\ls\p^a\wtf_a^{\hph{a}i}\p^b\wtf_{bi}+\wtf^{ai}\p_a^{\hph{a}b}\wtf_{bi}\rs,\non\\
y_{12}'\p^a\lp\wtf^{bi}\p_a\wtf_{bi}\rp &=& y_{12}'\ls\p^a\wtf^{bi}\p_a\wtf_{bi}+\wtf^{ai}\p^b_{\hph{b}b}\wtf_{ai}\rs,\non\\
y_{13}'\p^a\lp\wtf^{bi}\p_b\wtf_{ai}\rp &=& y_{13}'\ls\p^a\wtf^{bi}\p_b\wtf_{ai}+\wtf^{ai}\p_a^{\hph{a}b}\wtf_{bi}\rs,\non\\
y_{14}'\p^a\lp\wtf^{bi}\p_i\wtf_{ab}\rp &=& y_{14}'\ls\p^a\wtf^{bi}\p_i\wtf_{ab}-\wtf^{ai}\p^b_{\hph{b}i}\wtf_{ab}\rs,\non
\eea
\bea
y_{15}'\p^a\lp\wtH_a^{\hph{a}bi}\p^c\wtH_{bci}\rp &=& y_{15}'\ls\p^a\wtH_a^{\hph{a}bi}\p^c\wtH_{bci}+\wtH^{abi}\p_a^{\hph{a}c}\wtH_{bci}\rs,\non\\
y_{16}'\p^a\lp\wtH^{bci}\p_a\wtH_{bci}\rp &=& y_{16}'\ls\p^a\wtH^{bci}\p_a\wtH_{bci}+\wtH^{abi}\p^c_{\hph{c}c}\wtH_{abi}\rs,\non\\
y_{17}'\p^a\lp\wtH^{bci}\p_b\wtH_{aci}\rp &=& y_{17}'\ls\p^a\wtH^{bci}\p_b\wtH_{aci}-\wtH^{abi}\p_a^{\hph{a}c}\wtH_{bci}\rs,\non\\
y_{18}'\p^a\lp\wtH^{bci}\p_i\wtH_{abc}\rp &=& y_{18}'\ls\p^a\wtH^{bci}\p_i\wtH_{abc}+\wtH^{abi}\p^c_{\hph{c}i}\wtH_{abc}\rs,\non\\
y_{19}'\p^a\lp\wtH^{ijk}\p_a\wtH_{ijk}\rp &=& y_{19}'\ls\p^a\wtH^{ijk}\p_a\wtH_{ijk}+\wtH^{ijk}\p^a_{\hph{a}a}\wtH_{ijk}\rs,\non\\
y_{20}'\p^a\lp\wtH^{ijk}\p_i\wtH_{ajk}\rp &=& y_{20}'\ls\p^a\wtH^{ijk}\p_i\wtH_{ajk}+\wtH^{ijk}\p^a_{\hph{a}i}\wtH_{ajk}\rs,\non
\eea

Demanding that the sum of all these pieces vanish gives a system of linear equations for the coefficients $c_i$, $x_i'$, and $y_i'$.  The solution involves an arbitrary choice for $c_7$, $x_4'$, $x_5'$, $x_{12}'$, $x_{18}'$, and $x_{19}'$, and then all other coefficients are fixed (we of course omit $c_i$ corresponding to terms with derivatives of the dilaton, as these drop out of the twisted product),
\be
c_5=c_6=c_9=c_{12}=c_{13}=c_{25}=c_{26}=c_{28}=c_{33}=c_{34}=c_{37}=c_{42}=c_{43}=c_{48}=0,\non
\ee
\be
x_2=x_3=x_6=x_{11}=x_{14}=x_{16}=x_{17}=x_{20}=x_{21}=y_{15}=0,\non
\ee
\be
c_8=-c_7,\quad c_{10}=-\frac{2}{3}c_7,\quad c_{11}=\hlf c_7,\quad c_{14}=\frac{1}{6}c_7,\quad c_{27}=-4c_7,\quad c_{29}=4c_7,\non
\ee
\be
c_{30}=8c_7,\quad c_{31}=-6c_7,\quad c_{32}=2c_7,\quad c_{38}=-8c_7,\quad c_{39}=4c_7,\quad c_{40}=-4c_7,
\ee
\be
c_{41}=8c_7,\quad c_{45}=-8c_7,\quad c_{46}=-2c_7,\quad c_{47}=-\frac{2}{3}c_7,\non
\ee
\be
y_1'=4c_7-2x_{18}',\quad y_2'=-2c_7-x_{18}',\quad y_3'=8c_7,\quad y_4'=-2c_7-3x_{19}',\quad y_5'=-8c_7,\non
\ee
\be
y_6'=-12c_7,\quad y_7'=-6c_7-x_4',\quad y_8'=2x_4',\quad y_9'=-x_5',\quad y_{10}'=2x_5',\non
\ee
\be
y_{11}'=-4c_7,\quad y_{12}'=4c_7-x_{12}',\quad y_{13}'=x_{12}',\quad y_{14}'=4c_7-x_{12}',\quad y_{16}'=-x_{18}',\non
\ee
\be
y_{17}'=2x_{18}',\quad y_{18}'=x_{18}',\quad y_{19}'=-x_{19}',\quad y_{20}'=3x_{19}',\non
\ee
\be
x_1'=4c_7,\quad x_7'=6x_7+x_4',\quad x_8'=x_5',\quad x_9'=8c_7,\quad x_{10}'=4c_7,\non
\ee
\be
x_{13}'=-2c_7,\quad x_{15}'=4c_7-x_{12}',\quad x_{22}'=x_{18}',\quad x_{23}'=x_{19}'.\non
\ee

\newpage
\providecommand{\href}[2]{#2}\begingroup\raggedright\endgroup


\begin{thebibliography}{10}

\bibitem{Becker:1996gj}
K.~Becker and M.~Becker, ``{M theory on eight manifolds},''
  \href{http://dx.doi.org/10.1016/0550-3213(96)00367-7}{{\em Nucl.Phys.} {\bf
  B477} (1996)  155--167}, \href{http://arxiv.org/abs/hep-th/9605053}{{\tt
  arXiv:hep-th/9605053 [hep-th]}}.

\bibitem{Dasgupta:1999ss}
K.~Dasgupta, G.~Rajesh, and S.~Sethi, ``{M theory, orientifolds and G -
  flux},'' {\em JHEP} {\bf 9908} (1999)  023,
  \href{http://arxiv.org/abs/hep-th/9908088}{{\tt arXiv:hep-th/9908088
  [hep-th]}}.

\bibitem{McOrist:2012yc}
J.~McOrist and S.~Sethi, ``{M-theory and Type IIA Flux Compactifications},''
  \href{http://dx.doi.org/10.1007/JHEP12(2012)122}{{\em JHEP} {\bf 1212} (2012)
   122},
\href{http://arxiv.org/abs/1208.0261}{{\tt arXiv:1208.0261 [hep-th]}}.

\bibitem{Maxfield:2013wka}
T.~Maxfield, J.~McOrist, D.~Robbins, and S.~Sethi, ``{New Examples of Flux
  Vacua},'' \href{http://dx.doi.org/10.1007/JHEP12(2013)032}{{\em JHEP} {\bf
  1312} (2013)  032},
\href{http://arxiv.org/abs/1309.2577}{{\tt arXiv:1309.2577 [hep-th]}}.

\bibitem{MMQRSinprogress}
T.~Maxfield, J.~McOrist, C.~Quigley, D.~Robbins, and S.~Sethi work in progress.

\bibitem{Green:1996dd}
M.~B. Green, J.~A. Harvey, and G.~W. Moore, ``{I-brane inflow and anomalous
  couplings on D-branes},''
  \href{http://dx.doi.org/10.1088/0264-9381/14/1/008}{{\em Class. Quant. Grav.}
  {\bf 14} (1997)  47--52},
\href{http://arxiv.org/abs/hep-th/9605033}{{\tt arXiv:hep-th/9605033}}.

\bibitem{Cheung:1997az}
Y.-K.~E. Cheung and Z.~Yin, ``{Anomalies, branes, and currents},''
  \href{http://dx.doi.org/10.1016/S0550-3213(98)00115-1}{{\em Nucl. Phys.} {\bf
  B517} (1998)  69--91},
\href{http://arxiv.org/abs/hep-th/9710206}{{\tt arXiv:hep-th/9710206}}.

\bibitem{Scrucca:1999uz}
C.~A. Scrucca and M.~Serone, ``{Anomalies and inflow on D-branes and O -
  planes},'' \href{http://dx.doi.org/10.1016/S0550-3213(99)00357-0}{{\em
  Nucl.Phys.} {\bf B556} (1999)  197--221},
  \href{http://arxiv.org/abs/hep-th/9903145}{{\tt arXiv:hep-th/9903145
  [hep-th]}}.

\bibitem{Minasian:1997mm}
R.~Minasian and G.~W. Moore, ``{K-theory and Ramond-Ramond charge},'' {\em
  JHEP} {\bf 11} (1997)  002,
\href{http://arxiv.org/abs/hep-th/9710230}{{\tt arXiv:hep-th/9710230}}.

\bibitem{Stefanski:1998yx}
J.~Stefanski, Bogdan, ``{Gravitational couplings of D-branes and O-planes},''
  \href{http://dx.doi.org/10.1016/S0550-3213(99)00147-9}{{\em Nucl.Phys.} {\bf
  B548} (1999)  275--290}, \href{http://arxiv.org/abs/hep-th/9812088}{{\tt
  arXiv:hep-th/9812088 [hep-th]}}.

\bibitem{Craps:1998fn}
B.~Craps and F.~Roose, ``{Anomalous D-brane and orientifold couplings from the
  boundary state},''
  \href{http://dx.doi.org/10.1016/S0370-2693(98)01438-5}{{\em Phys.Lett.} {\bf
  B445} (1998)  150--159}, \href{http://arxiv.org/abs/hep-th/9808074}{{\tt
  arXiv:hep-th/9808074 [hep-th]}}.

\bibitem{Becker:2010ij}
K.~Becker, G.~Guo, and D.~Robbins, ``{Higher derivative brane couplings from
  T-duality},'' \href{http://dx.doi.org/10.1007/JHEP09(2010)029}{{\em JHEP}
  {\bf 1009} (2010)  029}, \href{http://arxiv.org/abs/1007.0441}{{\tt
  arXiv:1007.0441 [hep-th]}}.

\bibitem{Godazgar:2013bja}
H.~Godazgar and M.~Godazgar, ``{Duality completion of higher derivative
  corrections},'' \href{http://dx.doi.org/10.1007/JHEP09(2013)140}{{\em JHEP}
  {\bf 1309} (2013)  140},
\href{http://arxiv.org/abs/1306.4918}{{\tt arXiv:1306.4918 [hep-th]}}.

\bibitem{Meissner:1996sa}
K.~A. Meissner, ``{Symmetries of higher order string gravity actions},''
  \href{http://dx.doi.org/10.1016/S0370-2693(96)01556-0}{{\em Phys.Lett.} {\bf
  B392} (1997)  298--304},
\href{http://arxiv.org/abs/hep-th/9610131}{{\tt arXiv:hep-th/9610131
  [hep-th]}}.

\bibitem{Hohm:2013jaa}
O.~Hohm, W.~Siegel, and B.~Zwiebach, ``{Doubled $\alpha'$-Geometry},''
\href{http://arxiv.org/abs/1306.2970}{{\tt arXiv:1306.2970 [hep-th]}}.

\bibitem{Garousi:2012yr}
M.~R. Garousi, ``{T-duality of the Riemann curvature corrections to
  supergravity},'' \href{http://dx.doi.org/10.1016/j.physletb.2012.12.012}{{\em
  Phys.Lett.} {\bf B718} (2013)  1481--1488},
\href{http://arxiv.org/abs/1208.4459}{{\tt arXiv:1208.4459 [hep-th]}}.

\bibitem{Liu:2013dna}
J.~T. Liu and R.~Minasian, ``{Higher-derivative couplings in string theory:
  dualities and the B-field},''
\href{http://arxiv.org/abs/1304.3137}{{\tt arXiv:1304.3137 [hep-th]}}.

\bibitem{Garousi:2009dj}
M.~R. Garousi, ``{T-duality of curvature terms in D-brane actions},''
  \href{http://dx.doi.org/10.1007/JHEP02(2010)002}{{\em JHEP} {\bf 1002} (2010)
   002}, \href{http://arxiv.org/abs/0911.0255}{{\tt arXiv:0911.0255 [hep-th]}}.

\bibitem{Garousi:2010rn}
M.~R. Garousi, ``{T-duality of anomalous Chern-Simons couplings},''
  \href{http://dx.doi.org/10.1016/j.nuclphysb.2011.06.019}{{\em Nucl.Phys.}
  {\bf B852} (2011)  320--335},
\href{http://arxiv.org/abs/1007.2118}{{\tt arXiv:1007.2118 [hep-th]}}.

\bibitem{Becker:2011ar}
K.~Becker, G.~Guo, and D.~Robbins, ``{Four-Derivative Brane Couplings from
  String Amplitudes},'' \href{http://dx.doi.org/10.1007/JHEP12(2011)050}{{\em
  JHEP} {\bf 1112} (2011)  050},
\href{http://arxiv.org/abs/1110.3831}{{\tt arXiv:1110.3831 [hep-th]}}.

\bibitem{Garousi:2013gea}
M.~R. Garousi, A.~Ghodsi, T.~Houri, and G.~Jafari, ``{T-duality of D-brane
  action at order $\alpha'$ in bosonic string theory},''
  \href{http://dx.doi.org/10.1007/JHEP10(2013)103}{{\em JHEP} {\bf 1310} (2013)
   103},
\href{http://arxiv.org/abs/1308.4609}{{\tt arXiv:1308.4609 [hep-th]}}.

\bibitem{RWinprogress}
D.~Robbins and Z.~Wang work in progress.

\bibitem{Buscher:1987sk}
T.~H. Buscher, ``{A symmetry of the string background field equations},''
\href{http://dx.doi.org/10.1016/0370-2693(87)90769-6}{{\em Phys. Lett.} {\bf
  B194} (1987)  59}.

\bibitem{Bachas:1999um}
C.~P. Bachas, P.~Bain, and M.~B. Green, ``{Curvature terms in D-brane actions
  and their M-theory origin},'' {\em JHEP} {\bf 05} (1999)  011,
\href{http://arxiv.org/abs/hep-th/9903210}{{\tt arXiv:hep-th/9903210}}.

\bibitem{Garousi:1996ad}
M.~R. Garousi and R.~C. Myers, ``{Superstring scattering from D-branes},''
  \href{http://dx.doi.org/10.1016/0550-3213(96)00316-1}{{\em Nucl.Phys.} {\bf
  B475} (1996)  193--224}, \href{http://arxiv.org/abs/hep-th/9603194}{{\tt
  arXiv:hep-th/9603194 [hep-th]}}.

\bibitem{Becker:2011bw}
K.~Becker, G.-Y. Guo, and D.~Robbins, ``{Disc amplitudes, picture changing and
  space-time actions},'' \href{http://dx.doi.org/10.1007/JHEP01(2012)127}{{\em
  JHEP} {\bf 1201} (2012)  127},
\href{http://arxiv.org/abs/1106.3307}{{\tt arXiv:1106.3307 [hep-th]}}.

\end{thebibliography}



\end{document}